\documentclass[a4paper, 12pt]{iopart}

\usepackage[dvips]{graphicx}	
\usepackage{subfigure}
\expandafter\let\csname equation*\endcsname\relax
\expandafter\let\csname endequation*\endcsname\relax

\usepackage{amsmath}
\usepackage{iopams}
\usepackage{booktabs}
\usepackage{soul}
\usepackage{color}
\usepackage[bottom]{footmisc} 
\usepackage{hyperref}
\usepackage{mciteplus}
\usepackage{tabularx}
\usepackage{accents}
\usepackage{fancyhdr}
\usepackage{cite}
\usepackage{threeparttable}

\renewcommand{\sectionmark}[1]{\markright{\thesection\quad #1}{} }

\usepackage[margin=1in]{geometry}
\usepackage[framemethod=TikZ]{mdframed}

\mdfdefinestyle{MyFrame}{%
    linecolor=gray,
    outerlinewidth=0.4pt,
    roundcorner=4pt,
    innertopmargin=0.5\baselineskip,
    innerbottommargin=0.5\baselineskip,
    innerrightmargin=\parindent,
    innerleftmargin=\parindent,
    backgroundcolor=gray!05!white}

\hypersetup{colorlinks=true, linkcolor=black, citecolor=blue, urlcolor=blue}

\input{paperdef.tex}

\graphicspath{{figs/}}

\newcommand{\cf}{cf.~}
\newcommand{\mobs}{\hat m}
\newcommand{\mobsvec}{\mathbf{\hat m}}
\newcommand{\mvec}{\mathbf{m}}
\newcommand{\muobs}{\hat \mu}
\newcommand{\muobsvec}{\boldsymbol{\hat\mu}}
\newcommand{\muvec}{\boldsymbol{\mu}}
\newcommand{\dmuobs}{\Delta \hat \mu}
\newcommand{\muobssm}{\hat \mu^\mathrm{conv}}
\newcommand{\dmuobssm}{\Delta \hat \mu^\mathrm{conv}}
\newcommand{\mpeak}{\hat m_\alpha}

\newcommand{\dm}{\Delta m}
\newcommand{\dmth}{\Delta m}
\newcommand{\dmexp}{\Delta \hat{m}}
\newcommand{\mpred}{m}

\newcommand{\cmu}{\chi_\mathrm{\mu}}
\newcommand{\cmua}{\chi_{\mathrm{\mu},\alpha}}

\newcommand{\cbox}[1]{\\[.3em]\centerline{\tt #1}\\[.3em]}
\newcommand{\subroutine}[2]{\\ \begin{mdframed}[style=MyFrame]\small\begin{tabularx}{\linewidth}{r@{}X} \texttt{#1} & \texttt{#2} \end{tabularx}\end{mdframed}}
\newcommand{\mHlow}{\mathrm{low-}M_H}

\newcommand{\mhmodp}{m_h^{\mathrm{mod}+}}
\newcommand{\ATLAS}{{ATLAS}}
\newcommand{\CMS}{{CMS}}

\newlength{\dhatheight}
\newcommand{\doublehat}[1]{%
    \settoheight{\dhatheight}{\ensuremath{\hat{#1}}}%
    \addtolength{\dhatheight}{-0.35ex}%
    \hat{\vphantom{\rule{1pt}{\dhatheight}}%
    \smash{\hat{#1}}}}

\newcommand{\citeHiggsall}{\cite{Heinemeyer:2011aa, Giardino:2012ww, *Azatov:2012wq, *Carmi:2012zd, *Low:2012rj, *Corbett:2012dm, *Giardino:2012dp, *Ellis:2012hz,*Espinosa:2012im,*Carmi:2012in,*Banerjee:2012xc, *Bonnet:2012nm,*Dobrescu:2012td, *Cacciapaglia:2012wb,*Corbett:2012ja,*Masso:2012eq,*Azatov:2012qz,*Belanger:2012gc,*Cheung:2013kla,*Belanger:2013kya,*Falkowski:2013dza,*Alanne:2013dra,*Giardino:2013bma,*Djouadi:2013qya,*Chang:2013cia,*Dumont:2013wma, Djouadi:2012rh,Lafaye:2009vr, *Klute:2012pu, *Plehn:2012iz,Carmi:2012yp, *Azatov:2012bz, *Espinosa:2012ir,*Montull:2012ik, Altmannshofer:2012ar, *Chang:2012ve,*Celis:2013rcs, *Enberg:2013ara,*Coleppa:2013dya, Benbrik:2012rm,*Espinosa:2012in,*Cao:2012yn,*Arbey:2012bp,*Arbey:2013jla,*Scopel:2013bba,*Moretti:2013lya,Drees:2012fb, *Bechtle:2012jw, Han:2013ic, *Cao:2013wqa}}
\newcommand{\citeHiggsEffC}{\cite{Giardino:2012ww, *Azatov:2012wq, *Carmi:2012zd, *Low:2012rj, *Corbett:2012dm, *Giardino:2012dp, *Ellis:2012hz,*Espinosa:2012im,*Carmi:2012in,*Banerjee:2012xc, *Bonnet:2012nm,*Dobrescu:2012td, *Cacciapaglia:2012wb,*Corbett:2012ja,*Masso:2012eq,*Azatov:2012qz,*Belanger:2012gc,*Cheung:2013kla,*Belanger:2013kya,*Falkowski:2013dza,*Djouadi:2013qya,*Chang:2013cia,*Dumont:2013wma,*Alanne:2013dra, *Giardino:2013bma, Djouadi:2012rh, Lafaye:2009vr, *Klute:2012pu, *Plehn:2012iz}}
\newcommand{\citeCompHiggsAndEffC}{\cite{Carmi:2012yp, *Azatov:2012bz, *Espinosa:2012ir,*Montull:2012ik}}
\newcommand{\citeHiggsSUSY}{\cite{Benbrik:2012rm,*Espinosa:2012in,*Cao:2012yn,*Arbey:2012bp,*Arbey:2013jla,*Scopel:2013bba,*Moretti:2013lya,Drees:2012fb,*Bechtle:2012jw}}
\newcommand{\citeheavyH}{\cite{Heinemeyer:2011aa, Drees:2012fb, *Bechtle:2012jw}}
\newcommand{\citeTHDM}{\cite{Altmannshofer:2012ar, *Chang:2012ve,*Celis:2013rcs,*Enberg:2013ara,*Coleppa:2013dya}}
\newcommand{\citeSMextensions}{\cite{Han:2013ic,*Cao:2013wqa}}

\begin{document}

\preprint{BONN-TH-2013-07\\ DESY 13-078}
\title[\HS\ User Manual]{\HS: Confronting arbitrary Higgs sectors with measurements at the Tevatron and the LHC}
\author{Philip Bechtle$^1$, Sven Heinemeyer$^2$, Oscar St\r{a}l$^3$, Tim Stefaniak$^{1,4}$\\ and Georg Weiglein$^5$}
\address{$^1$ Physikalisches Institut der Universit\"at Bonn, Nu{\ss}allee 12, 53115 Bonn, Germany}
\address{$^2$ Instituto de F\'isica de Cantabria (CSIC-UC), Santander, Spain}
\address{$^3$ The Oskar Klein Centre, Department of Physics, Stockholm University, SE-106 91 Stockholm, Sweden}
\address{$^4$ Bethe Center for Theoretical Physics, University of Bonn, Nu{\ss}allee 12, 53115 Bonn, Germany}
\address{$^5$ Deutsches Elektronen-Synchrotron DESY, Notkestrasse 85, D-22607 Hamburg, Germany}
\eads{\mailto{bechtle@physik.uni-bonn.de}, \mailto{Sven.Heinemeyer@cern.ch}, \mailto{oscar.stal@fysik.su.se}, \mailto{tim@th.physik.uni-bonn.de}, \mailto{Georg.Weiglein@desy.de}}

\begin{abstract}

\texttt{HiggsSignals} is a \texttt{Fortran90} computer code that
allows to test the compatibility of Higgs sector predictions against
Higgs rates and masses measured at the LHC or the Tevatron. Arbitrary
models with any number of Higgs bosons can be investigated using a
model-independent input scheme based on \HB. The test
is based on the calculation of a $\chi^2$ measure from the predictions
and the measured Higgs rates and masses, with the ability of fully 
taking into account systematics and correlations for the signal rate
predictions, luminosity and Higgs mass predictions. It features two
complementary methods for the test. First, the peak-centered method,
in which each observable is defined by a Higgs signal rate measured at
a specific hypothetical Higgs mass, corresponding to a tentative Higgs
signal. Second, the mass-centered method, where the test is evaluated by
comparing the signal rate measurement to the theory prediction at the
Higgs mass predicted by the model. The program allows for the
simultaneous use of both methods, which is useful in testing models
with multiple Higgs bosons. The code automatically combines the
signal rates of multiple Higgs bosons if their signals cannot be
resolved by the experimental analysis. 
We compare results obtained with \HS\ to official ATLAS and CMS results
for various examples of Higgs property determinations and find very good
agreement. A few examples of \HS\ applications are provided, 
going beyond the scenarios investigated by the LHC collaborations. 
For models with more than one Higgs boson we recommend to use \HS\ and
\HB\ in parallel to exploit the full constraining power of Higgs search
exclusion limits and the measurements of the signal seen at
$m_H\approx125.5$~GeV.

\end{abstract}

\maketitle

\tableofcontents
\setcounter{footnote}{0}


\clearpage
\section{Introduction}
Searches for a Higgs boson~\cite{Higgs:1964ia,*Englert:1964et,*Higgs:1964pj,*Guralnik:1964eu,*Higgs:1966ev,*Kibble:1967sv} have been one of the driving factors behind
experimental particle physics over many years. Until recently, results
from these searches have always been in the form of exclusion limits,
where different Higgs mass hypotheses are rejected at a certain
confidence level (usually $95\%$) by the non-observation of any
signal. This has been the case for Standard Model (SM) Higgs searches at
LEP~\cite{Barate:2003sz}, the Tevatron \cite{TEVNPH:2012ab}, and (until
July 2012) also for the LHC experiments
\cite{Aad:2012an,*Chatrchyan:2012tx}. Limits have also been presented on
extended Higgs sectors in theories beyond the SM, where one prominent
example are the combined limits on the Higgs sector of the minimal
supersymmetric standard model (MSSM) from the LEP
experiments~\cite{Schael:2006cr,Abbiendi:2013hk}. To test the
predictions of models with arbitrary 
Higgs sectors consistently against all the available experimental data
on Higgs exclusion limits, we have presented the public tool \HB\ \cite{Bechtle:2011sb,*Bechtle:2008jh}, which
recently appeared in version \vers{4}{0}{0} \cite{Bechtle:2013wla,Bechtle:2013gu}. 

With the recent discovery of a new state---compatible with a SM Higgs
boson---by the LHC experiments \ATLAS\ \cite{ATLASDiscovery} and \CMS\ \cite{CMSDiscovery},
models with extended Higgs sectors are facing new constraints. It is no
longer sufficient to test for non-exclusion, but the model predictions
must be tested against the measured mass and
rates of the observed state, which contains more information. Testing
the model predictions of a Higgs sector with an arbitrary number of Higgs bosons against this Higgs
signal\footnote{Here, and in the following, the phrase \emph{Higgs signal} refers
to any hint or observation of a signal in the data of the Tevatron/LHC Higgs searches,
regardless of whether in reality this is due to the presence of a
Higgs boson. In fact, the user can directly define the Higgs signals, i.e. the signal strength at a given mass peak or as a function of Higgs 
masses, which should be considered as observables in \HS, see Sect.~\ref{Sect:expdata} for more details.} (and potentially against other signals of
additional Higgs states discovered in the future) is the purpose of a new public computer
program, \HS, which we present here. 

\HS\ is a \texttt{Fortran90/2003} code, which evaluates a $\chi^2$ measure to provide a quantitative
answer to the statistical question of how compatible the Higgs
  search data (measured signal strengths and masses) is with the model
  predictions. This $\chi^2$ value can be evaluated with two distinct
  methods, namely the \textit{peak-centered} and the
  \textit{mass-centered} $\chi^2$ method. In the \textit{peak-centered}
  $\chi^2$ method, the (neutral) Higgs signal rates and masses predicted
  by the model are tested against the various signal rate measurements
  published by the experimental collaborations for a fixed 
  Higgs mass hypothesis. This hypothetical Higgs mass is typically motivated by the
  signal ``peak'' observed in the channels with high mass resolution, \ie the
  searches for $H\to\gamma\gamma$ and $H\to ZZ^{(*)}\to 4\ell$. In this
  way, the model is tested \textit{at the mass
    position of the observed peak}. In the \textit{mass-centered} $\chi^2$ method on the other hand, \HS~tries
  to find for every neutral Higgs boson in the model the corresponding
  signal rate measurements, which are performed under the assumption of a
  Higgs boson mass equal to the predicted Higgs mass. Thus, the $\chi^2$
  is evaluated at the \textit{model-predicted mass position}. For this
  method to be applicable, the experimental measurements therefore have to be given for a
  certain mass range. 

The input from the user is given in the form of Higgs masses, production cross
sections, and decay rates in a format similar to that used in \HB.
 The experimental data from Tevatron and
LHC Higgs searches is provided with the program, so there is no need for the user to include these values
manually. However, it is possible for the user to modify or add to the data
at will.
Like \HB, the aim is to always keep \HS~updated with the latest experimental results. 

The usefulness of a generic code such as \HS~has become apparent in the
last year, given the intense work by theorists to use the new Higgs
measurements as constraints on the SM and theories for new 
physics~\citeHiggsall. With \HS, there now exists a public tool that can be used for both
model-independent and model-dependent studies of Higgs masses,
couplings, rates, etc.~in a consistent framework. The $\chi^2$ output of
\HS\ also makes it convenient to use it as direct input to global
fits, where a first example application can be found in Ref.~\cite{Bechtle:2013mda}.

This document serves both as an introduction to the physics and
statistical methods used by \HS~and as a technical manual for users of
the code. It is organized as follows. Sect.~\ref{sect:HiggsSearches}
contains a very brief review of Higgs searches at hadron colliders,
focusing on the published data which provides the key experimental
input for \HS\ and the corresponding theory predictions. In
Sect.~\ref{sect:Statistics} we present the \HS\ algorithms,
including the precise definitions of the two $\chi^2$ methods mentioned above. 
Sect.~\ref{sect:Manual} provides the technical
description (user manual) for how to use the code. We
discuss the performance of \HS~and validate with official fit results for Higgs coupling scaling factors from \ATLAS\ and \CMS\ in
Sect.~\ref{sect:Examples}. Furthermore, we give some examples
of fit results, which can be obtained by interpreting all presently
available Higgs measurements. We conclude in
Sect.~\ref{sect:Conclusions}. In the appendix, details are given on the
implementation of theory mass uncertainties in the mass-centered $\chi^2$ method.


\section{Higgs signals in collider searches}
\label{sect:HiggsSearches}

The experimental data used in \HS~is collected at hadron colliders, mainly the LHC, but there are also some complementary measurements from the
Tevatron collider. This will remain the case for the foreseeable future,
but the \HS\ methods can be easily extended to include data from, for instance, a
future $e^+e^-$ linear collider. In this section we give a very brief
review of Higgs searches at hadron colliders, focussing the description
on the experimental data that provides the basic input for \HS. For a more complete review see, e.g., Ref.~\cite{Djouadi:2005gi,*Djouadi:2005gj,*Dittmaier:2012nh}.

\begin{figure}
\subfigure[The best-fit signal strength $\muobs$ for the LHC Higgs process
$(pp)\to H\to ZZ^{(*)}\to 4\ell$, given as a function of the assumed Higgs
  mass $\mH$. The cyan band gives the $68\%~\mathrm{C.L.}$ uncertainty of the measurement.]
  {\includegraphics[width=0.42\textwidth]{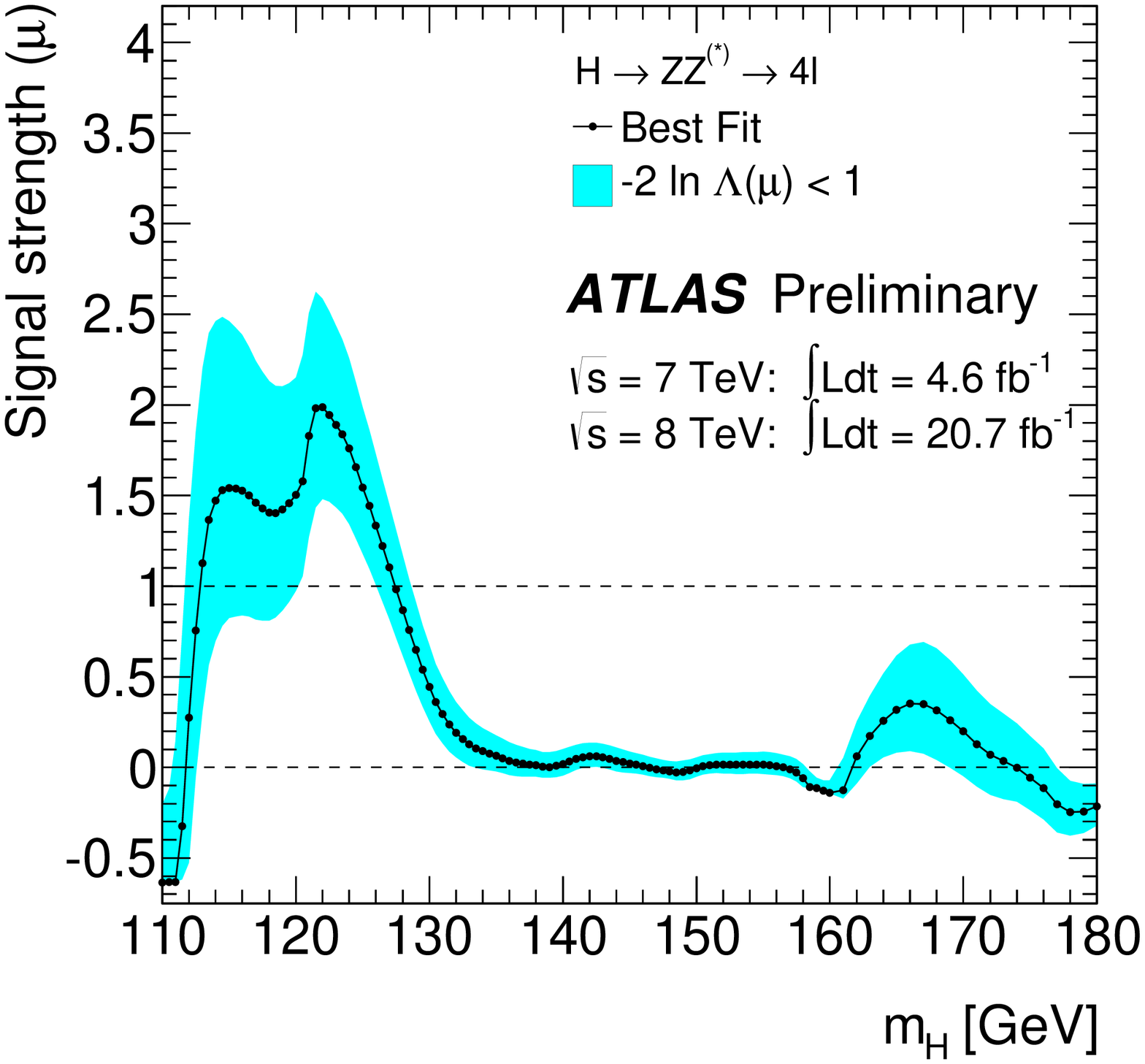}}\hfill 
\subfigure[The signal strength of various Higgs channels measured at
  a fixed hypothetical Higgs mass of $\mH=125.7\gev$. The combined signal
  strength scales all Higgs signal rates uniformly and is estimated to
  $\muobs_\mathrm{comb} = 0.80\pm 0.14$.]
  {\includegraphics[width=0.53\textwidth,clip, trim=-3cm 0cm -3cm 0cm]
  {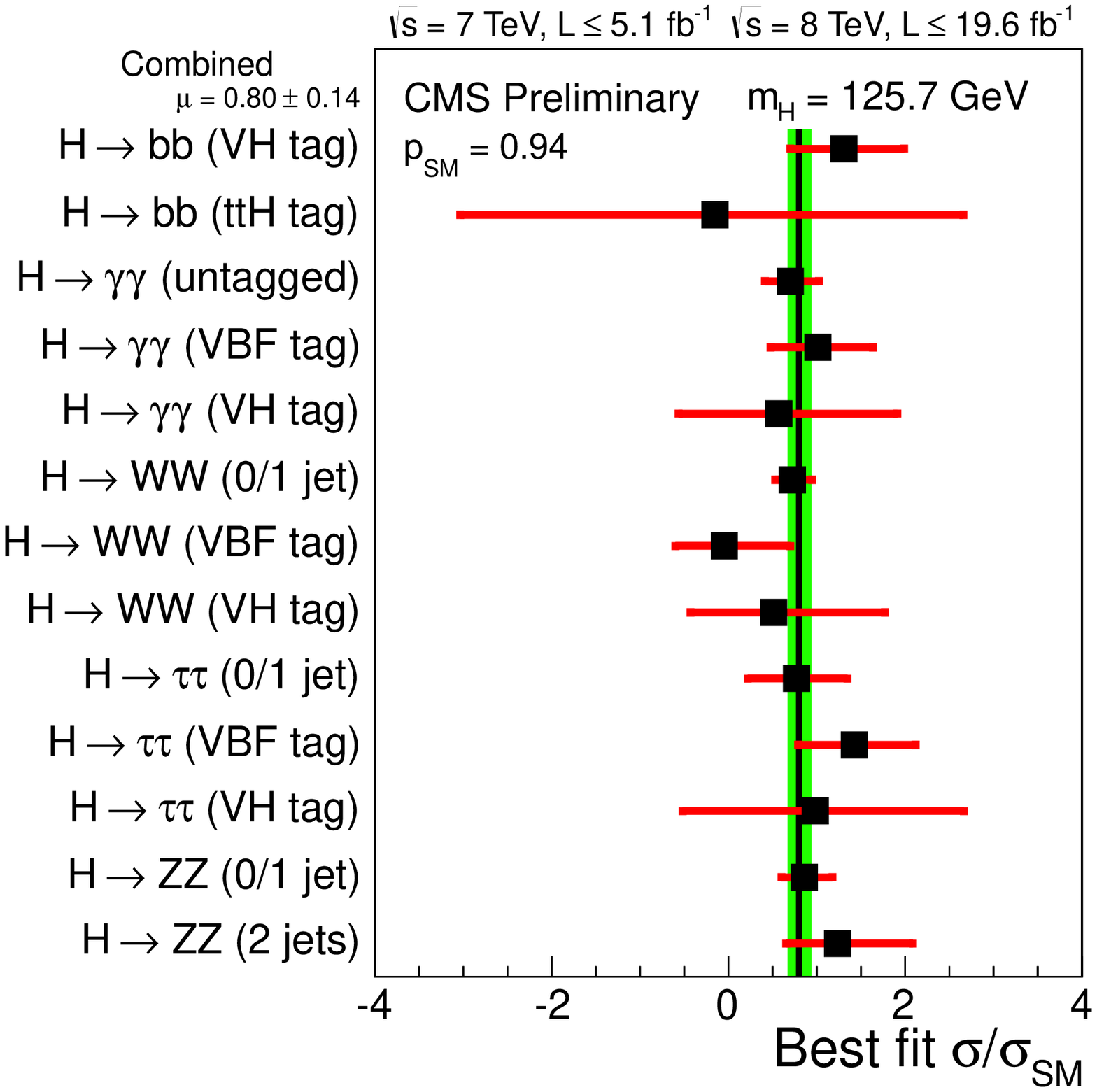}} 
\caption{Measured signal strength modifiers by \ATLAS\ in the search for
  $H\to ZZ^{(*)}\to 4\ell$ \cite{ATLAS-CONF-2013-013} (a), and the best fit rates
  (in all currently investigated Higgs decay channels) for a Higgs signal at $\mH=125.7\gev$
  according to CMS \cite{CMS-PAS-HIG-13-005} (b).} 
\label{fig:muplots}
\end{figure}

Most searches for Higgs bosons at the LHC are performed under the
assumption of the SM. This fixes completely the couplings of the Higgs
state to fermions and vector bosons, and both the cross sections and
branching ratios are fully specified as a function of the Higgs boson
mass, $\mH$. Most up-to-date predictions, including an extensive
list of references, can be found
in~\cite{Dittmaier:2011ti,Dittmaier:2012vm}.
This allows experiments to measure one-parameter scalings 
of the total SM rate of a certain (ensemble of) signal channel(s), so-called \emph{signal strength modifiers},
corresponding to the best fit to the data. These measurements are the
basic experimental input used by \HS. Two examples of this (from \ATLAS\ and \CMS) are shown in
Fig.~\ref{fig:muplots}. The left plot (taken from
\cite{ATLAS-CONF-2013-013}) shows the measured value of the signal
strength modifier, which we denote by $\muhat$, in the inclusive $pp\to~H\to ZZ^{(*)}\to 4\ell$
process as a function of $\mH$ (black line). The cyan band gives a
$\pm 1\,\sigma$ uncertainty on the measured rate. Since the signal
strength modifier is measured relative to its SM value ($\muhat=1$,
displayed in Fig.~\ref{fig:muplots} by a dashed line), this contains also
the theory uncertainties on the SM Higgs cross section and branching
ratios~\cite{Dittmaier:2011ti,Dittmaier:2012vm,Denner:2011mq}. As can be seen from Fig.~\ref{fig:muplots}, the measured value
of $\muhat$ is allowed to take on negative values. In the absence of
sizable signal-background interference---as is the case for the
SM---the signal model would not give $\muhat<0$. This must therefore be
understood as statistical downward fluctuations of the data w.r.t. the background expectation (the
average background-only expectation is $\muhat=0$). To keep $\muhat$ as
an unbiased estimator of the true signal strength, it is however
essential that the full range of values is retained. As we shall see in
more detail below, the applicability of \HS\ is limited to the mass
range for which measurements of $\muhat$ are reported. It is therefore
highly desirable that experiments publish this information even for
mass regions where a SM Higgs signal has been excluded. 

A second example of \HS\ input, this time from CMS, is shown in the right plot of
Fig.~\ref{fig:muplots} (from~\cite{CMS-PAS-HIG-13-005}). This figure
summarizes the measured signal strength modifiers for \emph{all}
relevant Higgs decay channels at an interesting value of the Higgs mass, here
$\mH=125.7\gev$. This particular value is typically
selected to correspond to the maximal significance for a signal seen in
the data. It is important to note that, once a value for
$\mH$ has been selected, this plot shows a compilation of  information
for the separate channels that is also available directly from the
mass-dependent plots (as shown in Fig.~\ref{fig:muplots}(a)). Again, the
error bars on the measured $\muhat$ values correspond to $1\sigma$
uncertainties that include both experimental (systematic and
statistical) uncertainties, as well as SM theory uncertainties. 

The idea of \HS\ is to compare the experimental measurements of signal
strength modifiers to the Higgs sector predictions in arbitrary
models. The model predictions must be provided by the user for each parameter point to be tested. 
To be able to do this consistently, we here
describe the basic definitions that we apply. The production of Higgs
bosons at hadron colliders can essentially proceed through five partonic
subprocesses: gluon fusion (ggf), vector boson fusion (vbf), associated
production with a gauge boson ($HW$/$HZ$), or associated
production with top quarks ($ttH$),
see~\cite{Dittmaier:2011ti,Dittmaier:2012vm} for details. 
In models with an
enhanced Higgs coupling to bottom quarks, the process $b\bar{b}\to H$ is
usually added. In this five-flavor scheme a $b$ quark parton
distribution describes the collinear gluon splitting to pairs of bottom
quarks inside the proton. This contribution should be matched
consistently, and in most cases, added to the gluon fusion
subprocess (as prescribed by the Santander matching procedure~\cite{Harlander:2011aa}). We therefore sometimes refer to the sum of the gluon
  fusion and $b\bar{b}\to H$ subprocesses as single Higgs production
  (singleH). Internally, \HS\ uses the same LHC cross sections for SM
Higgs production at $\sqrt{s}=7$ and $8\tev$ as \HBv{4}~\cite{Bechtle:2013wla}. The
same holds for the reference SM branching ratios, which follow the
prescription of the LHC Higgs Cross Section Working Group
\cite{Dittmaier:2011ti,Dittmaier:2012vm}, see
  also~\cite{Denner:2011mq} for more details. 
These branching ratios are the same as those used by the LHC experiments. 

The theory prediction for the signal strength modifier of one specific analysis,
from a single Higgs boson~$H$, is computed in \HS\ as 
\begin{equation}
\mu = \sum_{i} c_i\omega_i,
\label{Eq:mu}
\end{equation}
where the sum runs over all channels considered in this analysis. A
channel is characterized by one specific production \textit{and} one specific decay mode. The
individual \emph{channel signal strength} is given by 
\begin{equation}
c_i=\frac{\left[\sigma\times \mathrm{BR}\right]_i}{\left[\sigma_{\mathrm{SM}}\times \mathrm{BR}_{\mathrm{SM}}\right]_i},
\label{Eq:ci}
\end{equation}
and the \emph{SM channel weight} is
\begin{equation}
\omega_i=\frac{\epsilon_i\left[\sigma_\mathrm{SM}\times \mathrm{BR}_\mathrm{SM}\right]_i}{\sum_j\epsilon_j\left[\sigma_{\mathrm{SM}}\times \mathrm{BR}_{\mathrm{SM}}\right]_j}.
\label{Eq:omega}
\end{equation}
The SM weights contain the relative experimental \emph{efficiencies},
$\epsilon_i$, for the different channels. Unfortunately, these are
rarely quoted in experimental publications. If they are available, these numbers can be used by \HS, which leads to a
more reliable comparison between theory predictions and the
experimental data for these channels. In the case of unknown efficiencies, all channels considered by the analysis are treated equally, \ie we set all $\epsilon_i \equiv 1$. Note, however, that for many observables approximate numbers for the channel efficiencies can be inferred by reproducing official fit results on scale factors for production cross sections or coupling strengths, which will be further discussed in Section~\ref{Sec:ValidationOffEff}.

One final word of caution should be added here: If the model features a non-standard tensor structure for the particles, which should be confronted with the data, these interactions might lead to observable differences in the experimentally measured kinematic distributions and therefore to changes of the signal acceptance/efficiency of the Higgs analyses. In order to obtain reliable results from \HS\ for these types of models, one needs to check whether these effects are negligible. An interface for \HS, where the user can insert model signal efficiencies  for each analysis, which are changed with respect to the SM signal efficiencies, is a planned feature for future development. However, it is impossible to completely unfold this model dependence using only the currently available public information.


\section{Statistical approach in \HS}
\label{sect:Statistics}

As mentioned already in the introduction, \HS\ contains two different
statistical methods to test models against the experimental
data. These methods are complementary, and to provide a full model
test it is advisable in many situations to use both
simultaneously. Nevertheless, we leave the final choice of method to
the user, and we therefore first describe both methods separately,
before discussing their combination in Sect.~\ref{Sect:bothmethods}.

As already touched upon in the previous section, the search results of
\ATLAS\ and \CMS\ are reported in the form of the signal strength
modifier $\hat\mu$, the ratio of the best-fit signal strength to the
expected SM strength of a signal in a certain channel, and its
uncertainty $\Delta\hat\mu$. In the profile likelihood approach
\cite{Cowan:2010js} used by the experimental collaborations,
$\Delta\hat\mu$ is derived from the allowed variation of the signal
strength multiplier $\mu$ around the best fit value $\hat\mu$. This is
calculated using the likelihood ratio $\lambda(\mu)={\cal
  L}(\mu,\doublehat{\theta})/{\cal L}(\hat\mu,\hat\theta)$; the ratio
of the likelihood function ${\cal L}$ for a given $\mu$ with nuisance
parameters $\doublehat{\theta}$ optimized at the given value of $\mu$,
divided by ${\cal L}$ for $\hat\mu$ and $\hat\theta$ optimized
simultaneously (see \cite{Cowan:2010js} for more details).

The uncertainty of $\hat\mu$ is then calculated using a test
statistics based on $-2\ln\lambda(\mu)$. According to \cite{Wilks1938,
  Wald1943}, this can be expressed as
\begin{equation}
-2\ln\lambda(\mu)=\frac{(\mu-\hat\mu)^2}{\sigma^2}+{\cal
O}(1/\sqrt{N}),
\label{Eq:lnlambda}
\end{equation}
where $N$ is the data sample size.  Generally, as shown in
\cite{Cowan:2010js}, this converges quite quickly to a central or
non-central $\chi^2$ distribution, depending on the nuisance
parameters. If the test statistics follows a $\chi^2$ distribution,
the uncertainties of the measurement can generally be treated as
Gaussian, hence we interpret all uncertainties $\Delta\hat\mu$ as
Gaussian, and neglect the ${\cal O}(1/\sqrt{N})$ term. Looking at the
experimental results used in \HS\ and the available event sample sizes, this
is justified in almost all analyses, apart from $H\to ZZ^{*}$, where visible differences from the
  Gaussian approximation are still possible due to the small event sample
  size. The largest remaining effects of non-Gaussian distributions
  are taken into account in \HS\ by using asymmetric uncertainties on the measured signal strength in
  the $\chi^2$ calculation, if
  published as such by the collaborations.

While the $\chi^2$ calculated in \HS~can be expected to statistically
approximate the true $-2\ln\lambda$ distribution, \cf
Eq~\eqref{Eq:lnlambda}, there are three relevant experimental input
quantities which can systematically affect the accuracy of the
\HS~output in case they are not presented in a complete form in the 
publicly disclosed information: Firstly, the relative efficiencies
$\epsilon_i$ of the various Higgs channels/processes
considered in the (categories of a) Higgs
analysis, as introduced in Eq.~\eqref{Eq:omega}. Secondly, the correlations
of the relevant experimental systematic uncertainties (\eg of the jet
energy scale (JES), $e^\pm/\gamma$ identification and energy scale, tagging efficiencies, 
etc.) between different Higgs search analyses. 
 Thirdly, the use of continuous variables for
  classification of channels/production processes (e.g. by using multivariate techniques), which cannot be mapped
  directly onto signal strengths measurements for distinct categories used as experimental input for the $\chi^2$ fit in \HS. An example for this is the CMS $H\to ZZ^{*}\to 4\ell$ analysis~\cite{CMS-PAS-HIG-13-002}. The effects of such an approach and an approximate solution to this problem within \HS\ is
  discussed in Section~\ref{Sec:ValidationOffEff}.

While the signal efficiencies, $\epsilon_i$, could be provided straight-forwardly for every analysis as public information, the communication of the
  (correlated) systematics, both from experimental and theoretical sources, used in a given analysis is not common. However, within the Gaussian approximation these could in principle be taken into account~in \HS. 
 For the future it would be desirable if this information was provided in a model-independent way. Some ideas on how information on correlated systematic uncertainties in Higgs boson rate measurements could be communicated can be found in Ref.~\cite{CorrSystDoc}.
We discuss the possible impact of including this information in Section~\ref{Sec:ValidationOffEff} for a few relevant cases.
  
  The $\chi^2$ based approach in \HS\ could in principle be replaced
  by the use of likelihood curves from the collaborations, which are
  currently available in $(m_H,~\muobs)$ grids for a few analyses~\cite{CMS-PAS-HIG-13-001,CMS-PAS-HIG-13-005},
  albeit not for the categories individually. Once they are
  available for the majority of analyses and for every single
  (category of an) analysis, the $\chi^2$ could partly be replaced by the use of
  these likelihoods. However, significant modifications of the final
  likelihood by a tool like \HS\ would still be required to make it
  applicable to arbitrary Higgs sectors, due to potentially different signal compositions and hence changed theoretical rate uncertainties. Moreover, the necessity of incorporating correlated systematics, as mentioned above, remains also in this approach. Already with the currently available statistics the ignorance of efficiencies and correlations of experimental systematics are often the dominant effects for the typically small deviations between the official results by the collaborations and the \HS\ results. The assumption on the parabolic shape of the likelihood, on the other hand, has typically a relatively small impact. More details will be given in Section~\ref{Sec:ValidationOffEff}.


\subsection{The peak-centered $\chi^2$ method}
\label{Sect:pc_chisq}

The objective of this method is to perform a $\chi^2$ test for the
hypothesis that \textit{a local excess, ``signal'' (or ``peak
observable''), in the observed data at a specified mass is
generated by the model.} In short, this test tries to minimize the
total $\chi^2$ by assigning, to each Higgs signal in the experimental
dataset used, any number of Higgs bosons of the model. From each
signal, both the predicted signal strength modifiers and the
corresponding predicted Higgs masses (for channels with good mass
resolution) enter the total $\chi^2$ evaluation in a correlated way. Schematically, the total $\chi^2$ is given by
\begin{equation}
\chi^2_\mathrm{tot} = \chi^2_\mu + \sum_{i=1}^{N_H}\chi^2_{m_i},
\end{equation}
where $N_H$ is the number of (neutral) Higgs bosons of the model. The calculation of the individual contributions from the signal strength modifiers, $\chi^2_\mu$, and the Higgs masses, $\chi^2_{m_i}$, will be discussed below.

The input data used in this method is based on the prejudice that a Higgs signal has been observed at a particular Higgs mass value, which does not necessarily have to be the exact same value for all observables. Technically, each observable is defined by a single text file, which contains all relevant information needed by \HS. An experimental dataset\footnote{The most up-to-date experimental data is contained in the folder \texttt{Expt\_tables/latestresults}. A summary of these observables, as included in the \HSv{1.0.0} release, is given in \refse{sect:Examples}, \reffi{Fig:peakobservables}.} is then a collection of observables, whose text files are stored in a certain subdirectory of the \HS\ distribution. Users may add, modify or remove the experimental data for their own purposes, see Sect.~\ref{Sect:expdata} for more details.

Currently, an obvious and prominent application of the peak-centered $\chi^2$ method would be the test of a single Higgs boson against the rate and mass measurements performed at around $125$--$126$~GeV in all channels reported by the experimental collaborations at the LHC and Tevatron. This scenario will be discussed in detail in Sect.~\ref{sect:Examples}. However, \HS\ is implemented in a way that is much more general: Firstly, contributions from other Higgs bosons in the model to the Higgs signals will be considered, and if relevant, included in the test automatically. Secondly, the extension of this test to more Higgs signals (in other mass regions) can simply be achieved by the inclusion of the proper experimental data, or for a phenomenological study, the desired pseudo-data.

\subsubsection{Signal strength modifiers}
\label{Sect:chisq_mu}
For $N$ defined signal observables, the total $\chi^2$ contribution is given by
\begin{equation}
\cmu^2 = \sum_{\alpha=1}^{N} \cmua^2 = (\muobsvec - \muvec)^T \mathbf{C_\mu^{-1}} (\muobsvec - \muvec),
\label{Eq:chimu}
\end{equation}
where the observed and predicted signal strength modifiers are contained in the $N$-dimensional vectors $\muobsvec$ and $\muvec$, respectively. $\mathbf{C_\mu}$ is the signal strength covariance matrix.

The signal strength covariance matrix $\covmu$ is constructed in the
following way. The diagonal elements $(\covmu)_{\alpha\alpha}$
(corresponding to signal observable $\alpha$) should first of all contain the
intrinsic experimental (statistical and systematic) $1\,\sigma$
uncertainties on the signal strengths squared, denoted by $(\Delta
\hat\mu_\alpha^*)^2$. These will be treated as uncorrelated
uncertainties, since there is no information publicly available on their
correlations. We define these uncorrelated uncertainties by subtracting
from the total uncertainty $\Delta \hat\mu_\alpha$ (which is given
directly from the $1\,\sigma$ error band in the experimental data,
cf.~Fig.~\ref{fig:muplots}) the luminosity uncertainty as well as
the theory uncertainties on the predicted signal rate (which we
shall include later as correlated uncertainties). Hereby, we assume that these uncertainties can be treated as Gaussian errors. This gives 
\begin{equation}
(\Delta \hat\mu_\alpha^*)^2 = (\Delta \hat\mu_\alpha)^2 - (\Delta \mathcal{L} \cdot\hat\mu_\alpha)^2 - \sum_{a=1}^k (\omega_a^\alpha \Delta c_a^\mathrm{SM})^2 \cdot \hat\mu_\alpha^2.
\label{Eq:intrinsic_dmusq}
\end{equation}
Here, $\Delta \mathcal{L}$ is the relative uncertainty on the luminosity, and $\Delta c_a^\mathrm{SM}$ is the SM channel rate uncertainty (for a total of $k$ channels contributing to the analysis with signal $\alpha$) given by
\begin{equation}
(\Delta c_a^\mathrm{SM})^2 = (\Delta \sigma_a^\mathrm{SM})^2 + (\Delta \brat_a^\mathrm{SM})^2,
\end{equation}
where $\Delta \sigma_a^\mathrm{SM}$ and $\Delta \brat_a^\mathrm{SM}$ are
the relative systematic uncertainties of the production cross section
$\sigma_a$ and branching ratio $\brat_a$, respectively, of the channel $a$
in the SM. Their values are taken from the LHC Higgs Cross Section Working Group~\cite{Dittmaier:2011ti,Dittmaier:2012vm}, evaluated around $m_H\sim 125\gev$:
\begin{equation}
\begin{array}{l}
\Delta \sigma^\mathrm{SM}_\mathrm{ggf} = 14.7\%,\\
\Delta \sigma^\mathrm{SM}_\mathrm{VBF} = 2.8\%,\\ 
\Delta \sigma^\mathrm{SM}_\mathrm{WH} = 3.7\%,\\
\Delta \sigma^\mathrm{SM}_\mathrm{ZH} = 5.1\%, \\
\Delta \sigma^\mathrm{SM}_\mathrm{ttH} =12.0\%,
\end{array}
\qquad
\begin{array}{l}
\Delta \brat^\mathrm{SM}(H\to\gamma\gamma) = 5.4\%,\\
\Delta \brat^\mathrm{SM}(H\to WW) = 4.8\%,\\
\Delta \brat^\mathrm{SM}(H\to ZZ) = 4.8\%, \\
\Delta \brat^\mathrm{SM}(H\to \tau\tau) = 6.1\%,\\
\Delta \brat^\mathrm{SM}(H\to bb) = 2.8\%.
\end{array}
\label{Eq:SMrateuncertainties}
\end{equation}
The SM channel weights, $\omega_a$, have been defined in \refeq{Eq:omega}.

The advantage of extracting $(\Delta \hat\mu_\alpha^*)^2$ via
Eq.~\eqref{Eq:intrinsic_dmusq} over using the experimental values
$(\Delta \hat\mu_\alpha)^2$ directly is that it allows for the
correlations in the theory uncertainties on the different channel rates
to be taken into account. These are correlated to other signals which
use the same channels, and since we want to investigate other models
beyond the SM, the theory uncertainties on the channel rates are in
general different. The same applies for the relative luminosity
uncertainties, which can usually be taken equal for all analyses within
one collaboration, thus leading to manageable correlations in the signal
strength modifiers. 

In the next step, we insert these correlated uncertainties into the
covariance matrix. To each matrix element
$(\mathbf{C_\mu})_{\alpha\beta}$, including the diagonal, we add a term
$(\Delta \mathcal{L}_\alpha \hat\mu_\alpha)(\Delta \mathcal{L}_\beta
\hat\mu_\beta)$ if the signals $\alpha$ and $\beta$ are observed in
analyses from the same collaboration (note that usually the further
simplification $\Delta \mathcal{L}_\alpha = \Delta \mathcal{L}_\beta$
applies in this case). We then add the correlated theory uncertainties
of the signal rates, given by 
\begin{equation}
\left(\sum_{a=1}^{k_\alpha}\sum_{b=1}^{k_\beta} \left[ \delta_{p(a)p(b)}
  \Delta \sigma_{p(a)}^\mathrm{model}\Delta \sigma_{p(b)}^\mathrm{model}
  +  \delta_{d(a)d(b)} \Delta \brat_{d(a)}^\mathrm{model}\Delta
  \brat_{d(b)}^\mathrm{model} \right] \cdot  \omega_{a,\alpha}^\mathrm{model} \omega_{b,\beta}^\mathrm{model} \right) \mu_\alpha \mu_\beta.
\end{equation}
Here, $k_\alpha$ and $k_\beta$ are the respective numbers of Higgs
(production~$\times$~decay) channels considered in the experimental
analyses where the signals $\alpha$ and $\beta$ are observed. We use the
index notation $p(a)$ and $d(a)$, to map the channel $a$ onto its
production and decay processes, respectively. In other words, analyses
where the signals share a common production and/or decay mode have
correlated systematic uncertainties. These channel rate uncertainties are inserted in the covariance matrix according to their relative contributions to the total signal rate \textit{in the model}, i.e. via the channel weight evaluated from the model predictions,
\begin{equation}
\omega_i^\mathrm{model}=\frac{\epsilon_i\left[\sigma\times \mathrm{BR}\right]_i}{\sum_j\epsilon_j\left[\sigma\times \mathrm{BR}\right]_j}.
\label{Eq:omegamodel}
\end{equation}

If the theory uncertainties on the Higgs production and decay rates, as well as the channel weights of
the model under investigation, are equal to those in the SM, and also the predicted signal strength matches with the observed signal strength, the
uncertainties $(\Delta \hat\mu_\alpha)^2$ extracted from the
experimental data are exactly restored for the diagonal elements
$(\covmu)_{\alpha\alpha}$, \cf Eq.~\eqref{Eq:intrinsic_dmusq}.  
Finally, it is worth emphasizing again that this procedure only takes
into account the correlations of the luminosity and theoretical signal
rate uncertainties, whereas correlations between common experimental
uncertainties (energy scale uncertainties, etc.) are neglected. Since
this information is not publicly available so far, it could not be
included in \HS. 


\subsubsection{Higgs mass observables}\label{Sec:HiggsMassObs}

The other type of observables that give
contributions to the total $\chi^2$ in the peak-centered method is the
measured masses corresponding to the observed signals. Not all signals
come with a mass measurement; this is something which is specified
explicitly in the experimental input data. In general, a Higgs boson
in the model that is not \emph{assigned} to a signal (see below for
the precise definition), receives a zero $\chi^2$ contribution from
this signal. This would be the case, for example, for multiple Higgs
bosons that are not close in mass to the observed signal. 

\HS~allows the probability density function (pdf) for the Higgs boson masses to
be modeled either as a uniform distribution (box), as a Gaussian, or
as a box with Gaussian tails. In the Gaussian case, a full correlation
in the theory mass uncertainty is taken into account for a Higgs boson
that is considered as an explanation for two (or more) signal observables (which include a mass measurement).

Assume that a signal $\alpha$ is observed at the
mass $\mpeak$, and that a Higgs boson $h_i$ with a predicted mass
$\mpred_i$ (potentially with a theory uncertainty $\dmth_i$), is
assigned to this signal. Its $\chi^2$ contribution is then simply
given by
\begin{equation}
\chi_{m_{i},\alpha}^2 =
\left\{
\begin{array}{ll}
0 & \mbox{, for}~|\mpred_i - \mpeak| \le \dm_i,\\
\infty & \mbox{, otherwise}
\end{array}
\right.
~\quad\mbox{with}~\quad \dm_i = \dmth_i + \dmexp_\alpha,
\label{Eq:boxpdf}
\end{equation}
for a uniform (box) mass pdf, and
\begin{equation}
\chi_{m_{H,i},\alpha}^2 =
\left\{
\begin{array}{ll}
0 & \mbox{, for}~|\mpred_i - \mpeak| \le \dmth_i,\\
(\mpred_i - \dmth_i - \mpeak)^2/(\dmexp_\alpha)^2 & \mbox{, for}~\mpred_i - \dmth_i < \mpeak,\\
(\mpred_i + \dmth_i - \mpeak)^2/(\dmexp_\alpha)^2 & \mbox{, for}~\mpred_i + \dmth_i > \mpeak,
\end{array}
\right.
\end{equation}
for a box-shaped pdf with Gaussian tails. Here, we denote the
experimental uncertainty of the mass measurement of the analysis associated to signal
$\alpha$ by $\dmexp_\alpha$.
The use of a box-shaped mass pdf, Eq.~\eqref{Eq:boxpdf}, is not recommended in situations where the theory mass uncertainty is small compared to the experimental precision of the mass measurement (and in particular when $\Delta m_i = 0$), since this can lead to overly restrictive results in the assignment of the Higgs boson(s) to high-resolution channels. Moreover, a box-shaped pdf is typically not a good description of the experimental uncertainty of a mass measurement in general. We included this option mostly for illustrational purposes.

In the case of a Gaussian mass pdf the $\chi^2$
calculation is performed in a similar way as the calculation of
$\chi^2_\mu$ in Eq.~\eqref{Eq:chimu}. We define for each Higgs boson
$h_i$ 
\begin{equation}
\chi_{m_{i}}^2 = \sum_{\alpha=1}^{N} \chi_{m_{i},\alpha}^2 = (\mobsvec - \mvec_i)^T \mathbf{C_{m_{i}}^{-1}} (\mobsvec - \mvec_i),
\label{Eq:chisq_mh}
\end{equation}
where the $\alpha$-th entry of the predicted mass vector $\mvec_i$ is
given by $m_i$, if the of Higgs boson $h_i$ is assigned to the signal $\alpha$, or $\mobs_\alpha$ otherwise (thus leading to a zero $\chi^2$ contribution from \textit{this} observable and \textit{this} Higgs boson).
 As can be seen from Eq.~\eqref{Eq:chisq_mh}, we construct a mass covariance matrix
$\mathbf{C_{m_{i}}}$ for each Higgs boson $h_i$ in the model. The diagonal
elements $(\mathbf{C_{m_{i}}})_{\alpha\alpha}$ contain the experimental
mass resolution squared, $(\dmexp_\alpha)^2$, of the analysis in which
the signal $\alpha$ is observed. The squared theory mass uncertainty,
$(\dmth_i )^2$, enters all matrix elements
$(\mathbf{C_{m_{i}}})_{\alpha\beta}$ (including the diagonal)
where the Higgs boson $h_i$ is assigned to both signal observables $\alpha$ and
$\beta$. Thus, the theoretical mass uncertainty is treated as fully correlated.

The sign of this correlation depends on the relative position of the predicted Higgs boson mass, $m_i$, with respect to the two (different) observed mass values, $\hat{m}_{\alpha,\beta}$ (where we assume $\hat{m}_\alpha < \hat{m}_\beta$ for the following discussion): If the predicted mass lies outside the two measurements, i.e. $m_i < \hat{m}_\alpha, \hat{m}_\beta$ or $m_i > \hat{m}_\alpha, \hat{m}_\beta$, then the
correlation is assumed to be positive. If it lies in between the two mass measurements, $\hat{m}_\alpha < m_i < \hat{m}_\beta$, the correlation is negative (i.e. we have anti-correlated observables). The necessity of this sign dependence can be illustrated as follows: Let us assume the predicted Higgs mass is varied within its theoretical uncertainty. In the first case, the deviations of $m_i$ from the mass measurements $\hat{m}_{\alpha,\beta}$ both either increase or decrease (depending on the direction of the mass variation). Thus, the mass measurements are positively correlated. However, in the latter case, a variation of $m_i$ towards one mass measurement always corresponds to a larger deviation of $m_i$ from the other mass measurements. Therefore, the theoretical mass uncertainties for these observables have to be anti-correlated.

\subsubsection{Assignment of multiple Higgs bosons}
\label{Sect:peakassignment}

If a model contains an extended (neutral) Higgs sector, it is a priori not clear which Higgs boson(s) give the best explanation of the experimental observations. Moreover, possible superpositions of the signal strengths of the Higgs bosons have to be taken into account. Another (yet hypothetical) complication arises if \emph{more than one} Higgs signal has been discovered in the \emph{same} Higgs search, indicating the discovery of \emph{another} Higgs boson. In this case, care has to be taken that a Higgs boson of the model is only considered as an explanation of one of these signals.

In the peak-centered $\chi^2$ method, these complications are taken into account by the automatic \emph{assignment} of the Higgs bosons in the model to the signal observables.
In this procedure, \HS~tests whether the combined signal strength of several Higgs bosons might yield a better fit than the assignment of a single Higgs boson to one signal in an analysis. Moreover, based on the predicted and observed Higgs mass values, as well as their uncertainties, the program decides whether a comparison of the predicted and observed signal rates is valid for the considered Higgs boson. A priori, all possible Higgs combinations which can be assigned to the observed signal(s) of an analysis are considered. If more than one signal exists in \textit{one} analysis, it is taken care of that each Higgs boson is assigned to at most one signal to avoid double-counting. A signal to which no Higgs boson is assigned contributes a $\chi^2$ penalty given by Eq.~\eqref{Eq:chimu} with the corresponding model prediction ${\mu_\alpha=0}$. This corresponds to the case where an observed signal cannot be explained by any of the Higgs bosons in the model.

For each Higgs search analysis the best Higgs boson assignment is found in the following way: For every possible assignment $\eta$ of a Higgs boson combination to the signal $\alpha$ observed in the analysis, its corresponding tentative $\chi^2$ contribution, $\chi_{\alpha,\eta}^2$, based on both the signal strength and potentially the Higgs mass measurement, is evaluated. In order to be considered for the assignment, the Higgs combination has to fulfill the following requirements:
\begin{itemize}
\item Higgs bosons which have a mass $m_i$ close enough to the signal mass $\hat{m}_\alpha$, \ie
\begin{equation}
| m_i  - \mpeak | \le \Lambda  \sqrt{(\dmth_i)^2 + (\dmexp_\alpha)^2},
\label{Eq:massoverlap}
\end{equation}
are required to be assigned to the signal $\alpha$. Here, $\Lambda$ denotes the \textit{assignment range}, which can be modified by the user, see Sect.~\ref{Sect:subroutines} (the default setting is $\Lambda=1$).
\item If the $\chi^2$ contribution from the measured Higgs mass is
  \textit{deactivated} for this signal, combinations with a Higgs boson that 
  fulfills Eq.~\eqref{Eq:massoverlap} are taken into account for
  a possible assignment, and not taken into account otherwise.
\item If the $\chi^2$ contribution from the measured Higgs mass is \textit{activated}, combinations with a Higgs boson mass which does not fulfill Eq.~\eqref{Eq:massoverlap} are still considered. Here, the difference of the measured and predicted Higgs mass is automatically taken into account by the $\chi^2$ contribution from the Higgs mass, $\chi_m^2$.
\end{itemize}
In the case where multiple Higgs bosons are assigned to the same signal, the combined signal strength modifier $\mu$ is taken as the sum over their predicted signal strength modifiers (corresponding to incoherently adding their rates). The best Higgs-to-signals assignment $\eta_0$ in an analysis is that which minimizes the overall $\chi^2$ contribution, i.e.
\begin{equation}
\eta_0 = \eta,\quad\mbox{where}~\sum_{\alpha=1}^\mathrm{N_{signals}} \chi_{\alpha,\eta}^2~\mbox{is minimal}.
\end{equation}
Here, the sum runs over all signals observed \emph{within this particular
analysis}. In this procedure, \HS~only considers assignments $\eta$ where each
Higgs boson is not assigned to more than one signal within the same
analysis in order to avoid double counting. 

There is also the possibility to enforce that a collection of peak observables is either
assigned or not assigned in parallel. This can be useful if certain peak observables stem from the same Higgs analysis but correspond to measurements performed for specific tags or categories (e.g.~as presently used in $H\to \gamma\gamma$ analyses). See Sect.~\ref{Sect:expdata} for a description of these \textit{assignment groups}.

A final remark should be made on the experimental resolution, $\dmexp_\alpha$, which enters Eq.~\eqref{Eq:massoverlap}. In case the analysis has an actual mass measurement that enters the $\chi^2$ contribution from the Higgs mass, $\dmexp_\alpha$ gives the uncertainty of the mass measurement. If this is not the case, $\dmexp_\alpha$ is an estimate of the mass range in which two Higgs boson signals cannot be resolved. This is taken to be the mass resolution quoted by the experimental analysis. Typical values are, for instance, $10\%$ (for $VH\to V(b\bar{b})$~\cite{CMS-PAS-HIG-12-044}) and $20\%$ (for $H\to\tau\tau$~\cite{CMS-PAS-HIG-13-004} and $H\to WW^{(*)}\to \ell\nu\ell\nu$~\cite{CMS-PAS-HIG-13-003}) of the assumed Higgs mass. It should be kept in mind that the \HS\ procedure to automatically assign (possibly several) Higgs bosons to the signals potentially introduces sharp transitions from assigned to unassigned signals at certain mass values, see Section~\ref{Sect:pc_performance} for a further discussion. More detailed studies of overlapping signals from multiple Higgs bosons, where possible interference effects are taken into account, are desirable in case evidence for such a scenario emerges in the future data.

\subsection{The mass-centered $\chi^2$ method}
\label{Sect:mc_chisq}

The mass-centered $\chi^2$ method is complementary to the peak-centered
$\chi^2$ method, since it allows for a more general test of the model
against the experimental data without reference to particular
signals. This method uses the data where the measured best-fit signal
strength modifiers are published as a function of the Higgs mass over
the (full) investigated mass range, as shown in Fig.~\ref{fig:muplots}(a).\footnote{This is sometimes
  referred to as the ``cyan-band plot'', or alternatively the ``$\muobs$
  plot''.} A $\chi^2$ test can then be
performed directly at the predicted Higgs mass(es), $\mpred_i$, of the
model if these fall within the experimentally investigated mass range of
an analysis $a$ (denoted by $G_a$). For Higgs bosons that are outside
this mass range, \HS\ provides no information. Also in this method, like
in the peak-centered case, it can be necessary to consider the combined
rates of several Higgs bosons which are close in mass compared to the
experimental resolution. We begin with a general discussion of the
single Higgs (non-mass-degenerate) case, and outline the combination
scheme below.  


\subsubsection{Theory mass uncertainties}

In the $\muobs$ plot the experimental mass uncertainty is already taken into account in the experimental analysis. However,
we also want to take into account a possible theoretical
uncertainty on the predicted Higgs mass, $\dmth_i$. \HS~provides two
different methods to include theoretical Higgs mass uncertainties in the
mass-centered $\chi^2$ evaluation: 
\begin{itemize}
\item[\textit{(i)}] (\textit{default setting}) In the first method the
  predicted Higgs mass is varied around $\mpred_i$ within its
  uncertainties. We denote this varied mass by $m'$ in the
  following. For a uniform (box) parametrization of the theoretical mass
  uncertainty, we have the allowed mass range
\begin{equation}
m' \in \left[ \mpred_i - \dmth_i , \mpred_i + \dmth_i \right] \equiv M_i.
\label{Eq:massrange}
\end{equation}
A tentative $\chi^2$ distribution is evaluated as a function
of $m'$, which, in the uniform (box) parametrization, takes the form 
\begin{equation}
\chi_i^2 (m') = \sum_{a=1}^{n} \frac{\left[\mu_a(\mpred_i) - \muobs_a(m')\right]^2}{(\dmuobs_a(m'))^2} \qquad  (m' \in M_i).
\label{Eq:chisq_dmth_variation_box}
\end{equation}
For the Gaussian parametrization, we have
\begin{equation}
\chi_i^2 (m') = \sum_{a=1}^{n}\left( \frac{\left[\mu_a(\mpred_i) - \muobs_a(m')\right]^2}{(\dmuobs_a(m'))^2}\right) + \frac{\left[\mpred_i-m'\right]^2}{(\dmth_i)^2}\quad \mbox{with~}m' \in G_a.
\label{Eq:chisq_dmth_variation_gaussian}
\end{equation}
In these expressions, $n$ denotes the total number of considered
analyses. Note that the predicted signal strengths, $\mu_a$, are always
calculated at the predicted central values for the Higgs mass,
$\mpred_i$, (from the user input), and the signal strength is held fixed
in the mass variation. This is clearly an approximation, but for small
theory mass uncertainties $\dmth_i$ it is reasonable to treat resulting
variations in $\mu$ as a second-order effect.\footnote{This requirement
puts an upper limit on a reasonable theoretical mass uncertainty: it
should be smaller than the typical mass interval over which the rate
predictions vary significantly (in the relevant channels).}
From a
practical viewpoint, it also reduces significantly the amount of model
information that has to be supplied by the user.  

The final values for $\muobs$ and $\dmuobs$ are chosen for each
Higgs boson $h_i$ at the mass value
$m_i^0 = m'$, where $\chi_i^2(m')$ is minimized (i.e.\ for each
Higgs boson separately, but combining all channels). In this way, the
\textit{most conservative} value of the predicted Higgs mass, within its
theory uncertainty, is used to define the measured signal strength
modifiers for the final $\chi^2$ evaluation. 

\item[\textit{(ii)}] In the second approach to include theory mass
  uncertainties, \HS~convolves the experimentally measured signal
  strength modifier, $\muobs_a(m)$, with a theory mass pdf, $g(m',m)$,
  resulting in 
\begin{equation}
\muobssm_a(m) = \int_{G_a} \mathrm{d}m' \muobs_a(m') g(m',m).
\label{Eq:muobssm}
\end{equation}
The theory mass pdf $g(m',m)$ can again be chosen to be either a uniform
(box) distribution or a Gaussian, both centered around the predicted
mass value, $m$, and with a box width of $\pm\dmth$ or a
Gaussian width $\dmth$, respectively. The pdf is normalized to unity
over the mass range $G_a$ in order to preserve probability. In the case
of zero theoretical Higgs mass uncertainty,\footnote{This is, e.g., the case in the SM, where the Higgs mass is
a free parameter, or in the (low-energy) MSSM, where, for
instance, the mass of the 
pseudoscalar Higgs boson~$A$ can be chosen to be an input
parameter.} $g(m',m) = \delta (m'
- m)$ in either case. The model prediction is therefore tested directly
against the measured value $\muobs(m)$ at the predicted (exact) value
for the mass $m$. 

The observed signal strength modifier after convolution, $\muobssm_a$,
now includes contributions to the measured signal strength modifier from
the mass region close to the predicted Higgs mass (weighted by
$g(m',m)$). Similarly, the upper and lower experimental $1\sigma$
uncertainty (cyan) band values, $\dmuobs_a$, are smeared 
\begin{equation}
\dmuobssm_a(m) = \int_{G_a} \mathrm{d}m' \dmuobs_a(m') g(m',m).
\label{Eq:dmuobssm}
\end{equation}
In this case it is the smeared quantities, evaluated from
Eqs.~\eqref{Eq:muobssm} and \eqref{Eq:dmuobssm}, that enter the $\chi^2$
test.  
\end{itemize}


\subsubsection{The Stockholm clustering scheme}

If more than one neutral Higgs boson of the model has a mass in the
relevant region of an analysis, $m_i \in G_a$, possible superpositions
of their signal rates have to be taken into account without
double-counting. In order to determine the relevant combinations (out of
the potentially many options), we use a  prescription inspired by jet
clustering. In a similar spirit, we call this the \textit{Stockholm
clustering scheme}: 

\begin{enumerate}
\item[1.] Determine the nearest neighboring Higgs bosons $h_i$ and $h_j$ by
  their mass difference $\Delta m_{ij} = |m_i - m_j|$. If $\min(\Delta
  m_{ij})$ is larger than the experimental mass resolution of the
  analysis, the clustering is finished, and we proceed to step 4. If it
  is smaller, the two Higgs bosons $h_i$ and $h_j$ will be clustered
  (combined). 
\item[2.] The combination of two adjacent Higgs bosons $h_i$ and $h_j$
  defines a new \emph{Higgs cluster} $h_k$ with the following properties: 
\begin{itemize}
\item If both Higgs bosons $h_i$ and $h_j$ have non-zero theoretical mass
  uncertainties ($\dmth_i \ne 0$ and $\dmth_j \ne 0$) the combined mass
  is obtained from a Gaussian average (regardless of the choice for
  Higgs mass pdf), 
\begin{equation}
m_k = (\dmth_k)^2 \left( \frac{m_i}{(\dmth_i)^2} + \frac{m_j}{(\dmth_j)^2} \right),
\label{Eq:mc_cluster_m}
\end{equation}
with the combined theoretical mass uncertainty
\begin{equation}
\dmth_k =  \frac{\dmth_i \dmth_j}{\sqrt{(\dmth_i)^2 + (\dmth_j)^2}}.
\label{Eq:mc_cluster_dm}
\end{equation}
\item If either $m_i$ or $m_j$ is known exactly, for instance $\dmth_i = 0$,
  the mass of the new Higgs cluster is chosen equal to this mass,
  $m_k = m_i$, with zero combined theory mass uncertainty, 
  $\dmth_k = \dmth_i = 0.$ 
\item If both $m_i$ and $m_j$ are known exactly, $\dmth_i = \dmth_j = 0$,
  the Higgs cluster is assigned an averaged mass $m_k = (m_i + m_j)/2$,
  with $\dmth_k = 0$. 
\end{itemize}
\item[3.] The procedure is repeated from step 1. The entities considered
  for further clustering include both the unclustered (initial) Higgs
  bosons, as well as the already combined Higgs clusters. The single
  Higgs bosons which form part of a cluster are no longer present.  
\item[4.] Each single Higgs boson or Higgs cluster $h_k$ that remains
  after the clustering according to steps $1$--$3$ enters the mass-centered
  $\chi^2$ test. Their predicted signal strength modifiers are formed
  from the incoherent sum (again, neglecting interference effects)
  of the individual signal strength modifiers for the combined Higgs
  bosons, 
\begin{equation}
\mu_k(m_k) = \sum_i \mu_i(\mpred_i).
\label{Eq:mc_cluster_mu}
\end{equation}
\end{enumerate}

In this way, the predictions that are compared to \textit{one}
implemented analysis are determined. \HS~repeats this procedure for all
implemented experimental analyses. Since the experimental mass
resolution can vary significantly between different analyses, the
resulting clustering in each case may differ. 

The two different treatments of the theoretical mass uncertainties, as
discussed above, have to be slightly extended for the case of Higgs
clusters: 
\begin{itemize}
\item[\textit{(i)}] If the Higgs boson $h_i$ is contained within a Higgs
  cluster $h_k$ for one analysis $a$, the considered mass region for the
  variation of $m'$ in \refeq{Eq:chisq_dmth_variation_box} is now the
  overlap region $M_i \cap M_k$, with $M_i = \left[m_i - \dmth_i , m_i +
    \dmth_i \right] $ in the case of a uniform (box) Higgs mass
  pdf.\footnote{If $M_i \cap M_k = \varnothing$, we increase $M_k$ until
    there is a (minimal) overlap. This will effectively lead to an
    evaluation of the tentative $\chi^2$ at the boundary of $M_i$ which
    is closest to the mass $m_k$ of the Higgs cluster.} We denote the
  resulting tentative total $\chi^2$ from the variation of the mass of
  Higgs boson $h_i$ by $\chi^2_i$. The variation is done for every Higgs
  boson contained in the cluster $h_k$. When the cluster $h_k$ is evaluated
  against the observed results for analysis $a$, the observed
  values $\muobs_a$ and $\dmuobs_a$ are defined at the value of $m'$
  where the global $\chi^2$, composed of all $\chi_i^2$ distributions,
  is minimal.\footnote{The global $\chi^2$ is defined in the mass region
    $(M_i \cap M_k) \cup (M_j \cap M_k) \cup \dots$, when the Higgs
    bosons $h_i,~h_j,\dots$ are combined in the cluster $h_k$.} 

\item[\textit{(ii)}] In the second approach, the convolution of
  the experimental $\muobs$ values with theory uncertainties is
  performed separately for each Higgs boson, or Higgs cluster $k$, with
  the combined Higgs mass pdf 
\begin{equation}
g_k(m',m) = \frac{1}{N}\sum_i g_i(m',m).
\end{equation}
The normalization factor $N = \int_{M_k}\mathrm{d}m' g_k(m',m)$ to
preserve probability. The sum runs over all Higgs bosons which have been
combined for this cluster.  
\end{itemize}

Once all model predictions and \emph{mass-centered observables} have been defined, when necessary using
Stockholm clustering as discussed above, the total mass-centered
$\chi^2$ is evaluated with a signal strength vector\footnote{The length
  of this vector depends in this case on the Higgs masses and the result
  of the clustering. Each analysis may contribute any number of entries
  $\alpha$, where $0\leq \alpha \leq N_{\mathrm{Higgs}}$.} and
covariance matrix constructed analogously as in the peak-centered
$\chi^2$ method, \cf \refeq{Eq:chimu}. The uncertainties of production
cross sections, decay rates, and the luminosity are again treated as
fully correlated Gaussian errors. Note that, in this method, there is no
contribution from Higgs mass measurements to the total $\chi^2$, since
the evaluation is done directly against the experimental data at the
predicted Higgs mass values (within their uncertainties).

As a final remark, we would like to point out that the $\muobs$ plots necessary for this method are so far only published for a few selected analyses.\footnote{Currently, the $\muobs$ plots are published only for the $H\to\gamma\gamma$, $H\to ZZ^{(*)}$ and $H\to WW^{(*)}$ searches.} Thus, there is not (yet) a full coverage of the various Higgs signal topologies with the mass-centered $\chi^2$ method. Furthermore, the published results cover only a limited range in the Higgs mass, which is a further limit to its applicability.


\subsection{Simultaneous use of both methods}
\label{Sect:bothmethods}

Since the two methods presented here are complementary---they test
inherently different statistical hypotheses---\HS~allows for the
possibility to apply the peak-centered and mass-centered $\chi^2$
methods simultaneously. We present here one approach, which attempts to
make maximal use of the available experimental information when testing
models with multiple Higgs bosons. The user of \HS\ is of course 
free to use other combinations of the two results, which can be 
derived completely independently. 

In the provided combined approach, \HS~first runs the peak-centered
$\chi^2$ method and assigns the Higgs bosons to the observed signals,
tracing the assigned combination for each analysis. In the second step,
all remaining Higgs bosons (which have not been assigned) are
considered with the mass-centered $\chi^2$ method; their respective
(mass-centered) $\chi^2$ contributions are constructed. In this way, a possible double-counting, where a Higgs boson is tested with \textit{both} the peak- and mass-centered $\chi^2$ method against the same data, is avoided. In the last
step, the total $\chi^2$ is evaluated. Here, the Higgs mass $\chi^2$
from the (relevant) signals, as well as the $\chi^2$ from combined
signal strength vectors from both the peak-centered \textit{and}
the mass-centered approach, are evaluated with a full covariance
matrix. This method thus tests the model predictions against the data in
the maximal possible way, while ensuring that no Higgs boson is tested
more than once against the same experimental data.  

As a final recommendation, it should be noted that the mass ranges for the measured $\muhat$ values are still much smaller than the mass ranges for (SM) Higgs exclusion limits. To constrain theories with Higgs bosons outside this smaller range (or below the lower limit of the range currently considered by LHC searches), it is still highly recommended to run \HB\ \cite{Bechtle:2011sb,*Bechtle:2008jh,Bechtle:2013wla} in parallel to \HS.


\section{Using \HS}
\label{sect:Manual}


\subsection{Installation}

The latest version of \HS\ can be downloaded from the webpage\\[.3em]
\centerline{\url{http://higgsbounds.hepforge.org}}\\[.3em]
which is also the home of \HB. Since \HS\ depends on the \HB\ libraries,
this code (version 4.0.0 or newer) should be downloaded and installed as
well. For further detail on how to do this, we refer to the \HB\ manual
\cite{Bechtle:2011sb,*Bechtle:2008jh,Bechtle:2013wla}. Like \HB, \HS\ is written in
{\tt Fortran 90/2003}. Both codes can be compiled, for example,
using {\tt gfortran} (version 4.2 or higher). After unpacking the
downloaded source files, which should create a new directory for \HS,  
the user possibly needs to set the correct path to the \HB\ installation in the \texttt{configure} file. Optionally, the path to a \FH\ installation (version 2.9.4 or higher recommended)~\cite{Heinemeyer:1998yj,*Heinemeyer:1998np,*Degrassi:2002fi,*Frank:2006yh,*Hahn:2009zz} can be set in order to use some of the example programs which use \FH\ subroutines (see below).
Furthermore,
compiler flags necessary for specific platforms can be placed here.
Configuration and installation starts with running 
\cbox{./configure}
which will generate a \texttt{makefile} from the initial file
\texttt{makefile.in}. Once this is done, run 
\cbox{make}
to produce the \HS~Fortran library (called {\tt libHS.a}) and the command line executable. 
In addition, the user may conveniently use a bash script,
\cbox{./run\_tests.bat}
to build the \HS\ library and executable as well as the provided example programs (described in Sect.~\ref{Sect:examples}). The script will then perform a few test runs.


\subsection{Input and output}
\label{Sect:HS_io}

\HS~is designed to require mostly the same input as \HB, so that users
 already familiar with this code should be able to transfer their
existing analyses to also use \HS\ with a minimal amount of extra
work. There are two ways to run \HS: either from the command line, or via the subroutines contained in the \HS~library {\tt libHS.a}. For the command line version, the model predictions (Higgs masses,
their theory uncertainties, total widths,
production and decay rates) have to be specified in data files using the same format as \HB-4, see Ref.~\cite{Bechtle:2013wla}. The command line version of \HS~is presented in more detail in Sect.~\ref{Sect:HScmdline}.

In the subroutine version, the model predictions (which can be given as effective couplings, or as cross sections either at partonic or hadronic level) have to be provided via subroutines. Most of these subroutines are shared with the \HB~library (for details we refer again to \cite{Bechtle:2013wla}).  In addition to the \HB\ input, \HS\ requires input of the theoretical uncertainties on both the Higgs masses and the
rate predictions. Therefore, \HS~contains two additional input subroutines to set these quantities, see Sect.~\ref{Sect:subroutines} for more details. An accessible demonstration of how to use the \HS\ subroutines is provided by the example programs, discussed further in Sect.~\ref{Sect:examples}.  

As already mentioned, the required input of Higgs production and decay
rates can be given either as effective couplings, or as cross sections
at partonic or hadronic level.  For supersymmetric models there is an
option of using the SUSY Les Houches Accord
(SLHA)~\cite{Skands:2003cj,SLHA2} for input (either using data files
or subroutines).  In this case, the production rates are always
approximated using the effective couplings specified in the two
\HB~specific input SLHA blocks (as specified in Ref.~\cite{Bechtle:2013wla}),
whereas the Higgs branching ratios are taken directly from the
corresponding decay blocks. If present, the theoretical mass
  uncertainties are read in from the SLHA block \texttt{DMASS} (as
  available e.g.\ from \texttt{FeynHiggs}). Since there is no
  consensus yet on how to encode the theoretical rate uncertainties in
  the SLHA format, these have to be given to \HS\ explicitly by
  hand.\footnote{This can be done by either calling the
      subroutine \texttt{setup\_rate\_uncertainties} (see below) or by
      including the rate uncertainties directly in the file
      \texttt{usefulbits\_HS.f90} in case the subroutine cannot be
      used (i.e. if \HS\ is run on the command line). If the user does not specify the rate uncertainties (in either case), they are assumed to be identical to the SM rate uncertainties, Eq.~\eqref{Eq:SMrateuncertainties}.}

The main results from \HS~are reported in the form of a $\chi^2$ value
and the number of considered observables. For reference, the code also
calculates the $p$-value associated to the total $\chi^2$ and the
number of degrees of freedom $N$. The user may specify the number of
free model parameters $N_p$ (see below). Then, the number of degrees of
freedom is given by $N = N_\mathrm{obs} - N_p$, where $N_\mathrm{obs}$
is the total number of the included observables. Note that if the user does not specify $N_p$, the $p$-value is evaluated assuming $N_p = 0$.

In the case of running with input data files, the \HS\ output is written into new files as described in Sect.~\ref{Sect:HScmdline}. There also exist subroutines, see Sect.~\ref{Sect:subroutines}, to specify the extent of screen output and to retrieve many quantities of interest for further analysis.

If \HS~is run in the SLHA mode, the results can be appended to the SLHA file in
the form of new SLHA-inspired\footnote{These blocks deviate from the SLHA conventions~\cite{Skands:2003cj,SLHA2} in the way that they contain string values (without whitespaces), which are parenthesized by the symbols `\texttt{||}'.} blocks. The main results are then collected in  
\cbox{BLOCK HiggsSignalsResults,} 
as shown for a specific example in Tab.~\ref{Tab:Block_HSoutput}. The first entries of this \texttt{BLOCK} contain general
information on the global settings of the \HS~run, \ie the version
number, the experimental data set, the $\chi^2$ method and the Higgs
mass parametrization used. Moreover, it lists the number of analyzed
observables of the different types (\texttt{BLOCK} entries \texttt{4}--\texttt{6}), as well as the total number (\texttt{BLOCK} entry \texttt{7}). Next, it gives the corresponding
$\chi^2$ values separately from the signal strength peak observables (\texttt{BLOCK} entry \texttt{8}), the Higgs mass peak observables (\texttt{BLOCK} entry \texttt{9}), and the mass-centered observables (\texttt{BLOCK} entry \texttt{10}).  The total signal strength $\chi^2$ for both methods (the sum of \texttt{BLOCK} entries \texttt{8} and \texttt{10}) is provided (\texttt{BLOCK} entry \texttt{11}), as is the total $\chi^2$ sum (\texttt{BLOCK} entry \texttt{12}). The final element (\texttt{BLOCK} entry \texttt{13}) gives the reference $p$-value, as discussed above.

\begin{table}
\centering
\renewcommand{\arraystretch}{1.0}
{\tt\footnotesize
\begin{tabular}{rrl}
\br
\multicolumn{3}{l}{BLOCK HiggsSignalsResults} \\
    0	&                 ||1.0.0||		&               \# HiggsSignals version \\
    1   &      		||latestresults||	&               \# experimental data set\\
    2   &                      3              	& 		\# Chi-squared method (1:peak-c, 2:mass-c, 3:both)\\
    3   &                      2              	&  		\# Higgs mass pdf (1:box, 2:Gaussian, 3:box+Gaussian)\\
    4   &                     26           	&     		\# Number of signal strength peak observables\\
    5   &                    11         		&       	\# Number of Higgs mass peak observables\\
    6   &                      1          		&      		\# Number of mass-centered observables\\
    7   &                     38        		&        	\# Number of observables (total)\\
    8   &            29.08807277        &       	\# $\chi^2$ from signal strength peak observables\\
    9   &             1.61700565         	&       	\# $\chi^2$ from Higgs mass peak observables\\
   10  &              1.03688409      	&         	\# $\chi^2$ from mass-centered observables\\
   11  &             30.12495686      	&          	\# $\chi^2$ from signal strength (total)\\
   12  &             31.74196250      	&          	\# $\chi^2$ (total)\\
   13  &              0.37648524       	&         	\# Probability (total $\chi^2$, total number observables)\\
\br
\end{tabular}
}
\caption{Example for the SLHA block \texttt{HiggsSignalsResults} after a
  successful run of \HS. The number of observables and $\chi^2$
  contributions are given separately for the signal strength and mass
  parts in the peak-centered $\chi^2$ method, and also for the
  mass-centered $\chi^2$ method.} 
\label{Tab:Block_HSoutput}
\end{table}

Additional output specific to the peak-centered $\chi^2$ method is
collected in 
\cbox{BLOCK HiggsSignalsPeakObservables.}
\begin{table}
\centering
\renewcommand{\arraystretch}{1.0}
{\tt\footnotesize
\begin{tabular}{rrrl}
\br
\multicolumn{4}{l}{BLOCK HiggsSignalsPeakObservables} \\
\#  OBS	&   FLAG	&             VALUE 	&  \# DESCRIPTION \\
     1  &   1  &                    201215801  		& \# Analysis ID \\
     1  &   2   &  ||ATL-CONF-2012-158||       & \# Reference to publication \\
     1  &   3   &   ||(pp)->h->WW->lnulnu||     & \# Description (Search channel) \\
     1  &   4   &                        8.00  		& \# Center-of-mass energy (TeV) \\
     1  &   5   &                       13.00 		&  \# Luminosity (fb$^{-1}$)\\
     1  &   6   &                        3.60   		& \# Luminosity uncertainty (in \%) \\
     1  &   7   &                        8.00   		&\# Mass resolution (GeV) \\
     1  &   8   &                      126.00   		&\# Mass value at peak position (GeV) \\
     1  &   9  &                       1.3460  		& \# Observed signal strength modifier ($\muobs$)\\
     1  &  10 &                        0.5204 		&  \# Lower 68\% C.L. uncertainty on $\muobs$ \\
     1  &  11  &                       0.5710  		& \# Upper 68\% C.L. uncertainty on $\muobs$ \\
     1  &  12   &                          001 		&  \# Assigned Higgs combination \\
     1  &  13  &                             1  		& \# Index of dominant Higgs boson \\
     1  &  14   &                           25  		& \# PDG number of dominant Higgs boson \\
     1  &  15   &                     126.1133 		&  \# Mass of the dominant Higgs boson \\
     1  &  16   &                       0.3305		&   \# Signal strength modifier of dom. Higgs \\
     1  &  17   &                       0.3305  		& \# Total predicted signal strength modifier $\mu$ \\
     1  &  18  &                        1.6196 		&  \# $\chi^2$ from signal strength \\
     1  &  19  &                       0.0000  		& \# $\chi^2$ from Higgs mass \\
     1  &  20   &                       1.6196 		&  \# $\chi^2$ (total)\\
     1  &  21    &                      2.3514  		& \#  $\chi^2$ for no predicted signal ($\mu=0$)\\
     2  &    1	&                      201209202  	& \# Analysis ID\\
     2  &   2   &	  ||ATL-CONF-2012-092||  & \# Reference to publication\\
     2   &  3   &  ||(pp)->h+...->ZZ->4l||      	& \# Description (Search channel)     \\
     \vdots	& \vdots	&	\vdots	&\vdots\\
\br
\end{tabular}
}
\caption{Example for the SLHA block \texttt{HiggsSignalsPeakObservables}. The first column enumerates through all considered peak observables, as indicated by the dots at the bottom.}
\label{Tab:Block_HSpeakobs}
\end{table}
We show an excerpt from this extensive \texttt{BLOCK} for an example
(MSSM) parameter point in Tab.~\ref{Tab:Block_HSpeakobs}. The first
identifier, \texttt{OBS}, in the \texttt{BLOCK} enumerates the peak
observables, whereas the second number, \texttt{FLAG}, labels the specific quantity (for this peak observable). For every peak observable, the first entries (\texttt{FLAG=1-11}) give general information about the experimental data defining the observable. This is followed
by model-specific information and the results from the
\HS~run. \texttt{FLAG=12} displays a binary code representing the Higgs boson combination
which has been assigned to the signal. It has the same length as the number of Higgs bosons\footnote{For technical reasons, \HS\ is currently limited to models with $n_H \leq 9$ neutral Higgs bosons, but this could easily be extended if there is a demand for more.}, such that an assigned Higgs boson with index $k$ corresponds to the binary value $2^{k-1}$. A code of only zeroes means that no Higgs boson has been assigned to this peak observable.
In the specific example shown in Tab.~\ref{Tab:Block_HSpeakobs}, the lightest of the three neutral
Higgs bosons in the MSSM (with $k=1$) has been assigned. 

This \texttt{BLOCK} also contains additional information
(index $i$, Particle data group (PDG) number, mass, and signal strength contribution under \texttt{FLAG=13-16})
about the assigned Higgs boson that gives the largest contribution to
the total predicted signal strength.  The total
predicted signal strength is given by \texttt{FLAG=17}. The \HS\ results (\texttt{FLAG=18-20}) contain
the $\chi^2$ contribution from the signal strength and
Higgs mass test from this observable, as well as the total $\chi^2$
contribution obtained for the assigned Higgs boson combination. Finally, the $\chi^2$ obtained for the case with no predicted signal, $\mu=0$, is given for \texttt{FLAG=21}. It should be noted that the
quoted $\chi^2$ values correspond to intermediate results in the total $\chi^2$
evaluation, where correlated uncertainties are taken into account by the
covariance matrix. For instance, the signal strength $\chi^2$ (\texttt{FLAG=18})
corresponds to $\cmua^2$ in Eq.~\eqref{Eq:chimu}, where $\alpha$ is the
index of the peak observable given in the first column of the
\texttt{BLOCK}. Thus, this quantity differs from the na\"ively
calculated $\chi^2 = (\mu-\muobs)^2/(\dmuobs)^2$, and might in the extreme case even be
negative due to the impact of correlated uncertainties. 

The results from the mass-centered $\chi^2$ method are summarized in 
\cbox{BLOCK HiggsSignalsMassCenteredObservables} 
in a similar way as in \texttt{BLOCK HiggsSignalsPeakObservables}. An example is given in Tab.~\ref{Tab:Block_HSmassobs}. The model-independent information about the observable (\texttt{FLAG=1-7}) is identical to the corresponding information in \texttt{BLOCK HiggsSignalsPeakObservables}. However, since the evaluated experimental quantities of the mass-centered observable depend on the model prediction, \cf~Sect.~\ref{Sect:mc_chisq}, we give the information of the tested Higgs boson (cluster) at first (\texttt{FLAG=8-10}), corresponding to Eqs.~\eqref{Eq:mc_cluster_m}--\eqref{Eq:mc_cluster_mu}. The number and binary code of the combined Higgs bosons, which form a Stockholm Higgs cluster, is given by \texttt{FLAG=11} and \texttt{12}, respectively. 
From the experimental data is given the mass position (\texttt{FLAG=13}), and the measured signal strength with its lower and upper uncertainties (\texttt{FLAG=14-16}). Finally, the resulting $\chi^2$ contribution from this mass-centered observable is given at \texttt{FLAG=17}.

Note that there is also the possibility to create a new SLHA file with the \HS\ output blocks even if the input was not provided in SLHA format. Moreover, \HS~can give an extensive screen output with similar information as encoded in the three SLHA output blocks. The level of information that is desired should then be specified before the \HS~run via the subroutine \texttt{setup\_output\_level}. See Sect.~\ref{Sect:subroutines} for more details.

\begin{table}
\centering
\renewcommand{\arraystretch}{1.0}
{\tt\footnotesize
\begin{tabular}{rrrl}
\br
\multicolumn{4}{l}{BLOCK HiggsSignalsMassCenteredObservables} \\
\#  OBS   &	FLAG                      &   VALUE  	&\# DESCRIPTION	\\
     1  &   1  &                    201215801   & \# Analysis ID				\\
     1  &   2   &  ||ATL-CONF-2012-158||        & \# Reference to publication	\\
     1  &   3   &    ||(pp)->h->WW->lnulnu||      & \# Description (Search channel)	\\
     1  &   4    &                         8.00   & \# Center-of-mass energy (TeV)	\\
     1  &   5     &                       13.00   & \# Luminosity	(fb$^{-1}$)\\
     1  &   6    &                         3.60   & \# Luminosity uncertainty (in \%)	\\
     1  &   7    &                         8.00   & \# Mass resolution (GeV)	\\
     1  &   8   &                        122.65   & \# Mass of tested Higgs boson (GeV)	\\
     1  &   9    &                         2.00   & \# Mass uncertainty of tested Higgs boson (GeV)	\\
     1  &  10    &                       0.7379   & \# Signal strength of tested Higgs boson(s)	\\
     1  &  11  &                            1   & \# Number of combined Higgs bosons	\\
     1	&  12	&	 001	& \# Combined Higgs boson code \\ 
     1  &  13    &                       122.90   & \# Observed mass value (GeV)	\\
     1  &  14    &                       1.8269   & \# Observed signal strength $\muobs$ 	\\
     1  &  15    &                       0.6822   & \# Lower 68\% C.L. uncertainty on $\muobs$	\\
     1  &  16    &                       0.7462   & \# Upper 68\% C.L. uncertainty on $\muobs$	\\
     1  &  17      &                     2.9617   & \# $\chi^2$ (total)	\\
     2   &    1  &                    201209202   & \# Analysis ID	\\
     2   &    2     &  ||ATL-CONF-2012-092||        & \# Reference to publication	\\
     2   &    3     &  ||(pp)->h+...->ZZ->4l||      & \# Description (Search channel)	\\
     \vdots	& \vdots	&	\vdots	&\vdots\\     
\br
\end{tabular}}
\caption{Example for the SLHA output block \texttt{HiggsSignalsMassCenteredObservables} containing information about the observables and results from the mass-centered $\chi^2$ method.}
\label{Tab:Block_HSmassobs}
\end{table}

\subsection{Running \HS~on the command line}
\label{Sect:HScmdline}

\HS~can be run on the command line as follows:
\cbox{\small./HiggsSignals <expdata> <mode> <pdf> <whichinput> <nHzero> <nHplus> <prefix>}
This command line call is very similar to the one of \HB~and the last four arguments have been directly taken over from \HB. The user may consult the \HB~manual~\cite{Bechtle:2013wla} for more details on these arguments. The number of neutral and charged Higgs bosons of the model are specified by \texttt{nHzero} and \texttt{nHplus}, respectively. As in \HB, the model predictions are read in from the data files specified by \texttt{prefix}. Which data files are required as input depends on the argument \texttt{whichinput}, which can take the string values \texttt{effC}, \texttt{part}, \texttt{hadr} and \texttt{SLHA} for the various input formats. The theory mass uncertainties are read in from the data file \texttt{<prefix>MHall\_uncertainties.dat} for both the neutral and charged Higgs bosons. If this file is absent these uncertainties are set to zero. For more information of the data file structure we refer to the \HB-4 manual~\cite{Bechtle:2013wla}. Note that for \texttt{whichinput=SLHA}, all the input is read in from the SLHA input file which, like the ordinary data files, should be specified by \texttt{<prefix>}. 

The first three arguments are intrinsic \HS~options. The string \texttt{<expdata>} specifies which experimental data  set should be used. \HS~will read in the observables found in the directory \texttt{Expt\_tables/<expdata>}. The second argument, \texttt{<mode>}, specifies which $\chi^2$ method should be used; it can take the string values \texttt{peak} (for the peak-centered $\chi^2$ method, described in Sect.~\ref{Sect:pc_chisq}), \texttt{mass} (for the mass-centered $\chi^2$ method, see Sect.~\ref{Sect:mc_chisq}), or \texttt{both} (for the simultaneous use of both methods, as described in Sect.~\ref{Sect:bothmethods}).
Finally, the \texttt{<pdf>} argument takes an integer selecting the parametrization for the Higgs mass uncertainty as either  \texttt{1} (box), \texttt{2} (Gaussian), or \texttt{3} (box+Gaussian) pdf.

As an example, the user may run \cbox{./HiggsSignals latestresults
  peak 2 effC 3 1 example\_data/mhmax/mhmax\_} which runs the
  peak-centered $\chi^2$ method on the provided parameter points in the
  $(M_A,~\tb)$ plane of the $m_h^\mathrm{max}$ benchmark
  scenario~\cite{Carena:2002qg} of the
  MSSM, using the most recent Higgs data contained in the directory
  \texttt{Expt\_tables/latestresults/}.

The \HS~output from a successful command line run is collected in the data file \texttt{<prefix>HiggsSignals\_results.dat}, except for the case \texttt{whichinput=SLHA}, where the results are attached as SLHA output blocks to the SLHA file, \cf Sect.~\ref{Sect:HS_io}. 
The SUSY spectrum generator \texttt{SPheno}~\cite{Porod:2003um,*Porod:2011nf}, used in conjunction with the model building tool \texttt{SARAH}~\cite{Staub:2008uz,*Staub:2009bi,*Staub:2010jh}, can write directly the \HB~(and thus \HS) data files for input in the effective couplings format.

\subsection{\HS~subroutines}
\label{Sect:subroutines}
In this section we present the subroutines needed for the use of \HS. First, we go step-by-step through the user subroutines encountered during a normal run of \HS. Then, we list additional (optional) subroutines for specific applications of \HS, and for a convenient handling of the output.

\subsubsection*{\bf Main user subroutines}
The subroutine that is usually called first is
\subroutine{initialize\_HiggsSignals(}{\textit{int} nHzero, \textit{int} nHplus, \textit{char*} expdata)}
which sets up the \HS~framework: It allocates internal arrays according to the number of neutral (\texttt{nHzero}) and charged\footnote{At this point, there are no measurements available of signal strength quantities for charged Higgs bosons, which are therefore not considered in any way by \HS.} (\texttt{nHplus}) Higgs bosons in the model and reads in the tables for the SM branching ratios in the same way as done in \HB. Furthermore, it calls the subroutine \texttt{setup\_observables}, which reads in the experimental data contained in the directory \texttt{Expt\_tables/}(\texttt{expdata}). The user may create a new directory in \texttt{Expt\_tables/} containing the relevant observables for his study, see Sect.~\ref{Sect:expdata} for more details. For convenience, we also provide a wrapper subroutine
\subroutine{initialize\_HiggsSignals\_latestresults(}{\textit{int} nHzero, \textit{int} nHplus)}

which does not require the third argument but uses the experimental data from the folder \texttt{Expt\_tables/}\texttt{latestresults/}.
\subroutine{setup\_pdf(}{\textit{int} pdf)}

The next step is to specify the probability density function (pdf) for the Higgs masses, which is done using {\tt setup\_pdf}. Available settings are $\texttt{pdf}=1$ for a uniform (box-shaped) distribution, $\texttt{pdf}=2$ for a Gaussian, and  $\texttt{pdf}=3$ for a box-shaped pdf with Gaussian tails. The impact of this choice has been discussed in detail in Sect.~\ref{sect:Statistics} and will furthermore be demonstrated in Sect.~\ref{sect:Examples}. With the subroutine
\subroutine{HiggsSignals\_neutral\_input\_MassUncertainty(}{\textit{double(nHzero)} dMh)}

values for the theory mass uncertainties $\dmth_i$ can be specified.
This subroutine sets the theoretical uncertainties of the neutral Higgs boson masses (in GeV) of the model via the array \texttt{dMh}. The default values (in case this subroutine is not invoked) is for all uncertainties to be zero. Note that \HB-4~also contains a similar subroutine (\texttt{set\_mass\_uncertainties}) to set theoretical mass uncertainties of the neutral and charged Higgs bosons. These uncertainties are taken into account via mass variation in the \HB~run. Since the treatment of these uncertainties is intrinsically different between the two codes, we allow the user to set the theoretical mass uncertainties for \HS\ independently using this subroutine.\footnote{The use of different theoretical mass uncertainties in \HB~and \HS~is restricted to the subroutine version. In the command line version of both programs, the theoretical uncertainties will be read in from the same data file, namely \texttt{<prefix>MHall\_uncertainties.dat}.}
\subroutine{setup\_rate\_uncertainties(}{\textit{double(5)} dCS, \textit{double(5)} dBR)}

For models with different uncertainties on the Higgs production cross sections and branching ratios than those for a SM Higgs boson, these should be specified using this subroutine, which sets the theoretical uncertainties of the production and decay rates (in \%) in the considered model. In the current implementation, LHC and Tevatron channels are considered to have the same relative rate uncertainties, and the rate uncertainties are assumed to be the same for all neutral Higgs bosons, independent of their masses. The input arrays should follow the structure of Table~\ref{tab:uncorder}.

The remaining required input (Higgs boson masses, total widths, branching ratios, cross sections) is identical to the \HB~input and should be set via the \HB~input subroutines, \cf Ref.~\cite{Bechtle:2013wla}.
\begin{table}[t]
\centering
\footnotesize
\begin{tabular}{c | c c c c c}
\br
Array	&		\multicolumn{5}{c}{Element}	\\
				&					1	&	2	&	3	&	4	&	5	\\
\mr				
\texttt{dCS}		&	singleH	&	VBF		&	$HW$		&	$HZ$		&	$t\bar t H$	\\
\texttt{dBR}		&	$H\to\gamma\gamma$	&	$H\to WW$		&	$H\to ZZ$		&	$H\to \tau\tau$		&	$H\to b\bar b$	\\
\br
\end{tabular}
\caption{Ordering of the elements of the input arrays \texttt{dCS} and \texttt{dBR} for the relative uncertainties of the hadronic production cross sections and branching ratios, respectively. Recall that the hadronic production mode ``singleH'' usually contains both the partonic processes $gg\to H$ and $b\bar{b}\to H$, currently assuming equal experimental efficiencies. The latter can change in the future once search categories with $b$-tags are included. This table will possibly be extended once measurements in new channels (e.g. $H\to Z\gamma$) are performed.}
\label{tab:uncorder}
\end{table}
\subroutine{setup\_nparam(}{\textit{int} Np)}
In order to evaluate a meaningful $p$-value during the \HS\ run, the program has to know the number of free model parameters, $N_p$, \cf Sect.~\ref{Sect:HS_io}. This number is specified by the subroutine \texttt{setup\_nparam}. If this subroutine is not called before the main \HS\ run, the code assumes no free model parameters, $N_p = 0$.
\subroutine{run\_HiggsSignals(}{\textit{int} mode, \textit{double} csqmu, \textit{double} csqmh, \textit{double} csqtot, \textit{int} nobs, \textit{double} Pvalue)}

Once all the input has been specified, the main \HS\ evaluation can be run by calling the {\tt run\_HiggsSignals} subroutine to start the $\chi^2$ evaluation. The \texttt{mode} flag specifies the $\chi^2$ method which is used in the following evaluation process. Possible values are $\texttt{mode}=\texttt{1}$ (peak-centered method, \cf Sect.~\ref{Sect:pc_chisq}), $\texttt{mode}=\texttt{2}$ (mass-centered method, \cf Sect.~\ref{Sect:mc_chisq}), or $\texttt{mode}=\texttt{3}$ (simultaneous use of both methods, \cf Sect.~\ref{Sect:bothmethods}).
After a successful run, this subroutine returns the $\chi^2$ contribution from the signal strength measurements (\texttt{csqmu}),\footnote{If $\texttt{mode}=\texttt{3}$, \texttt{csqmu} contains the contributions from peak and mass-centered observables.} the $\chi^2$ contribution from the Higgs mass measurements (\texttt{csqmh}), and the total $\chi^2$ value (\texttt{csqtot}). It also returns the number of observables involved in the $\chi^2$ evaluation (\texttt{nobs}). If the mass-centered $\chi^2$ method is employed, it is important to realize that \texttt{nobs} can depend on many parameters, such as the Higgs boson masses of the model (which may be inside or outside the range of an analysis). The Stockholm clustering can also affect the number of observables that are evaluated in the final $\chi^2$ calculation. Finally, the associated $p$-value (\texttt{Pvalue}) for the total $\chi^2$ with \texttt{nobs}$-N_p$ degrees of freedom is calculated. 
\subroutine{finish\_HiggsSignals}{()}

 At the end of a \HS\ run, the user should call this routine to deallocate all internal arrays. 
\subsubsection*{\bf Specific user subroutines}
This section provides a list (alphabetically ordered) of subroutines handling more special features of \HS.
\subroutine{assign\_toyvalues\_to\_peak(}{\textit{int}~obsID, \textit{double}~mu\_obs, \textit{double}~mh\_obs)}

If the user wants to perform a dedicated statistical study using pseudo-measurements (also called toy-measurements) for the Higgs signal rates and mass measurements, they can be set via this subroutine for the peak observable with the identification number \texttt{obsID}. This \textit{observable ID} is unique to the peak observable and is encoded in the experimental data, see Sect.~\ref{Sect:expdata} for more details. After a (dummy) run of \HS~the observable ID can also be read out with the subroutine \texttt{get\_ID\_of\_peakobservable} (see below).
The arguments \texttt{mu\_obs} and \texttt{mh\_obs} are the pseudo-measured values for the signal strength modifier $\muobs$ and the Higgs mass $\mobs$. Note that the uncertainties are kept at their original values.
\subroutine{assign\_rate\_uncertainty\_scalefactor\_to\_peak(}{\textit{int}~obsID, \textit{double}~scale\_mu)}

If the user wants to scale the uncertainties of the Higgs signal rate and mass measurements, this can be done via this subroutine in an analogous way as setting the toy measurements (using \texttt{assign\_toyvalues\_to\_peak}). Here, \texttt{scale\_mu} is the scale factor for the experimental uncertainty on the signal strength of the peak with identification number \texttt{obsID}. The theoretical rate uncertainties, which can be set independently via the subroutine \texttt{setup\_rate\_uncertainties} (see above), are unaffected by this scale factor. In this way, \HS~allows the user to scale the experimental and theoretical rate uncertainties independently. This is useful if the user is interested in a future projection of the compatibility between the model and the experimental data, assuming that a certain improvement in the precision of the measurements and/or theoretical predictions can be achieved.

After the \HS~run the user can employ the following ``\texttt{get\_}'' subroutines to obtain useful information from the \HS~output. The following three subroutines are contained in the Fortran module \texttt{io}.
\subroutine{get\_ID\_of\_peakobservable(}{\textit{int}~i, \textit{int}~obsID)}

If the peak-centered $\chi^2$ method is used, the peak observables are internally enumerated in \HS~based on their alphabetical appearance in the directory \texttt{Expt\_tables/(expdata)} of the used experimental dataset. This ordering is reflected \eg in the screen output and the SLHA output. However, a safer way to access the peak observables (for instance to set toy observables) is to use the unique observable ID of the peak observable. For this, the user may call this subroutine
which returns the observable ID \texttt{obsID} internally structured at the position~$i$. 
\subroutine{get\_number\_of\_observables(}{\textit{int}~ntotal, \textit{int}~npeakmu, \textit{int}~npeakmh, \textit{int}~nmpred, \textit{int}~nanalyses)}

This subroutine returns the total number of various observables: \texttt{ntotal} is the total number of observables, \texttt{npeakmu} and \texttt{npeakmh} are the number of signal strength and Higgs mass observables entering the peak-centered $\chi^2$ method, respectively, \texttt{nmpred} is the number of observables considered in the mass-centered $\chi^2$ method, and \texttt{nanalyses} gives the number of implemented analyses. Note that several mass-centered and peak observables can in general exist for each experimental analysis. 
\subroutine{get\_peakinfo\_from\_HSresults(}{\textit{int} obsID, \textit{double(npeak)}~mupred, \textit{int(npeak)}~domH, \textit{int(npeak)}~nHcomb)}

More information about the \HS\ result can be obtained by calling this subroutine. It returns the total predicted signal strength modifier, the index of the dominantly contributing Higgs boson and the number of combined Higgs bosons for the peak observable with observable identifier \texttt{obsID} as \texttt{mupred}, \texttt{npeak} and \texttt{nHcomb}, respectively.%
\subroutine{get\_Pvalue(}{\textit{int} Np, \textit{double} Pvalue)}
The user may apply the subroutine \texttt{get\_Pvalue} to evaluate
  the $p$-value again after \texttt{run\_HiggsSignals}, with the possibility
  to vary $N_p$. The result is based on the total $\chi^2$ and the
  total number of observables from the last \HS\ run as well as the
  number of free parameters, \texttt{Np}, which are passed as input to
  this subroutine.  \subroutine{get\_rates(}{\textit{int}~i,
  \textit{int}~collider, \textit{int}~Nchannels,
  \textit{int(Nchannels)}~IDchannels, \textit{double}~rate)}

This subroutine allows the user to read out the predicted signal rate for an arbitrary channel combination. This channel combination is specified by the number of combined channels, \texttt{Nchannels}, and the array \texttt{IDchannels}, which contains the two-digit IDs of these channels as specified in \cf~Tab.~\ref{tab:codes}. The output (\texttt{rate}) is the combined rate. It is more general than {\tt get\_Rvalues} (see below).
\begin{table}
\centering
\footnotesize
\begin{tabular}{c | c || c | c}
\br
$1^\mathrm{st}$ digit	&	production mode	& $2^\mathrm{nd}$ digit	&	decay mode \\
\mr
1					&	singleH  	&	1				&	$H\to \gamma\gamma$	\\
2					&	VBF				&	2				&	$H\to WW$	\\
3					&	$HW$			&	3				&	$H\to ZZ$	\\
4					&	$HZ$			&	4				&	$H\to \tau\tau$	\\
5					&	$t\bar tH$			&	5				&	$H\to b\bar b$	\\
\br
\end{tabular}
\caption{Channels codes used for Higgs production and decay modes, for example by the \texttt{get\_rates} subroutine (see text for details).}
\label{tab:codes}
\end{table}
\subroutine{get\_Rvalues(}{\textit{int}~i, \textit{int}~collider, \textit{double}~R\_H\_WW, \textit{double}~R\_H\_ZZ, \textit{double}~R\_H\_gaga, \textit{double}~R\_H\_tautau, \textit{double}~R\_H\_bb, \textit{double}~R\_VH\_bb)}

This returns the model-predicted signal rates (normalized to the SM signal rates) of Higgs boson \texttt{i} for the six different processes listed in Tab.~\ref{Tab:Rvalues}. These signal rates are calculated via Eq.~\eqref{Eq:mu}, assuming that all channels have the same relative efficiency, $\epsilon_i=1$. These quantities are evaluated either for the Tevatron or LHC with $\sqrt{s} = 7\tev$ or $8\tev$, as specified by the argument \texttt{collider}, taking the values \texttt{1}, \texttt{2} or \texttt{3} for Tevatron, LHC7 or LHC8, respectively.
\begin{table}[ht]
\centering
\footnotesize
\begin{tabular}{l|c|c}
\br
Argument	&	Production modes	&	Decay mode	\\
\mr
\texttt{R\_H\_WW} & $\mbox{singleH},~\mathrm{VBF},~HW,~HZ,~t\bar{t}H$ & $H \to WW$	\\
\texttt{R\_H\_ZZ} & $\mbox{singleH},~\mathrm{VBF},~HW,~HZ,~t\bar{t}H$ & $H \to ZZ$	\\
\texttt{R\_H\_gaga} & $\mbox{singleH},~\mathrm{VBF},~HW,~HZ,~t\bar{t}H$ & $H \to \gamma\gamma$	\\
\texttt{R\_H\_tautau} & $\mbox{singleH},~\mathrm{VBF},~HW,~HZ,~t\bar{t}H$ & $H \to \tau\tau$	\\
\texttt{R\_H\_bb} & $\mbox{singleH},~\mathrm{VBF},~HW,~HZ,~t\bar{t}H$ & $H \to b\bar{b}$	\\
\texttt{R\_VH\_bb} & $HW,~HZ$ & $H \to b\bar{b}$	\\
\br
\end{tabular}
\caption{Production and decay modes considered in the signal rate ratio quantities which are returned by the subroutine \texttt{get\_Rvalues}.}
\label{Tab:Rvalues}
\end{table}

In order to write the \HS\ SLHA output blocks, we provide three different SLHA output subroutines, contained in the Fortran module \texttt{io}. For more information about these output blocks, see Sect.~\ref{Sect:HS_io}.
\subroutine{HiggsSignals\_create\_SLHA\_output}{(\textit{char*} filename, \textit{int} detailed)}
If the user does not use the SLHA input format of \HS, or rather wants to write the output into a different file, this subroutine can be used to create a new file as specified by the argument \texttt{filename}. If this file already exists, \HS\ will \textit{not} overwrite this file but give a warning. The integer argument \texttt{detailed} takes values of \texttt{0} or \texttt{1}, determining whether only the block \texttt{HiggsSignalsResults} or all possible output blocks (i.e. also the block \texttt{HiggsSignalsPeakObservables} and/or \texttt{HiggsSignalsMassCenteredObservables}), respectively, are written to the file. The wrapper subroutine
\subroutine{HiggsSignals\_create\_SLHA\_output\_default}{(\textit{int} detailed)}
does the same but for the default filename called \texttt{HS-output.slha}.
\subroutine{HiggsSignals\_SLHA\_output}{(\textit{int} detailed)}
If \HS~is run on an SLHA input file, the subroutine {\tt HiggsSignals\_SLHA\_output} appends the \HS~results as blocks to the SLHA input file. 

The following ``\texttt{setup\_}'' subroutines can be used to change the default settings of the \HS\ run. Thus, they should be called before the subroutine \texttt{run\_HiggsSignals}.
\subroutine{setup\_assignmentrange(}{\textit{double} Lambda)}

This subroutine can be used to change the mass range, in which a Higgs boson is forced to be assigned to a peak observable, see Sect.~\ref{Sect:peakassignment}. The value \texttt{Lambda} corresponds to $\Lambda$ in Eq.~\eqref{Eq:massoverlap}.
\subroutine{setup\_correlations(}{\textit{int} corr\_mu, \textit{int} corr\_mh)}

The subroutine can be used to switch off (on) the correlations among the systematic uncertainties in the $\chi^2$ evaluation of the signal strength [Higgs mass] part by setting \texttt{corr\_mu} [\texttt{corr\_mh}]\texttt{ = 0}~(\texttt{1}). If this subroutine is not called, the default is to evaluate the $\chi^2$ \textit{with} correlated uncertainties (\texttt{corr\_mu = corr\_mh =  1}).
\subroutine{setup\_mcmethod\_dm\_theory(}{\textit{int}~mode)}

If the mass-centered $\chi^2$ method is used, the treatment of the Higgs mass theory uncertainty can be set by calling this subroutine with \texttt{mode}=\texttt{1} to use the mass variation (default), or \texttt{mode}=\texttt{2} for convolving the theory mass uncertainty with the $\muobs$ plot. See Sect.~\ref{Sect:mc_chisq} for more details of these methods.
\subroutine{setup\_output\_level(}{\textit{int} level)}
The user may control the screen output from the \HS~run with the subroutine, 
where \texttt{level} takes values from $0$ to $3$, corresponding to the following output:
\begin{itemize}
\item[0] Silent mode (suitable for model parameter scans, etc.) (\textit{default}),
\item[1] Screen output for each analysis with its peak and/or mass-centered observables. The channel signal strength modifiers and SM channel weights, \cf~Eq.~\eqref{Eq:ci} and \eqref{Eq:omega}, respectively, are given for all channels considered by the analysis.
\item[2] Screen output of the essential experimental data of the peak observables and/or implemented $\muobs$ plots (as used for the mass-centered $\chi^2$ method). For each observable, the signal channels are listed with the implemented efficiencies.
\item[3] Creates text files holding essential information about the experimental data and the model predictions for each observable. In the peak-centered $\chi^2$ run mode, the files \texttt{peak\_information.txt} and \texttt{peak\_massesandrates.txt} are created. The first file lists all peak observables, including a description and references to the publications, whereas the second file gives the observed and model-predicted values for the Higgs mass\footnote{If multiple Higgs bosons are assigned to the peak, we give the mass of the Higgs boson contributing dominantly to the signal rate.} and signal rates and their corresponding pull values, which we define as:
\begin{equation}
\mbox{pull value} = \frac{\mbox{predicted value} - \mbox{observed value}}{\mbox{(Gaussian combined) uncertainty}}
\end{equation}
Note that in this expression the effect of correlated uncertainties is not taken into account. In the mass-centered $\chi^2$ run mode, the files \texttt{mctables\_information.txt} and \texttt{mcobservables\_information.txt} are created. The first file gives general information about the analyses with an implemented $\muobs$-plot. The second file lists all mass-centered observables, which have been constructed during the \HS\ run, including the mass position, the observed and predicted signal strength values as well as their pull values.
\end{itemize}
For any of the options \texttt{level}~$=1-3$, the main \HS~results are printed to the screen at the end of the run.


\subsection{Example programs}
\label{Sect:examples}
\HS~provides the seven example programs \texttt{HSeffC}, \texttt{HShadr}, \texttt{HSwithSLHA}, \texttt{HBandHSwithSLHA}, \texttt{HSwithToys}, \texttt{HS\_scale\_uncertainties}, and \texttt{HBandHSwithFH}. They are contained in the subfolder 
\cbox{./example\_programs/}
of the main \HS\ distribution and can be compiled all together (except \texttt{HBandHSwithFH}) by running
\cbox{make HSexamples}
or separately by calling:
\cbox{make <name of example program>}
The first program, \texttt{HSeffC}, considers a model with one neutral Higgs boson and uses the effective couplings input subroutines of \HB~to set the input. It demonstrates how to scan over a certain Higgs mass range and/or over various effective couplings while calculating the total $\chi^2$ for every scan point. The code furthermore contains two functions: \texttt{get\_g2hgaga}, which calculates the loop-induced $H\gamma\gamma$ effective coupling from the effective (tree-level) Higgs couplings to third generation fermions and gauge bosons~\cite{LHCHiggsCrossSectionWorkingGroup:2012nn} (assuming a Higgs boson mass of $126\gev$), and a second function which interpolates the cross section uncertainty of the composed single Higgs production from the uncertainties of the gluon fusion and $b\bar{b}\to H$ processes using the effective $Hgg$ and $Hb\bar{b}$ couplings. This can be relevant if the Higgs coupling to bottom quark is strongly enhanced.

The second example program, \texttt{HShadr}, performs a two dimensional scan over common scale factors of the hadronic production cross sections of $p\accentset{(-)}{p} \to H$ and $p\accentset{(-)}{p} \to t\bar{t}H$ on the one side, denoted by $\mu_{ggf+ttH}$, and of $p\accentset{(-)}{p} \to q\bar{q}H$, $p\accentset{(-)}{p} \to WH$ and $p\accentset{(-)}{p} \to ZH$ on the other side, denoted by $\mu_{\mathrm{VBF}+VH}$. The Higgs branching ratios are kept at their SM values.

The third example program, \texttt{HSwithSLHA}, uses the SLHA input of \HB, \ie~an SLHA file which contains the two special input blocks for \HB. It can be executed with
\cbox{./HSwithSLHA <number of SLHA files> <SLHA filename>}
The program can test several SLHA files in one call. The total number of SLHA files must therefore be given as the first argument. The SLHA files must all have the same name, and should be enumerated by \texttt{SLHA\_filename.x}, where \texttt{x} is a number. Running, for example,
\cbox{./HSwithSLHA 2 SLHAexample.fh}
would require the two SLHA files \texttt{SLHAexample.fh.1} and \texttt{SLHAexample.fh.2} to be present. The output is written as SLHA blocks, \cf Sect.~\ref{Sect:HS_io}, which are appended to each input SLHA file. The example program \texttt{HBandHSwithSLHA} can be run in an analogous way. It employs both \HB\ and \HS\ on the provided SLHA file(s), demonstrating how these two codes can be run together efficiently.

The example program \texttt{HSwithToys} demonstrates how to set new values (corresponding to pseudo-measurements) for $\hat\mu$ and $\hat{m}$ for each signal. In the code, \HS~is first run on the SM with a Higgs mass around $126\gev$ using the effective couplings input format. Then, the predicted signal strengths are read out from the \HS~output and set as pseudo-measurements. A second \HS~run on these modified observables then results in a total $\chi^2$ of zero.

The example program \texttt{HS\_scale\_uncertainties} also runs on the SM with a Higgs mass around $126\gev$. It scans over a universal scale factor for \textit{(i)} the experimental uncertainty of the signal strength $\muobs$ only, \textit{(ii)} the theoretical uncertainties of the production cross sections and branching ratios only, and finally \textit{(iii)} both experimental and theoretical uncertainties. The output of each scan is saved in text files. In this way, rough projections of the model compatibility to a more accurate measurement in the future (with the same central values) can be made.

The last example, \texttt{HBandHSwithFH}, demonstrates how to run \HB\ and \HS\ simultaneously on a realistic model, in this case the MSSM. Here, \FH~\cite{Heinemeyer:1998yj,*Heinemeyer:1998np,*Degrassi:2002fi,*Frank:2006yh} is used to calculated the MSSM predictions needed as input for \HB\ and \HS.


\subsection{Input of new experimental data into \HS}
\label{Sect:expdata}
The ambition with \HS\ is to always keep the code updated with the latest experimental results. Nevertheless, there are several situations when a user may want to manually add new data (or pseudo-data) to the program, for example to assess the impact of a hypothetical future measurement. For advanced users, we therefore provide a full description of the data file format used by \HS.

For each observable that should be considered by \HS, there must exist a textfile (file suffix: \texttt{.txt}). This file should be placed in a directory
\cbox{Expt\_tables/(expdata)/}
where \texttt{(expdata)} is the name identifying the new (or existing) experimental dataset.\footnote{The identifier \texttt{(expdata)} is the argument which has to be passed to \texttt{initialize\_HiggsSignals} at initialization, \cf~Sect.~\ref{Sect:subroutines}.} All analysis files in this directory will then be read in automatically by \HS~during the initialization. 

As an example we show in Tab.~\ref{Tab:datafile} and~\ref{Tab:datafile2} the two data files for the inclusive measurement of the ATLAS $H\to ZZ^{(*)}\to 4\ell$ analysis~\cite{ATLAS-CONF-2013-013}, which define a peak observable and provide the full $\muobs$ plot as needed by the mass-centered $\chi^2$ method, respectively. The first $11$ rows of these files encode general information about the analysis and the observable (each row is required), as described in Tab.~\ref{tab:input}. Comments can be included in the top rows if they are starting with a \texttt{\#} symbol. Note that the \textit{observable ID} must be unique, whereas the \textit{analysis ID} must be the same for (peak- or mass-centered) observables, which correspond to the same analysis and where a multiple assignment of the same Higgs boson to the corresponding observables shall be avoided. In the (yet hypothetical) case that two distinct signals have been observed within the same analysis, their peak observables thus need to have the same analysis ID, otherwise a Higgs boson might be assigned to both signals. All integers should not have more than 10 digits.
%

\begin{table}
\renewcommand{\arraystretch}{1.0}
\centering
{\tt\footnotesize
\begin{tabular}{p{2.2cm}p{2.2cm}p{2.2cm}p{2.2cm}p{2.2cm}}
\br
2013013101	&	201301301	&	1	&	&	\\
\multicolumn{5}{l}{ATL-CONF-2013-013}	\\
LHC, &	ATL,&	 ATL	&	&	\\
\multicolumn{5}{l}{(pp)->h->ZZ->4l} 	\\
8 	&	25.3	&	0.036	&	&	\\
1	&	1		&			&	&	\\
1.1		&	&	&	&			\\
124.3	& 124.3 &	0.1	&	&	\\
4		&-1	&	&	&	\\
13	& 23 	&33 	&43 &	\\
& & & & \\
124.3	&	1.293	&	  1.697	&  2.194	&	\\
\br
\end{tabular}
}
\caption{Example file for an implemented peak observable. This file is located in the observable set \texttt{Expt\_tables/latestresults-1.0.0\_inclusive/} (with name \texttt{ATL\_H-ZZ-4l\_7-8TeV\_4.6fb-1\_20.7fb-1\_124.3GeV\_2013013101.txt}) and contains the information from the \texttt{ATLAS} search for the SM Higgs boson in the channel $H\to ZZ^{(*)}\to 4 \ell$~\cite{ATLAS-CONF-2013-013}. For a detailed description of each line in the file, see Tab.~\ref{tab:input}.}
\label{Tab:datafile}
\end{table}

 
 \begin{table}
\renewcommand{\arraystretch}{1.0}
\centering
{\tt\footnotesize
\begin{tabular}{p{2.2cm}p{2.2cm}p{2.2cm}p{2.2cm}p{2.2cm}}
\br
2013013201	&	201301301	&	2	&	&	\\
\multicolumn{5}{l}{ATL-CONF-2013-013}	\\
LHC, &	ATL,&	 ATL	&	&	\\
\multicolumn{5}{l}{(pp)->h->ZZ->4l} 	\\
8 	&	25.3	&	0.036	&	&	\\
1	&	1		&			&	&	\\
1.1	&	&	&	&	\\
110.0	&	 180.0	&	 0.1	&	&	\\
4		&	-1	&	&	&	\\
13	& 23 	&33 	&43	 &	\\
	&		&		& & \\
       110.0 &        -0.6568   &      -0.6395 &        -0.1845    \\
       110.1 &        -0.6563   &     -0.6384  &       -0.1730    \\
       110.2 &        -0.6558   &      -0.6372 &        -0.1615    \\
       110.3 &        -0.6552   &      -0.6361 &        -0.1499    \\
       110.4 &        -0.6547   &      -0.6349 &        -0.1384    \\
       \vdots	&       \vdots	&       \vdots	&       \vdots	&       	\\
\br
\end{tabular}
}
\caption{Example file for an analysis with a full $\muobs$ plot as needed for the mass-centered $\chi^2$ method. This file is located in \texttt{Expt\_tables/latestresults-1.0.0\_inclusive/} (with name \texttt{ATL\_H-ZZ-4l\_7-8TeV\_4.6fb-1\_20.7fb-1\_2013013201.txt}). It is the same analysis for which we already defined a peak observable in Tab.~\ref{Tab:datafile}. For a detailed description of each line in the file, see Tab.~\ref{tab:input}.}
\label{Tab:datafile2}
\end{table}

\begin{table}
\renewcommand{\arraystretch}{1.0}
\centering
\footnotesize
\begin{tabular}{cl}
\br
Row & Description\\
\mr
1 &	Observable ID, Analysis ID, Observable type (1: peak, 2: mass)\\
2 &	Publication reference\\
3 &	Collider ID, Collaboration ID, Experiment ID\\
4 &	Description of the search channel\\
5 &	CM energy (TeV), Integrated luminosity (fb$^{-1}$), Relative luminosity uncertainty\\
6 &	Higgs boson type (1: neutral, 2: charged), Enable $\chi^2$ from $m_H$ (0: no, 1: yes)\\
7 &	Mass resolution of analysis (GeV), assignment group (optional string without whitespaces) \\
8 &	Lowest Higgs mass, highest Higgs mass, Higgs mass interval  (of the following datatable)\\
9 &	Number of search channels, reference mass for efficiencies (-1: no efficiencies given)\\
10 & Search channel codes (see Tab.~\ref{tab:codes}) (\# entries must equal \# channels))\\
11 &	Channel efficiencies (\# entries must equal \# channels)\\
\br
\end{tabular}
\caption{Input format for general analysis information encoded in the
    first 11 rows of the experimental data file.}
\label{tab:input}
\end{table}

The channel codes in the $10^\mathrm{th}$ row are given as two-digit integers, where the first digit encodes the production mode, and the second digit the decay mode. The corresponding numbers are given in Tab.~\ref{tab:codes}. For example, the channel code of $(pp) \to HW \to (b\bar b) W$ is 35. In the example of Tab.~\ref{Tab:datafile}, we thus consider all five production modes, but only a single decay mode, \ie $H\to \gamma\gamma$.

Channel efficiencies can be included in the $11^\mathrm{th}$ row. They correspond to the channels as defined by the channel codes on the previous row, and thus have to be given in the same order. If the experimental channel efficiencies are unknown (as in the given example of an inclusive measurement), the reference mass in the $9^\mathrm{th}$ row should be set equal to $-1$, in which case the $11^\mathrm{th}$ row will be ignored. Since it must still be present, it could be left blank for the sake of clarity. Note that the channel efficiencies are defined as the fraction of events passing the analysis cuts, and \textit{not} the relative contribution of this channel to total signal yield. The latter would use information about the channel cross section, which in our case is already taken care of by the channel weights $\omega$, cf.~Eq.~\eqref{Eq:omega}. Furthermore, it is only the relative efficiencies among the channels that are important, and not their overall normalization (for the same reason). We therefore typically normalize the relative efficiencies such that the first element in the $11^\mathrm{th}$ row is equal to 1. As an example, the user may investigate one of the category measurements provided in the folder \texttt{Expt\_tables/latestresults-1.0.0/}.

From the $12^\mathrm{th}$ row onwards, the signal strength data is listed. Each row contains four values: the Higgs mass, the measured signal strength modifier at the lower edge of the $1\sigma$ uncertainty (``cyan'') band, $\hat\mu-\Delta\hat{\mu}$, the central value (best-fit) $\hat\mu$, and finally the signal strength modifier at the upper edge of the $1\,\sigma$ uncertainty band, $\hat\mu+\Delta\hat{\mu}$. In the case of a peak observable definition, as in Tab.~\ref{Tab:datafile}, the data file ends after the $12^\mathrm{th}$ row, since the signal strength is only measured at a single Higgs mass value (corresponding to the signal). In contrast, for the construction of mass-centered observables, the data is listed here for the full investigated mass range, which is typically extracted from the corresponding $\muobs$ plot using \texttt{EasyNData}~\cite{Uwer:2007rs}.

As a further remark, we point out a general limitation in the implementation of experimental data: some results from the LHC experiments are given for the combination of data collected at different center-of-mass energies, \eg at $7\tev$ and $8\tev$. These results cannot be disentangled by \HS. Therefore, these observables are implemented as if the data was collected at the center-of-mass energy, which can be assumed to be dominating the experimental data. This approximation is valid, since both the observed and the predicted signal strengths are treated as SM normalized quantities. The only remaining inaccuracy lies in the SM channel weights, Eq.~\eqref{Eq:omega}, which depend on the center-of-mass energy.

A complication arises in the assignment of Higgs observables if an
analysis with one measured mass peak value is split up in several
categories, each containing an individual signal rate measurement, see e.g.~\cite{ATLAS-CONF-2013-012,CMS-PAS-HIG-13-001}. In this case, each category result defines a peak observable, however only one of these observables can be associated with the mass
measurement from the analysis, which is going to contribute to the $\chi^2$. In all other categories this contribution has to be switched off. Nevertheless, this difference in the implementation can lead to inconsistent assignments of the Higgs boson(s) to the category observables. In order to enforce a consistent
assignment, peak observables can build an \textit{assignment group}. This enforces that the Higgs boson(s) are assigned to either all or none of the observables in this group, judged by the assignment status of the observable containing the mass measurement. For each peak observable, the assignment group can be specified in the experimental table, cf.
Tab.~\ref{tab:input}. Note that the analysis IDs of the category peak observables have to be different from each other.


\section{\HS\ applications}
\label{sect:Examples}

In this section we discuss a few example applications which demonstrate
the performance of \HS. Most of the examples are chosen such that their
results can be validated with official results from ATLAS and CMS. The
quality of agreement of the reproduced \HS~results with the official
results justifies the Gaussian limit approximation in
the statistical approach of \HS. Note that to a certain extent (which
is difficult to estimate), the accuracy of the reproduced results
suffers from the lack of publicly available information of the
analysis efficiencies on the various production modes. 
At the end of this section, we briefly discuss a few \HS\ example applications, where the results incorporate all presently available Higgs data from the LHC and the Tevatron. Another example application of \HS~within the context of the MSSM was presented in Ref.~\cite{Bechtle:2013gu}.


\clearpage
\subsection{Performance studies of \HS}

\subsubsection{The peak-centered $\chi^2$ method for a SM-like Higgs boson}
\label{Sect:pc_performance}
As a first application we discuss the performance of the peak-centered
$\chi^2$ method on a SM-like Higgs boson. As already shown in
Fig.~\ref{fig:muplots}(b), a simple one parameter fit can be performed
to the signal strength modifier $\mu$, which scales the predicted signal
rates of all investigated Higgs channels uniformly. In this fit the
Higgs mass is held fixed at \eg $\mH=125.7\gev$. Using the signal strength
measurements of the individual search channels obtained by the
CMS collaboration \cite{CMS-PAS-HIG-13-005}, as given in Fig.~\ref{fig:muplots}(b), the best-fit signal strength reconstructed with \HS~is $\muobs_\mathrm{comb} = 0.77\pm 0.14$. This agrees well with the official CMS result, $\muobs_\mathrm{comb}^\mathrm{CMS} = 0.80\pm 0.14$ \cite{CMS-PAS-HIG-13-005}. Using \HS\ with similar data from ATLAS \cite{ATLAS-CONF-2013-034}, where the experimental results for all categories are unfortunately not available at a common value for the Higgs mass, the published value of $\muobs_\mathrm{comb}^\mathrm{ATLAS}=1.30\pm 0.20$ at $m_H = 125.5\gev$ \cite{ATLAS-CONF-2013-034} is nevertheless reproduced reasonably well by $\muobs_\mathrm{comb} = 1.24\pm 0.20$.

\begin{figure}[t]
\centering
\includegraphics[width=0.9\textwidth]{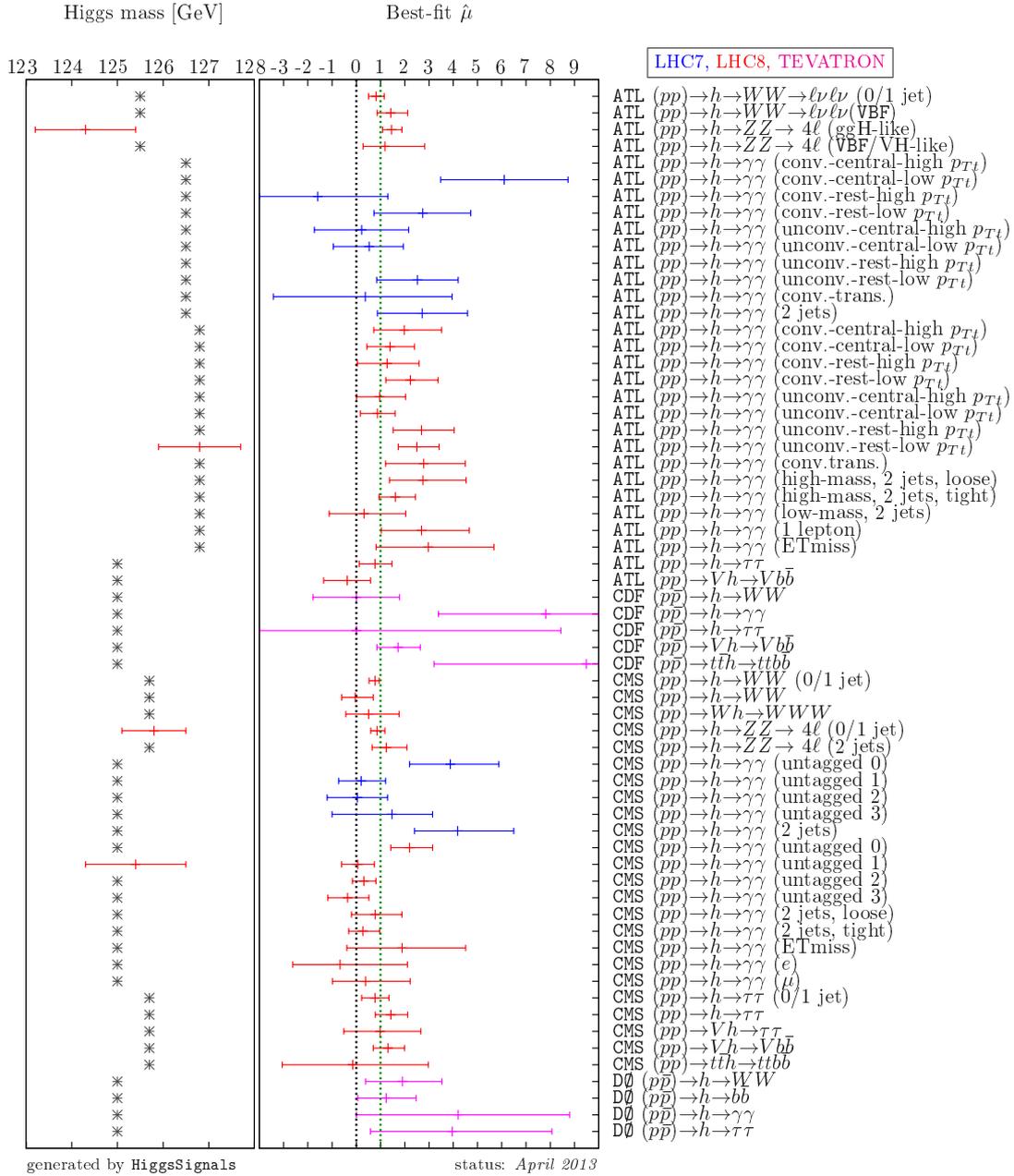}
\caption{Overview of the Higgs signal rate and mass measurements
(\textit{status shortly after the Moriond conference 2013}) from
ATLAS~\cite{ATLAS-CONF-2012-091,ATLAS-CONF-2012-160,ATLAS-CONF-2012-161,ATLAS-CONF-2012-170,ATLAS-CONF-2013-012,ATLAS-CONF-2013-013,ATLAS-CONF-2013-030,ATLAS-CONF-2013-034,Aad:2013wqa}, CMS~\cite{CMS-PAS-HIG-11-024,*CMS-PAS-HIG-12-042,CMS-PAS-HIG-12-015,CMS-PAS-HIG-12-039,CMS-PAS-HIG-12-044,CMS-PAS-HIG-12-045,CMS-PAS-HIG-13-001,CMS-PAS-HIG-13-002,CMS-PAS-HIG-13-003,CMS-PAS-HIG-13-004,CMS-PAS-HIG-13-005,Chatrchyan:2013yea}
and the Tevatron experiments CDF~\cite{Aaltonen:2013ipa} and D\O~\cite{Abazov:2013zea}, as they are implemented in
\HSv{1.0.0}~as \textit{peak observables}. The left panel shows the Higgs
mass value for which the signal strength was measured. A value with
error bars indicates that the mass value is treated as a Higgs mass
observable in the peak-centered $\chi^2$ method, whereas a gray asterisk
only serves as an indication of the Higgs mass value, which was assumed in the rate measurement. This value does not enter directly the total $\chi^2$. For some
LHC analyses, measurements for both the $7\tev$ and $8\tev$ data exist,
shown in blue and red, respectively. If the measurement is based on the
combined $7/8\tev$ dataset, we treat it as an $8\tev$ measurement only. For the $H\to\gamma\gamma$ analyses from ATLAS and CMS, the special tagged categories were implemented as separate peak observables, including their efficiencies, but collected together in assignment groups. In total there are 4 Higgs mass observables and 63 Higgs signal rate observables. This data is used for the performance scans in Fig.~\ref{Fig:SM} and the example applications in Sect.~\ref{sect:Combinedfits}.}
\label{Fig:peakobservables}
\end{figure}

Now, we collect as peak observables the measured signal rates from the LHC experiments
ATLAS~\cite{ATLAS-CONF-2012-091,ATLAS-CONF-2012-160,ATLAS-CONF-2012-161,ATLAS-CONF-2012-170,ATLAS-CONF-2013-012,ATLAS-CONF-2013-013,ATLAS-CONF-2013-030,ATLAS-CONF-2013-034,Aad:2013wqa}
and CMS~\cite{CMS-PAS-HIG-11-024,*CMS-PAS-HIG-12-042,CMS-PAS-HIG-12-015,CMS-PAS-HIG-12-039,CMS-PAS-HIG-12-044,CMS-PAS-HIG-12-045,CMS-PAS-HIG-13-001,CMS-PAS-HIG-13-002,CMS-PAS-HIG-13-003,CMS-PAS-HIG-13-004,CMS-PAS-HIG-13-005,Chatrchyan:2013yea}, as
well as the Tevatron experiments CDF~\cite{Aaltonen:2013ipa} and D\O~\cite{Abazov:2013zea}, as
summarized in Fig.~\ref{Fig:peakobservables}. If possible, we implement
results from the $7\tev$ and $8\tev$ LHC runs as separate observables.
However, if the only quoted result is a combination of both
center-of-mass energies we implement it as an $8\tev$ result. As mentioned in Sect.~\ref{sect:Statistics}, we employ the quoted asymmetric uncertainties to account for the dominant effects of potentially remaining non-Gaussian behavior of the measurements. The $H\to \gamma \gamma$ and
$H\to ZZ^{(*)}\to 4\ell$ analyses of ATLAS and CMS have a
rather precise mass
resolution, thus we treat the implemented mass value of their signal as
a measurement which enters the Higgs mass part of the total $\chi^2$,
\cf Sect.~\ref{Sect:pc_chisq}. Note however that the implemented mass
value is not necessarily the most precise measurement of the Higgs mass
but rather the mass value for which the signal strength was published
by the experimental analysis. The Higgs mass can be determined more
accurately from a simultaneous fit to the mass and the signal
strength. This can be done with the mass-centered $\chi^2$ method, as
discussed in the next subsection. Note also that the Higgs mass values assumed in the signal strength measurements can differ by up to $\sim 2.5\gev$. It would be desirable if the experiments would present their best-fit signal strengths for all available channels (including specially tagged categories) also for a common Higgs mass (equal or close to the Higgs mass value preferred by the combined data) once a combination of different channels is performed. In the present case, global fits combining the signal strength measurements performed at different Higgs masses rely on the assumption that these measurements do not vary too much within these mass differences.

It can nevertheless be interesting to discuss the total $\chi^2$ distribution
obtained in the peak-centered $\chi^2$ method as a function of the
Higgs mass, $m_H$. This
serves as a demonstration of the three different Higgs mass
uncertainty parametrizations (box, Gaussian, box+Gaussian pdfs), as
well as the implications of taking into account the correlations among
the systematic uncertainties in the $\chi^2$ calculation. Furthermore,
features of the automatic assignment of the Higgs boson to the peak
observables can be studied. 
In the following example, we set the predicted signal strength for all Higgs channels to their SM values ($\mu_i \equiv 1$) and set the production and decay rate uncertainties to the values given in Eq.~\eqref{Eq:SMrateuncertainties}, as recommended by the LHC Higgs Cross Section Working Group for the SM Higgs boson around $m_H \simeq 125\gev$. We then evaluate the total peak-centered $\chi^2$ for each Higgs boson mass $m_H \in [110,~140]\gev$ using the peak observables presented in Fig.~\ref{Fig:peakobservables}. In the SM the Higgs mass is treated as a free parameter, which corresponds to setting the theory mass uncertainty to zero. In order to illustrate the effects of a non-zero theory mass uncertainty, we also consider a model with SM-like Higgs couplings, but which has a $2\gev$ theory uncertainty on the predicted Higgs mass.
\begin{figure}[t]
\centering
\subfigure[No correlations, $\Delta m^\mathrm{th} = 0$ GeV.\label{Fig:SM_corr0dm0}]{
\includegraphics[angle=270, width=0.47\textwidth]{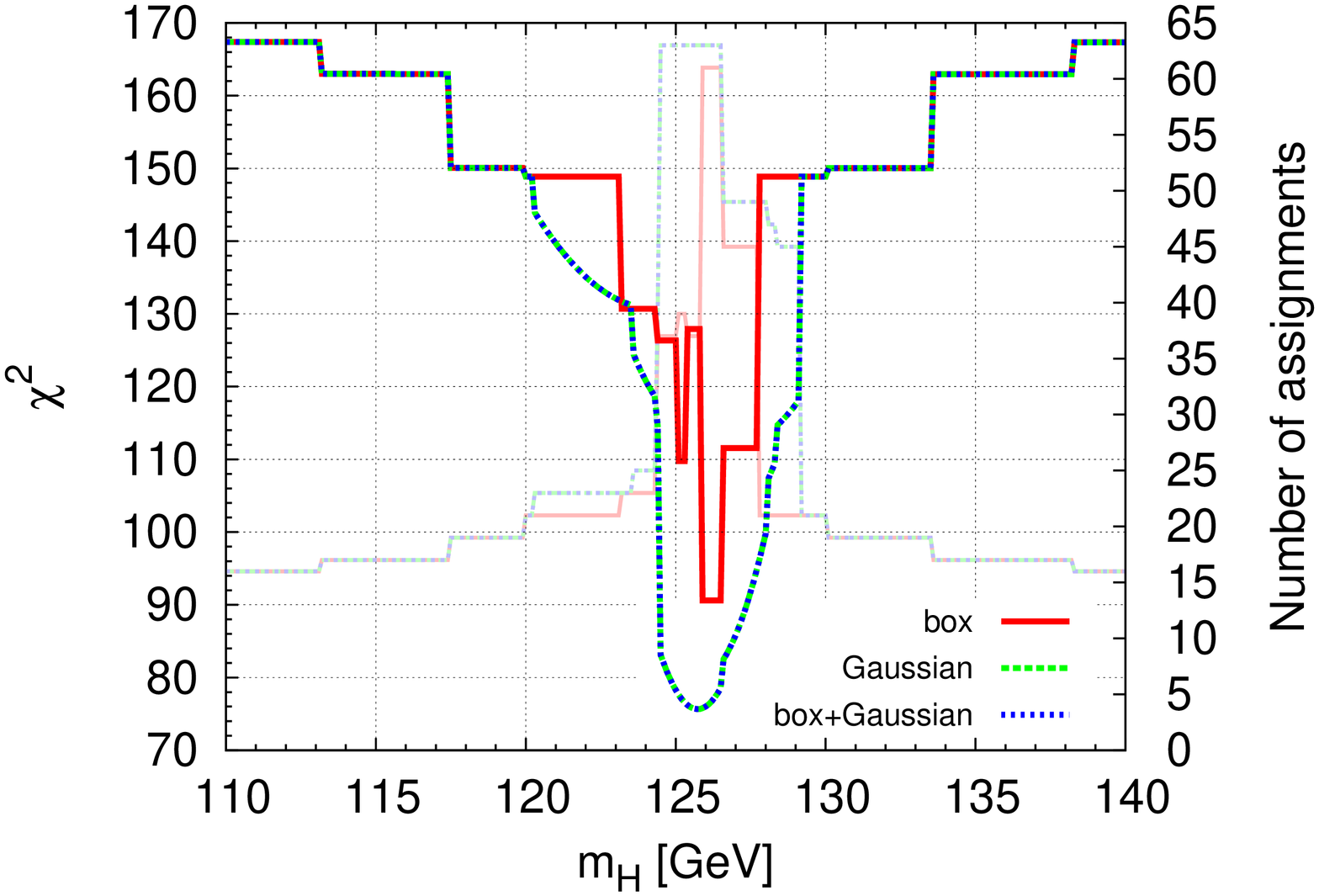}
}\hfill
\subfigure[No correlations, $\Delta m^\mathrm{th} = 2$ GeV.\label{Fig:SM_corr0dm2}]{
\includegraphics[angle=270, width=0.47\textwidth]{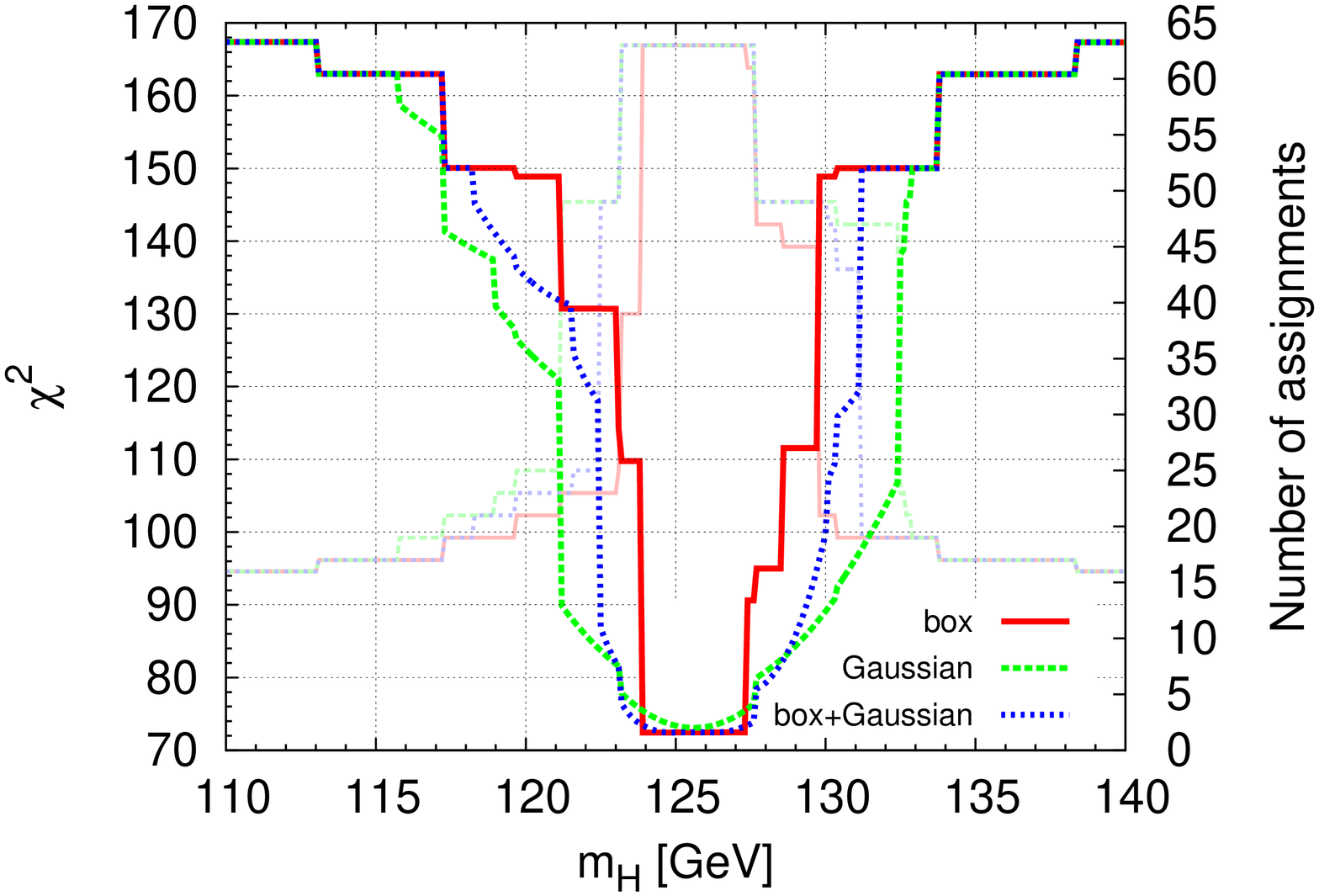}
}
\subfigure[With correlations, $\Delta m^\mathrm{th} = 0$ GeV.\label{Fig:SM_corr1dm0}]{
\includegraphics[angle=270, width=0.47\textwidth]{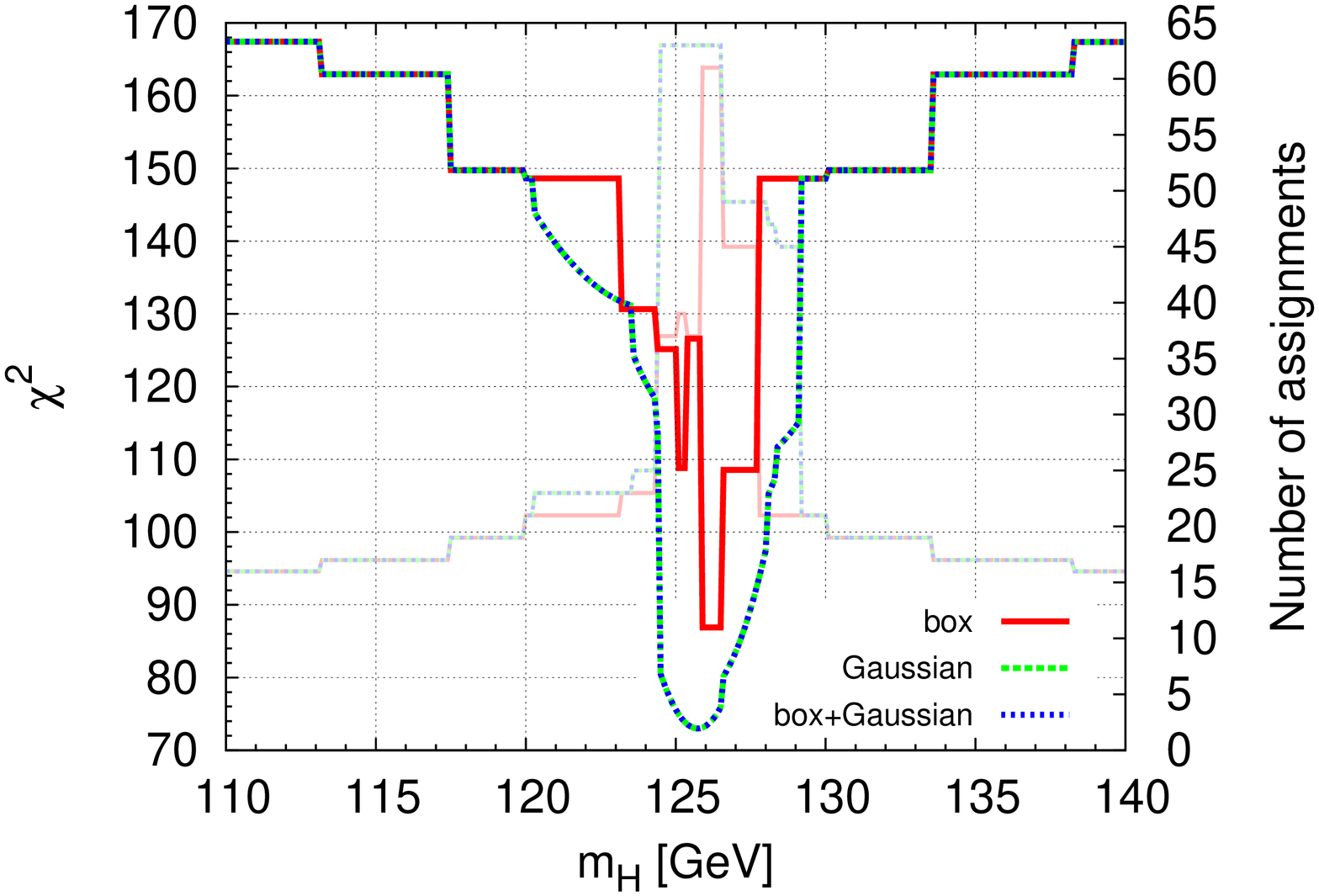}
}\hfill
\subfigure[With correlations, $\Delta m^\mathrm{th} = 2$ GeV.\label{Fig:SM_corr1dm2}]{
\includegraphics[angle=270, width=0.47\textwidth]{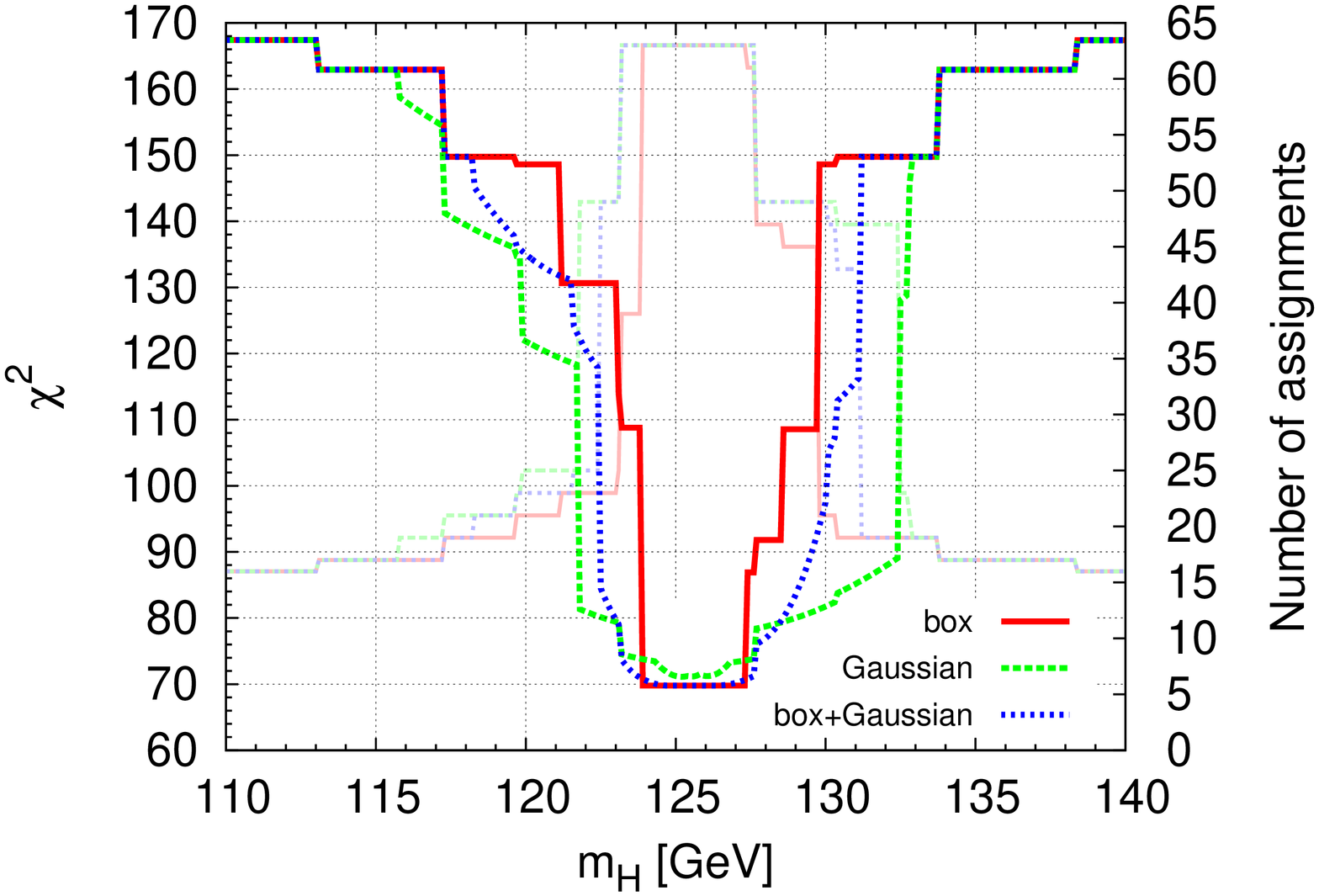}
}
\caption{Total $\chi^2$ distribution obtained by the peak-centered
$\chi^2$ method for a SM Higgs boson with mass $m_H$ obtained from the 63 peak observables (status: April 2013) shown in~Fig.~\ref{Fig:peakobservables}. In (a, b), the total $\chi^2$ is evaluated without taking into account the correlations among the systematic uncertainties, whereas they are fully included in (c, d). In (a, c) no theoretical mass uncertainty $\Delta m$ is assumed (like in the SM) whereas in (b, d) we set $\Delta m=2\gev$. For each setting, we show the total $\chi^2$ obtained for all three parametrizations of the theoretical Higgs mass uncertainty: box (solid red), Gaussian (dashed green) and box+Gaussian (dotted blue) pdf. For each case, we also give the total number of peak observables, which have been assigned with the Higgs boson, depicted by the corresponding faint lines.} 
\label{Fig:SM}
\end{figure}

The total $\chi^2$ mass distribution is shown in Fig.~\ref{Fig:SM} for
four different cases: In Fig.~\ref{Fig:SM}(a,b) the correlations among
the systematic uncertainties of the signal rates, luminosity and Higgs mass predictions are neglected, whereas they are taken into
account in Fig.~\ref{Fig:SM}(c,d). In order to demonstrate
the difference between the three parametrizations of the Higgs mass
uncertainty we show the $\chi^2$ distribution assuming a theoretical Higgs mass
uncertainty of $\dmth=0\gev$ in Fig.~\ref{Fig:SM}(a,c) and
$\dmth=2\gev$ in Fig.~\ref{Fig:SM}(b,d), respectively. Furthermore, Fig.~\ref{Fig:SM} includes the number of peak observables, which have been assigned with the Higgs boson, as a function of the Higgs mass. These are depicted by the faint graphs for each Higgs mass uncertainty parametrization.

The discontinuous shape of the $\chi^2$ distribution is caused by
changes in the Higgs boson assignment to the individual
observables. Recall that, if the Higgs mass $m_H$ is too far away from
the implemented mass position of the peak observable, 
the Higgs boson is not assigned to the signal. This yields a $\chi^2$ contribution
corresponding to no predicted signal, $\mu=0$,
\cf Sect.~\ref{Sect:pc_chisq}. Most of the peak observables have
different mass resolutions, therefore the $\chi^2$ distribution has a
staircase-like shape. At each step, the total number of peak observable assignments changes.

As can be seen in Fig.~\ref{Fig:SM} all three 
parametrizations of the theoretical Higgs mass uncertainty
yield the same total $\chi^2$ values if the Higgs
mass $m_H$ is far away from the implemented signal mass position,
because typically observables which enter the Higgs mass part of the
$\chi^2$ in the Gaussian parametrization exhibit a decent mass
resolution, and the Higgs boson is only assigned if this $\chi^2$ is
low, \ie $m_H \approx \mobs$. Conversely, at the $\chi^2$ minimum at a Higgs mass $m_H\sim 125-126\gev$, we obtain slightly different $\chi^2$ values for the three parametrizations: Firstly, assuming that every observable is assigned with the Higgs boson, the minimal $\chi^2$ is in general slightly higher in the Gaussian case than in the box and box+Gaussian case if the Higgs mass measurements do not have the same central values for all (mass sensitive) peak observables. In that case, there will always be a non-zero $\chi^2$ contribution from the Higgs mass measurements for any predicted value of the Higgs mass. Secondly, in the case of no theoretical mass uncertainty, the box parametrization does not exhibit a full assignment of all currently implemented peak observables at any Higgs mass value. This is because the mass measurements of the ATLAS $H\to\gamma\gamma$~\cite{ATLAS-CONF-2013-012} and $H\to ZZ^{(*)}\to4\ell$~\cite{ATLAS-CONF-2013-013} observables have a mass difference of $2.5\gev$, which corresponds to a discrepancy of around $2.5~\sigma$~\cite{ATLAS-CONF-2013-014}. Thus, the Higgs boson is only assigned to either of these (groups of) observables, receiving a maximal $\chi^2$ penalty from the other observable (group). In fact, we observe a double minimum structure in Fig.\ref{Fig:SM}(a,c), because for a Higgs mass $m_H \in [125.4,~125.8]\gev$, neither the ATLAS $H\to\gamma\gamma$ nor the $H\to ZZ^{(*)}\to4\ell$ observables are assigned with the Higgs boson, leading to a large total $\chi^2$. This illustrates that the box-shaped pdf is an inappropriate description of the Higgs mass likelihood in the absence of sizable theoretical mass uncertainties.

A difference between the Gaussian and
the theory box with experimental Gaussian (box+Gaussian)
parametrization appears only for non-zero $\dmth$. For $\dmth=2\gev$
the minimal $\chi^2$ is obtained for a plateau $m_H \approx (124.8 -
126.5)\gev$ in the box+Gaussian case, whereas in the Gaussian case we
have a non-degenerate minimum at $m_H=125.7\gev$. However, outside this plateau the
$\chi^2$ shape of the box+Gaussian increases faster than in the
Gaussian case, since the uncertainty governing this Gaussian slope is
smaller.

For the Gaussian parametrization of the theoretical Higgs mass uncertainty and no theoretical mass uncertainty
the minimal $\chi^2$ at $m_H=125.7\gev$ changes from $75.7$ to $73.0$ (for $63$ signal strength observables and $4$ mass observables)
 if we include the correlations among the systematic
uncertainties in the $\chi^2$ evaluation. 
In the case of a non-zero theoretical mass uncertainty, also the shape of the total $\chi^2$ distribution can be affected when the correlations are taken into account. Recall that only in the Gaussian parametrization the correlations of the theoretical mass uncertainties enter the $\chi^2$ evaluation, featuring a sign dependence on the relative position of the predicted Higgs mass value with respect to the two observed Higgs mass values, cf. Sect.~\ref{Sec:HiggsMassObs}. This results in a shallower slope of the $\chi^2$ distribution at Higgs masses larger than all mass measurements, $m_H \gtrsim 126.8\gev$, since all mass observables are positively correlated in this case.

In conclusion we would like to emphasize that, although the direct $\chi^2$ contribution from (the few) mass measurements to the total $\chi^2$ might appear small in comparison to the $\chi^2$ contribution from (many) signal strength measurements, the automatic assignment of Higgs boson(s) to the peak observables introduces a strong mass dependence, even for peak observables without an implemented mass measurement. Hereby, the procedure tries to ensure that a comparison of the predicted and observed signal strength is valid for each observable (depending on the mass resolution of the corresponding Higgs analysis), or otherwise considers the signal as not explainable by the model.

\subsubsection{Combining search channels with the mass-centered $\chi^2$ method}


\begin{figure}[t]
\centering
\subfigure[Simultaneous evaluation of $7$ and $8\tev$ results from the ATLAS SM $H\to \gamma\gamma$ search~\cite{ATLAS-CONF-2012-091}.]{\includegraphics[width=0.47\textwidth]{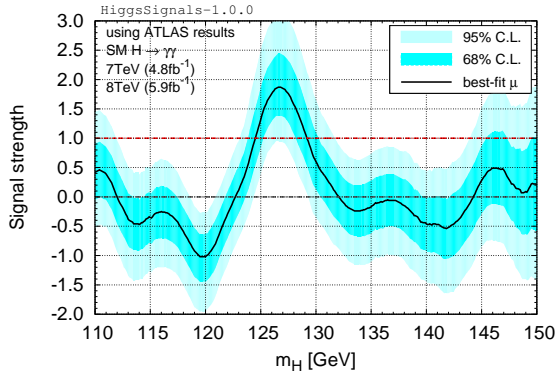}}\hfill
\subfigure[Simultaneous evaluation of ATLAS searches for $H\to \gamma\gamma,~ZZ$ and $WW$~\cite{ATLAS-CONF-2012-091,ATLAS-CONF-2012-098,ATLAS-CONF-2012-092}.]{\includegraphics[width=0.47\textwidth]{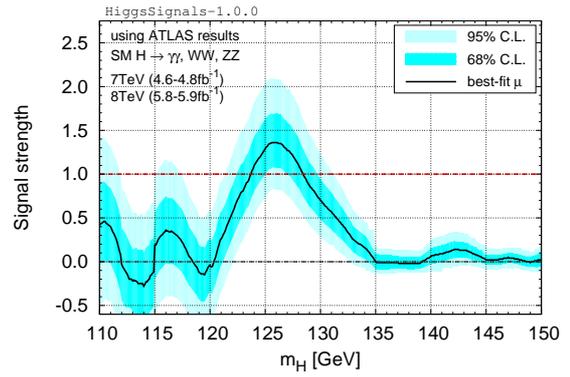}}
\subfigure[Official ATLAS combination of $7$ and $8\tev$ results from the ATLAS SM $H\to \gamma\gamma$ search~\cite{ATLAS-CONF-2012-091}.]{\includegraphics[width=0.47\textwidth]{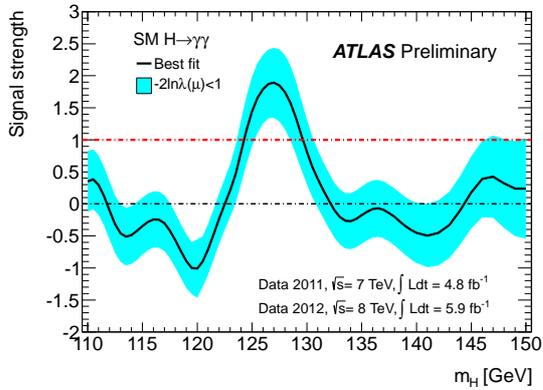}}\hfill
\subfigure[Official ATLAS combination of the SM $H\to \gamma\gamma,~ZZ,~WW,~b\bar{b}$ and $\tau^+\tau^-$ searches~\cite{ATLAS:2012gk}.]{\includegraphics[width=0.47\textwidth]{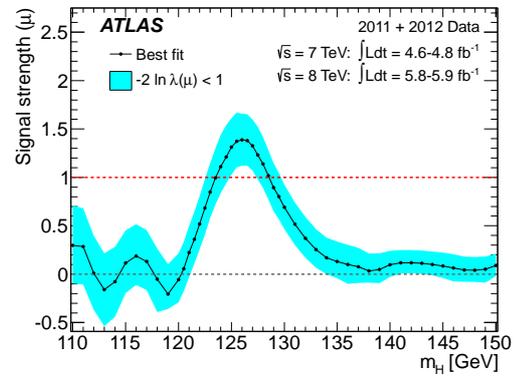}}
\caption{Reconstruction of the combined best-fit signal strength from the results of the individual dataset / channels with the mass-centered $\chi^2$ method (a, b). For comparison, we give the official ATLAS results in (c, d).
}
\label{Fig:mc_comb}
\end{figure}



\begin{figure}[t]
\centering
\includegraphics[width=0.8\textwidth]{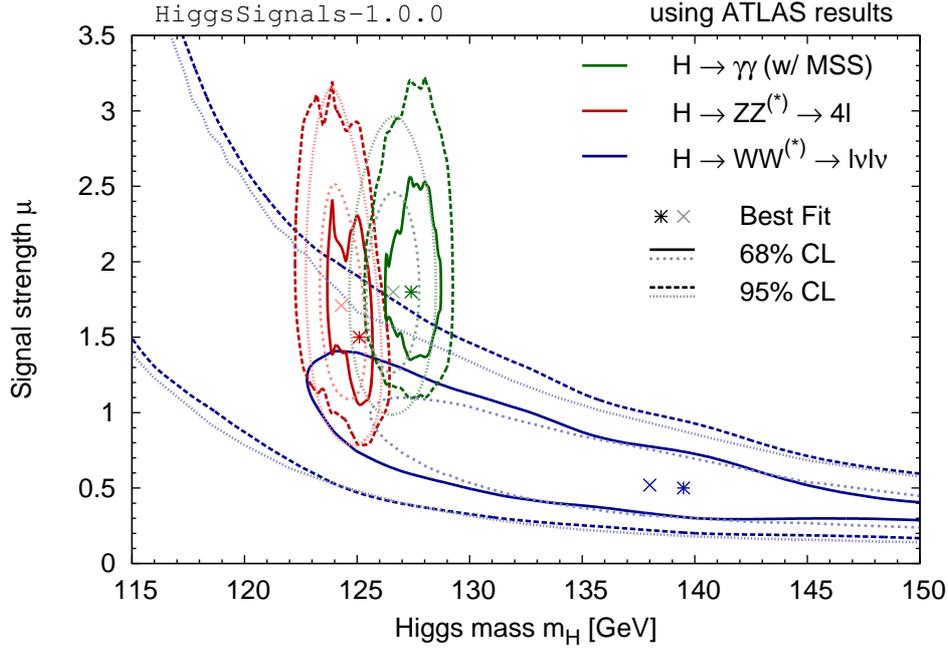}
\caption{Results from a simultaneous fit to the Higgs mass and signal strength using the experimental data from the ATLAS searches $H\to\gamma\gamma$~\cite{ATLAS-CONF-2012-168}, $H\to WW^{(*)}\to \ell\nu\ell\nu$~\cite{ATLAS-CONF-2013-030} and $H\to ZZ^{(*)}\to 4\ell$~\cite{ATLAS-CONF-2013-013}. The corresponding results from ATLAS are overlaid as faintly colored contours.}
\label{Fig:mhmufit_Moriond2013}
\end{figure}


As a first demonstration of the mass-centered $\chi^2$ method we evaluate simultaneously the $7\tev$ and $8\tev$ results from ATLAS for the Higgs searches $H\to\gamma\gamma$~\cite{ATLAS-CONF-2012-091}, as well as its evaluation together with the $H\to WW^{(*)}\to \ell\nu\ell\nu$~\cite{ATLAS-CONF-2012-098} and $H\to ZZ^{(*)}\to 4\ell$~\cite{ATLAS-CONF-2012-092} searches. This is possible because the full $\muobs$ plot was published for these analyses for $7\tev$ and $8\tev$, except for the $H\to ZZ^{(*)}\to 4\ell$ search where only the combined $7/8\tev$ result is available.\footnote{Since it is not possible to disentangle this result into $7\tev$ and $8\tev$, we implemented this observable as $8\tev$ only data in \HS.}

We scan the relevant Higgs mass range $m_H=(110 - 150)\gev$, as well as
the signal strength $\mu$, and at each point ($m_H,~\mu$) evaluate the
mass-centered $\chi^2$ using the corresponding $\muobs$ plots as
\textit{mass-centered observables}. We then find the best-fit $\mu$
value (and the corresponding $1\sigma$ and $2\sigma$ regions) by
minimizing the $\chi^2$ (finding $\Delta \chi^2 = 1$ and $\Delta \chi^2
= 4$, respectively) for a fixed Higgs mass $m_H$. This is shown in
Fig.~\ref{Fig:mc_comb}(a) and~\ref{Fig:mc_comb}(b) for the
$H\to\gamma\gamma$ channel and the combination of $H\to\gamma\gamma$,
$H\to WW^{(*)}\to \ell\nu\ell\nu$ and $H\to ZZ^{(*)}\to 4\ell$,
respectively. These results nicely agree with the corresponding official
ATLAS results~\cite{ATLAS-CONF-2012-091, ATLAS:2012gk}, which are shown
in Fig.~\ref{Fig:mc_comb}(c,d) for comparison. Especially at the signal
around $\simeq 126\gev$ the Gaussian limit approximation works very well
due to the relatively large number of events (in the
$H\to\gamma\gamma$ analysis). Note 
that in Fig.~\ref{Fig:mc_comb}(d) also the channels $H\to\tau\tau$ and
$VH \to b\bar{b}$ are included, however, these observables are rather
insignificant for this result due to large uncertainties on the signal
strength measurement as well as a poor mass resolution. 

Instead of minimizing the $\chi^2$ for a fixed Higgs mass $m_H$, we now perform a two parameter fit to $m_H$ and $\mu$, using the latest currently available $\muobs$ plots from the ATLAS searches\footnote{ATLAS did not include a new $\muobs$ plot in their $H\to\gamma\gamma$ search update at the Moriond 2013 conference~\cite{ATLAS-CONF-2013-012}. Therefore, we have to use an older result here. We use the $\muobs$ plot from~\cite{ATLAS-CONF-2012-168} which includes the mass scale systematic (MSS) uncertainty.} $H\to\gamma\gamma$~\cite{ATLAS-CONF-2012-168}, $H\to WW^{(*)}\to \ell\nu\ell\nu$~\cite{ATLAS-CONF-2013-030} and $H\to ZZ^{(*)}\to 4\ell$~\cite{ATLAS-CONF-2013-013}. For a given signal hypothesis, ($m_H,~\mu)$, we scan the full mass range, $m_H'\in [115,~150]\gev$ with a step size of $0.1\gev$, and the signal strength modifier $\mu'$ in steps of $0.05$.
 For each scanning point we evaluate the mass-centered $\chi^2$ value, $\chi^2_\mathrm{MC}$, for the hypothesis ($m_H',~\mu'$), where
\begin{equation}
\mu' = \left\{ 	\begin{array}{ll} 	
				\mu 	& \mbox{if}\quad m_H' = m_H, \\
				0	& \mbox{if}\quad m_H' \ne m_H. \\ 
	 	  	\end{array}
	   \right.
\end{equation}
The obtained $\chi^2$ values from this scan are summed and associated
with the point ($m_H,~\mu)$. 
Thus we test the combined hypothesis of having a
Higgs boson at $m_H$ with signal strength $\mu$, and no signal elsewhere.
The procedure is then repeated for all points in the two-dimensional
($m_H, \mu$) plane to obtain the 2D $\chi^2$ likelihood map. The results are
shown in Fig.~\ref{Fig:mhmufit_Moriond2013} for each Higgs decay mode separately. For
comparison, we also show the official ATLAS
results~\cite{ATLAS-CONF-2013-030,ATLAS-CONF-2012-168, Aad:2013wqa} as faintly colored contours. Qualitatively, the obtained
$68\%$ and $95\%$ C.L. regions (corresponding to $\Delta\chi^2=2.30$ and $\Delta \chi^2=5.99$, respectively) agree fairly well for $H\to ZZ$ and $H\to WW$, whereas the $H\to \gamma\gamma$ result is shifted towards larger Higgs masses by around $0.8\gev$. A potential reason for this discrepancy is that effects of the mass scale systematic (MSS) uncertainty are only indirectly taken into account in \HS\ by simply using the corresponding plateau-shaped $\muobs$ plot~\cite{ATLAS-CONF-2012-168} instead of including the MSS uncertainty in the profile likelihood as a nuisance. Nevertheless, the $68\%$ and $95\%$ C.L. regions still have a large overlap. Note also that the spiky structures of the contour ellipses
  in Fig.~\ref{Fig:mhmufit_Moriond2013} are rather an artifact of
  our data extraction with \texttt{EasyNData}~\cite{Uwer:2007rs} than a physical
  effect.\footnote{It would therefore be desirable if the experimental
 collaborations published the data of the $\muobs$ plots also in
 tabular form in accurate precision.} 

A simultaneous fit to the ATLAS Higgs channels $H\to\gamma\gamma$~\cite{ATLAS-CONF-2012-168}, $H\to ZZ^{(*)}\to4\ell$~\cite{ATLAS-CONF-2013-013} and $H\to WW^{(*)}\to \ell \nu \ell \nu$~\cite{ATLAS-CONF-2013-030} can also be performed. The best fit point of such a combination is found at 
\begin{equation}
m_H = 125.4\substack{+0.2\\-0.4} \gev,\quad \mu = 1.4 \substack{+ 0.3 \\ -0.2},
\end{equation}
where the uncertainties given refer to the 1D profiled $68\%$ confidence interval. We have verified that these results remain stable when varying the step sizes in the scan.

The two discussed examples show the usefulness of the mass-centered
$\chi^2$ method. We focussed here on the validation of the method by
comparing with official results from ATLAS. It is however easy to go
beyond that and take all available data from ATLAS and CMS (and 
the Tevatron) into account for a simultaneous analysis. This we leave for a future
study. However, we would like to emphasize again that the usefulness
of this method strongly depends on the information (here in particular
the $\muobs$ plots for the individual channels) the experimental
collaborations decide to publish.


\subsection[Validation with official fit results for Higgs coupling scaling factors]{Validation with official fit results for Higgs coupling scaling factors\sectionmark{Validation with official fit results\dots}}\label{Sec:ValidationOffEff}
\sectionmark{Validation with official fit results\dots}

A major task after the discovery of a Higgs-like state is the
determination of its coupling properties and thus a thorough test of its
compatibility with the SM. Both ATLAS~\cite{ATLAS-CONF-2013-034,ATLAS-CONF-2012-127} and
CMS~\cite{CMS-PAS-HIG-12-045,CMS-PAS-HIG-13-005} have obtained results for Higgs 
coupling scaling factors in the
framework of restricted benchmark models proposed by the LHC Higgs Cross
Section Working Group~\cite{LHCHiggsCrossSectionWorkingGroup:2012nn}. 
Numerous other studies have been performed, both for Higgs coupling
scaling factors
~\citeHiggsEffC\ as well as for particular models, including composite Higgs scenarios~\citeCompHiggsAndEffC, Two Higgs Doublet Models (2HDMs)~\citeTHDM, supersymmetric models~\citeHiggsSUSY\ as well as other, more exotic extensions of the SM~\citeSMextensions.
 Here, we want to focus on the reproduction of the official
ATLAS and CMS results using the Higgs coupling scaling factors as
defined in the benchmark models of Ref.~\cite{LHCHiggsCrossSectionWorkingGroup:2012nn} in order to validate the \HS\ implementation.

\begin{table}
\renewcommand{\arraystretch}{1.0}
\centering
\footnotesize
 \begin{tabular}{lccrrrrr}
\br
 Higgs search channel & energy $\sqrt{s}$ & $\muobs \pm \dmuobs $ & \multicolumn{5}{c}{SM signal composition [in \%]} \\
& & & ggH & VBF & WH &  ZH & $t\bar{t}H$ \\
 \mr
$ H\to WW^{(*)}\to \ell\nu\ell\nu$ (0/1 jet)~\cite{ATLAS-CONF-2013-030,Aad:2013wqa} & $ 7/8\tev$ &$  0.82\substack{+  0.33\\ -  0.32}$ & $  97.2$ & $   1.6$ & $   0.7$ & $   0.4$ & $   0.1$\\ 
$ H\to WW^{(*)}\to \ell\nu\ell\nu$ (2 jet)~\cite{ATLAS-CONF-2013-030,Aad:2013wqa} & $ 7/8\tev$ &$  1.42\substack{+  0.70\\ -  0.56}$ & $  19.8$ & $  80.2$ & $   0.0$ & $   0.0$ & $   0.0$\\ 
$ H\to ZZ^{(*)}\to 4\ell$ (ggH-like)~\cite{ATLAS-CONF-2013-013,Aad:2013wqa} &$ 7/8\tev$ & $  1.45\substack{+  0.43\\ -  0.37}$ & $  92.5$ & $   4.5$ & $   1.9$ & $   1.1$ & $   0.0$\\ 
$ H\to ZZ^{(*)}\to 4\ell$ (VBF/VH-like)~\cite{ATLAS-CONF-2013-013,Aad:2013wqa} &$ 7/8\tev$ & $  1.18\substack{+  1.64\\ -  0.90}$ & $  36.8$ & $  43.1$ & $  12.8$ & $   7.3$ & $   0.0$\\ 
$ H\to \gamma\gamma$ (unconv.-central-low $p_{Tt}$)~\cite{ATLAS-CONF-2012-091} & $ 7\tev$ &$  0.53\substack{+  1.41\\ -  1.48}$ & $  92.9$ & $   3.8$ & $   2.0$ & $   1.1$ & $   0.2$\\ 
$ H\to \gamma\gamma$ (unconv.-central-high $p_{Tt}$)~\cite{ATLAS-CONF-2012-091} &$ 7\tev$ & $  0.22\substack{+  1.94\\ -  1.95}$ & $  65.5$ & $  14.8$ & $  10.8$ & $   6.2$ & $   2.7$\\ 
$ H\to \gamma\gamma$ (unconv.-rest-low $p_{Tt}$)~\cite{ATLAS-CONF-2012-091} & $ 7\tev$ &$  2.52\substack{+  1.68\\ -  1.68}$ & $  92.6$ & $   3.7$ & $   2.2$ & $   1.2$ & $   0.2$\\ 
$ H\to \gamma\gamma$ (unconv.-rest-high $p_{Tt}$)~\cite{ATLAS-CONF-2012-091} & $ 7\tev$ &$ 10.44\substack{+  3.67\\ -  3.70}$ & $  64.4$ & $  15.2$ & $  11.8$ & $   6.6$ & $   2.0$\\ 
$ H\to \gamma\gamma$ (conv.-central-low $p_{Tt}$)~\cite{ATLAS-CONF-2012-091}& $ 7\tev$ &$  6.10\substack{+  2.63\\ -  2.62}$ & $  92.7$ & $   3.8$ & $   2.1$ & $   1.1$ & $   0.2$\\ 
$ H\to \gamma\gamma$ (conv.-central-high $p_{Tt}$)~\cite{ATLAS-CONF-2012-091} & $ 7\tev$ &$ -4.36\substack{+  1.80\\ -  1.81}$ & $  65.7$ & $  14.4$ & $  11.0$ & $   6.2$ & $   2.8$\\ 
$ H\to \gamma\gamma$ (conv.-rest-low $p_{Tt}$)~\cite{ATLAS-CONF-2012-091}  & $ 7\tev$ &$  2.74\substack{+  1.98\\ -  2.01}$ & $  92.7$ & $   3.6$ & $   2.2$ & $   1.2$ & $   0.2$\\ 
$ H\to \gamma\gamma$ (conv.-rest-high $p_{Tt}$)~\cite{ATLAS-CONF-2012-091} &$ 7\tev$ & $ -1.59\substack{+  2.89\\ -  2.90}$ & $  64.4$ & $  15.1$ & $  12.1$ & $   6.4$ & $   2.0$\\ 
$ H\to \gamma\gamma$ (conv.-trans.)~\cite{ATLAS-CONF-2012-091} & $ 7\tev$ &$  0.37\substack{+  3.58\\ -  3.79}$ & $  89.2$ & $   5.0$ & $   3.7$ & $   1.9$ & $   0.3$\\ 
$ H\to \gamma\gamma$ (2 jet)~\cite{ATLAS-CONF-2012-091} & $ 7\tev$ &$  2.72\substack{+  1.87\\ -  1.85}$ & $  23.3$ & $  75.9$ & $   0.5$ & $   0.2$ & $   0.1$\\ 
$ H\to \gamma\gamma$ (unconv.-central-low $p_{Tt}$)~\cite{ATLAS-CONF-2013-012} & $ 8\tev$ &$  0.87\substack{+  0.73\\ -  0.70}$ & $  92.0$ & $   5.0$ & $   1.7$ & $   0.8$ & $   0.5$\\ 
$ H\to \gamma\gamma$ (unconv.-central-high $p_{Tt}$)~\cite{ATLAS-CONF-2013-012} &$ 8\tev$ & $  0.96\substack{+  1.07\\ -  0.95}$ & $  78.6$ & $  12.6$ & $   4.7$ & $   2.6$ & $   1.4$\\ 
$ H\to \gamma\gamma$ (unconv.-rest-low $p_{Tt}$)~\cite{ATLAS-CONF-2013-012}& $ 8\tev$ &$  2.50\substack{+  0.92\\ -  0.77}$ & $  92.0$ & $   5.0$ & $   1.7$ & $   0.8$ & $   0.5$\\ 
$ H\to \gamma\gamma$ (unconv.-rest-high $p_{Tt}$)~\cite{ATLAS-CONF-2013-012} &$ 8\tev$ & $  2.69\substack{+  1.35\\ -  1.17}$ & $  78.6$ & $  12.6$ & $   4.7$ & $   2.6$ & $   1.4$\\ 
$ H\to \gamma\gamma$ (conv.-central-low $p_{Tt}$)~\cite{ATLAS-CONF-2013-012} & $ 8\tev$ &$  1.39\substack{+  1.01\\ -  0.95}$ & $  92.0$ & $   5.0$ & $   1.7$ & $   0.8$ & $   0.5$\\ 
$ H\to \gamma\gamma$ (conv.-central-high $p_{Tt}$)~\cite{ATLAS-CONF-2013-012} & $ 8\tev$ &$  1.98\substack{+  1.54\\ -  1.26}$ & $  78.6$ & $  12.6$ & $   4.7$ & $   2.6$ & $   1.4$\\ 
$ H\to \gamma\gamma$ (conv.-rest-low $p_{Tt}$)~\cite{ATLAS-CONF-2013-012}  & $ 8\tev$ &$  2.23\substack{+  1.14\\ -  1.01}$ & $  92.0$ & $   5.0$ & $   1.7$ & $   0.8$ & $   0.5$\\ 
$ H\to \gamma\gamma$ (conv.-rest-high $p_{Tt}$)~\cite{ATLAS-CONF-2013-012} &$ 8\tev$ & $  1.27\substack{+  1.32\\ -  1.23}$ & $  78.6$ & $  12.6$ & $   4.7$ & $   2.6$ & $   1.4$\\ 
$ H\to \gamma\gamma$ (conv.trans.)~\cite{ATLAS-CONF-2013-012} & $ 8\tev$ &$  2.78\substack{+  1.72\\ -  1.57}$ & $  92.0$ & $   5.0$ & $   1.7$ & $   0.8$ & $   0.5$\\ 
$ H\to \gamma\gamma$ (high-mass, 2 jet, loose)~\cite{ATLAS-CONF-2013-012} &$ 8\tev$ & $  2.75\substack{+  1.78\\ -  1.38}$ & $  45.3$ & $  53.7$ & $   0.5$ & $   0.3$ & $   0.2$\\ 
$ H\to \gamma\gamma$ (high-mass, 2 jet, tight)~\cite{ATLAS-CONF-2013-012} & $ 8\tev$ &$  1.61\substack{+  0.83\\ -  0.67}$ & $  27.1$ & $  72.5$ & $   0.3$ & $   0.1$ & $   0.0$\\ 
$ H\to \gamma\gamma$ (low-mass, 2 jet)~\cite{ATLAS-CONF-2013-012}  & $ 8\tev$ &$  0.32\substack{+  1.72\\ -  1.44}$ & $  38.0$ & $   2.9$ & $  40.1$ & $  16.9$ & $   2.1$\\ 
$ H\to \gamma\gamma$ ($E_T^\mathrm{miss}$ sign.)~\cite{ATLAS-CONF-2013-012} &$ 8\tev$ & $  2.97\substack{+  2.71\\ -  2.15}$ & $   4.4$ & $   0.3$ & $  35.8$ & $  47.4$ & $  12.2$\\ 
$ H\to \gamma\gamma$ ($1\ell$)~\cite{ATLAS-CONF-2013-012} & $ 8\tev$ &$  2.69\substack{+  1.97\\ -  1.66}$ & $   2.5$ & $   0.4$ & $  63.3$ & $  15.2$ & $  18.7$\\ 
$ H\to \tau\tau$~\cite{ATLAS-CONF-2013-034,ATLAS-CONF-2012-160} & $ 7/8\tev$ &$  0.77\substack{+  0.70\\ -  0.65}$ & $  88.1$ & $   7.1$ & $   3.1$ & $   1.7$ & $   0.0$\\ 
$ VH\to V(bb)$~\cite{ATLAS-CONF-2013-034,ATLAS-CONF-2012-161} &$ 7/8\tev$ & $ -0.38\substack{+  0.97\\ -  0.97}$ & $   0.0$ & $   0.0$ & $  63.8$ & $  36.2$ & $   0.0$\\ 
 \br
 \end{tabular}
\caption{Signal strength measurements, $\muobs$, from various ATLAS Higgs searches implemented in \HS\ as peak observables. Results from combined $7/8\tev$ data are implemented as $8\tev$-only in \HS. The $H\to\gamma\gamma$ measurements where performed at a Higgs mass of $m_H = 126.5\gev$ [$126.8\gev$] for the $7\tev$ [$8\tev$] results, while the remaining channels are measured at $m_H = 125.5\gev$. In the last columns, we give the assumed signal composition for a SM Higgs boson.}
\label{Tab:ATLAS_peakobs}
\end{table}

\begin{table}
\renewcommand{\arraystretch}{1.0}
\centering
\begin{threeparttable}[b]
\footnotesize
 \begin{tabular}{lccrrrrr}
 \br
Higgs search channel &  energy $\sqrt{s}$ & $\muobs \pm \dmuobs $ & \multicolumn{5}{c}{SM signal composition [in \%]} \\
& & & ggH & VBF & WH &  ZH & $t\bar{t}H$ \\
 \mr
$H\to WW^{(*)} \to \ell\nu\ell\nu$ (0/1 jet)~\cite{CMS-PAS-HIG-13-003} & $7/8\tev$ & $  0.77\substack{+  0.17\\ -  0.24}$ & $  95.0$ & $   5.0$ & $   0.0$ & $   0.0$ & $   0.0$\\ 
$H\to WW^{(*)}\to \ell\nu\ell\nu$ (VBF)~\cite{CMS-PAS-HIG-11-024,*CMS-PAS-HIG-12-042} &$7/8\tev$ & $ -0.05\substack{+  0.75\\ -  0.55}$ & $  38.2$ & $  61.8$ & $   0.0$ & $   0.0$ & $   0.0$\\ 
$WH\to W(WW^{(*)}) \to 3\ell3\nu$~\cite{CMS-PAS-HIG-13-009} &$7/8\tev$ & $  0.51\substack{+  1.26\\ -  0.94}$ & $   0.0$ & $   0.0$ & $  100.0$\tnote{1} & $   0.0$ & $   0.0$\\ 
$H\to ZZ^{(*)}\to 4\ell$ (0/1 jet)~\cite{CMS-PAS-HIG-13-002} & $7/8\tev$ & $  0.86\substack{+  0.32\\ -  0.26}$ & $  89.8$ & $  10.2$ & $   0.0$ & $   0.0$ & $   0.0$\\ 
$H\to ZZ^{(*)}\to 4\ell$ (2 jet)~\cite{CMS-PAS-HIG-13-002} & $7/8\tev$ &$  1.24\substack{+  0.85\\ -  0.58}$ & $  71.2$ & $  28.8$ & $   0.0$ & $   0.0$ & $   0.0$\\ 
$H\to \gamma\gamma$ (untagged 0)~\cite{CMS-PAS-HIG-12-015,CMS-PAS-HIG-13-001} &$7\tev$ & $  3.88\substack{+  2.00\\ -  1.68}$ & $  61.4$ & $  16.9$ & $  12.0$ & $   6.6$ & $   3.1$\\ 
$H\to \gamma\gamma$ (untagged 1)~\cite{CMS-PAS-HIG-12-015,CMS-PAS-HIG-13-001} &$7\tev$ & $  0.20\substack{+  1.01\\ -  0.93}$ & $  87.7$ & $   6.2$ & $   3.6$ & $   2.0$ & $   0.5$\\ 
$H\to \gamma\gamma$ (untagged 2)~\cite{CMS-PAS-HIG-12-015,CMS-PAS-HIG-13-001} &$7\tev$ & $  0.04\substack{+  1.25\\ -  1.24}$ & $  91.4$ & $   4.4$ & $   2.5$ & $   1.4$ & $   0.3$\\ 
$H\to \gamma\gamma$ (untagged 3)~\cite{CMS-PAS-HIG-12-015,CMS-PAS-HIG-13-001} &$7\tev$ & $  1.47\substack{+  1.68\\ -  2.47}$ & $  91.3$ & $   4.4$ & $   2.6$ & $   1.5$ & $   0.2$\\ 
$H\to \gamma\gamma$ (2 jet)~\cite{CMS-PAS-HIG-12-015,CMS-PAS-HIG-13-001} & $7\tev$ &$  4.18\substack{+  2.31\\ -  1.78}$ & $  26.7$ & $  72.6$ & $   0.4$ & $   0.2$ & $   0.0$\\ 
$H\to \gamma\gamma$ (untagged 0)~\cite{CMS-PAS-HIG-13-001} & $8\tev$ &$  2.20\substack{+  0.95\\ -  0.78}$ & $  72.9$ & $  11.7$ & $   8.2$ & $   4.6$ & $   2.6$\\ 
$H\to \gamma\gamma$ (untagged 1)~\cite{CMS-PAS-HIG-13-001} &$8\tev$ & $  0.06\substack{+  0.69\\ -  0.67}$ & $  83.5$ & $   8.5$ & $   4.5$ & $   2.6$ & $   1.0$\\ 
$H\to \gamma\gamma$ (untagged 2)~\cite{CMS-PAS-HIG-13-001} &$8\tev$ & $  0.31\substack{+  0.50\\ -  0.47}$ & $  91.5$ & $   4.5$ & $   2.3$ & $   1.3$ & $   0.4$\\ 
$H\to \gamma\gamma$ (untagged 3)~\cite{CMS-PAS-HIG-13-001} &$8\tev$ & $ -0.36\substack{+  0.88\\ -  0.81}$ & $  92.5$ & $   3.9$ & $   2.1$ & $   1.2$ & $   0.3$\\ 
$H\to \gamma\gamma$ (2 jet, tight)~\cite{CMS-PAS-HIG-13-001} &$8\tev$ & $  0.27\substack{+  0.69\\ -  0.58}$ & $  20.6$ & $  79.0$ & $   0.2$ & $   0.1$ & $   0.1$\\ 
$H\to \gamma\gamma$ (2 jet, loose)~\cite{CMS-PAS-HIG-13-001} & $8\tev$ &$  0.78\substack{+  1.10\\ -  0.98}$ & $  46.8$ & $  51.1$ & $   1.1$ & $   0.6$ & $   0.5$\\ 
$H\to \gamma\gamma$ ($\mu$)~\cite{CMS-PAS-HIG-13-001} & $8\tev$ &$  0.38\substack{+  1.84\\ -  1.36}$ & $   0.0$ & $   0.2$ & $  50.4$ & $  28.6$ & $  20.8$\\ 
$H\to \gamma\gamma$ ($e$)~\cite{CMS-PAS-HIG-13-001} & $8\tev$ &$ -0.67\substack{+  2.78\\ -  1.95}$ & $   1.1$ & $   0.4$ & $  50.2$ & $  28.5$ & $  19.8$\\ 
$H\to \gamma\gamma$ ($E_T^\mathrm{miss}$)~\cite{CMS-PAS-HIG-13-001} &$8\tev$ & $  1.89\substack{+  2.62\\ -  2.28}$ & $  22.1$ & $   2.6$ & $  40.6$ & $  23.0$ & $  11.7$\\ 
$H\to \tau\tau$ (0/1 jet)~\cite{CMS-PAS-HIG-13-004} &$7/8\tev$ & $  0.77\substack{+  0.58\\ -  0.55}$ & $  95.0$ & $   5.0$ & $   0.0$ & $   0.0$ & $   0.0$\\ 
$H\to \tau\tau$ (VBF)~\cite{CMS-PAS-HIG-13-004} &$7/8\tev$ & $  1.42\substack{+  0.70\\ -  0.64}$ & $  19.8$ & $  80.2$ & $   0.0$ & $   0.0$ & $   0.0$\\ 
$VH\to V(\tau\tau)$~\cite{CMS-PAS-HIG-13-004, CMS-PAS-HIG-12-053} & $7/8\tev$ &$  0.98\substack{+  1.68\\ -  1.50}$ & $   0.0$ & $   0.0$ & $  17.2$ & $   9.8$ & $   0.0$\\ 
$VH\to V(bb)$~\cite{CMS-PAS-HIG-13-001,CMS-PAS-HIG-12-044} &$7/8\tev$ & $  1.30\substack{+  0.73\\ -  0.63}$ & $   0.0$ & $   0.0$ & $  63.8$ & $  36.2$ & $   0.0$\\ 
$ttH\to tt(bb)$~\cite{Chatrchyan:2013yea} & $7/8\tev$ &$ -0.15\substack{+  3.12\\ -  2.90}$ & $   0.0$ & $   0.0$ & $   0.0$ & $   0.0$ & $ 100.0$\\ 
\br
 \end{tabular}
 \begin{tablenotes}
 \footnotesize
 \item[1] The signal is contaminated to $12.0\%$ by $WH\to W(\tau\tau)$.
 \end{tablenotes}
 \end{threeparttable}
\caption{Signal strength measurements, $\muobs$, from various CMS Higgs searches implemented in \HS\ as peak observables. Results from combined $7/8\tev$ data are implemented as $8\tev$-only in \HS. The $H\to\gamma\gamma$ measurements where performed at a Higgs mass of $m_H = 125.0\gev$, while the remaining channels are measured at $m_H = 125.7\gev$. In the last columns, we give the assumed signal composition for a SM Higgs boson.}
\label{Tab:CMS_peakobs}

\end{table}

We validate with the ATLAS and CMS results, as presented at the Moriond 2013 conference~\cite{ATLAS-CONF-2013-034,CMS-PAS-HIG-12-045}. The measurements from ATLAS and CMS, which are used as observables for our reproduced fits, are summarized in Tabs.~\ref{Tab:ATLAS_peakobs} and~\ref{Tab:CMS_peakobs}, respectively.
In the ATLAS fits of Higgs coupling scaling factors the Higgs mass is assumed to be $m_H =125.5\gev$.
However, for a Higgs mass of $125.5\gev$ there
are no signal strengths measurements for the $H\to\gamma\gamma$
categories available in the literature. Instead, we use the $\muobs$
measurements performed at $126.5\gev$ and $126.8\gev$ for the $7$ and $8\tev$ data, respectively~\cite{ATLAS-CONF-2012-091,ATLAS-CONF-2013-012}, keeping in mind that this might lead to
some inaccuracies. The ATLAS $H\to WW^{(*)}\to \ell\nu\ell\nu$ and $H\to ZZ^{(*)}\to 4\ell$ signal strength measurements
were extracted from Ref.~\cite{Aad:2013wqa}. Note that for the remaining channels, $H\to \tau \tau$ and $VH \to Vb\bar{b}$, only the inclusive $\muobs$ measurements are available in the literature, whereas the ATLAS fit also includes information of their sub-channels~\cite{ATLAS-CONF-2013-034}. In the CMS fits of Higgs coupling scaling factors a Higgs mass of $m_H=125.7\gev$ is assumed. All signal strength measurements, as listed in Tab.~\ref{Tab:CMS_peakobs}, have been performed for this assumed Higgs mass value, except for the $H\to\gamma\gamma$ categories being measured at $m_H =125.0\gev$.


\begin{figure}[t]
\centering
\subfigure[Comparison with ATLAS results~\cite{ATLAS-CONF-2013-034,Aad:2013wqa}. Both the $68\%$ and $95\%$ C.L. regions are shown.]{\includegraphics[trim=0cm 0cm 0cm 0cm, width=0.49\textwidth]{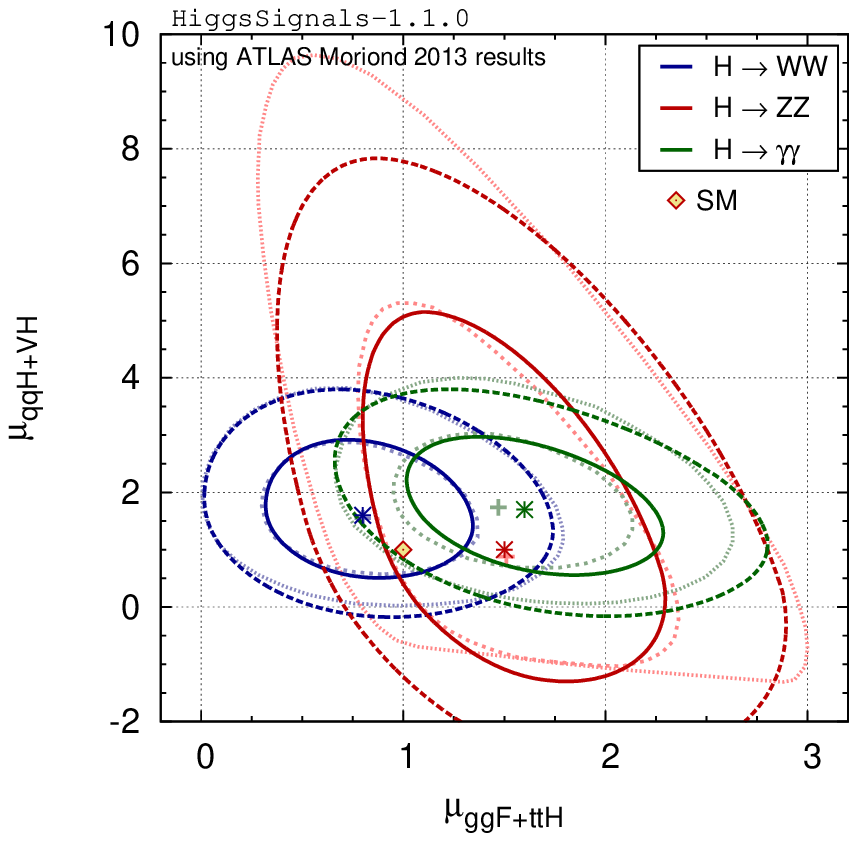}}\hfill
\subfigure[Comparison with CMS results~\cite{CMS-PAS-HIG-13-005}. Only the $68\%$ C.L. regions are shown.]{\includegraphics[trim=0cm 0cm 0cm 0cm,width=0.49\textwidth]{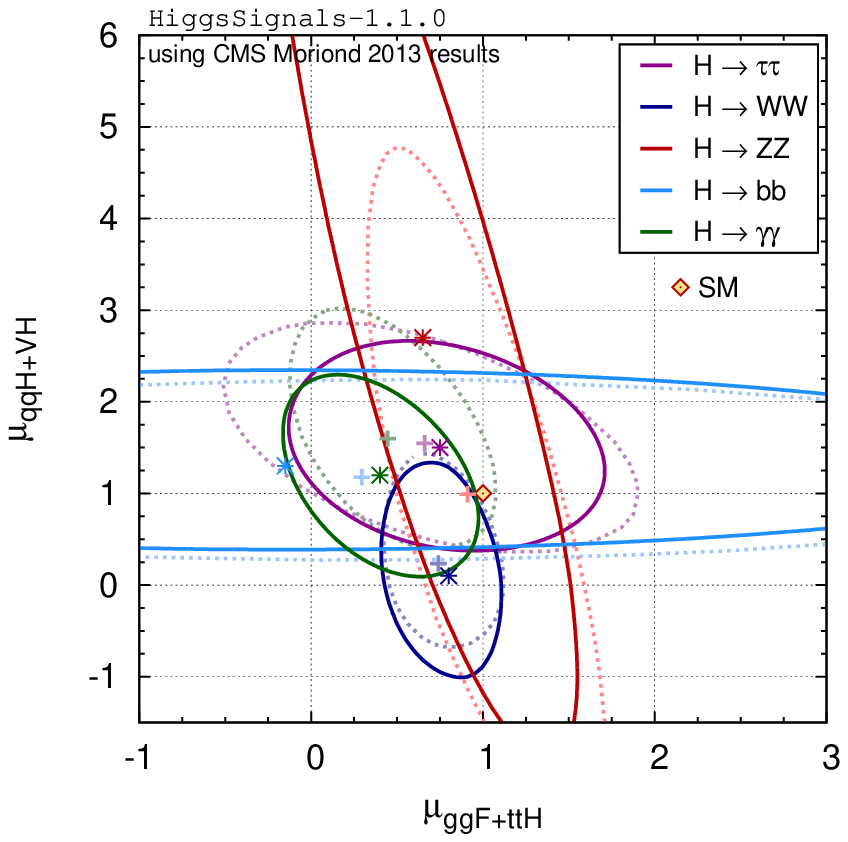}}
\caption{Comparison of fit results for the universal scale factors for the production cross sections of gluon-gluon fusion (ggf) and top quark pair associated Higgs production (ttH), $\mu_\mathrm{ggf+ttH}$, and of vector boson fusion (qqH) and vector boson associated Higgs production (VH), $\mu_\mathrm{qqH+VH}$, using the individual Higgs search channel results from ATLAS [in (a)] and CMS [in (b)]. The $68\%$ ($95\%$) C.L. regions are shown as deep colored, solid (dashed) and faintly colored, dotted (fine-dotted) contours for the \HS\ results and official ATLAS/CMS result, respectively. The best fit points are given by the asterisk [plus sign] for the \HS\ [official] result.
 }
\label{Fig:CSscaling}
\end{figure}


Before we discuss the benchmark fits of Higgs coupling scale factors, we look at  ATLAS and CMS fits that explicitly target the different production modes by combining channels with a particular decay mode. These fits allow to investigate sources of potential deviations between the official and the reproduced \HS\ results separately for each Higgs boson decay mode. Furthermore, unknown channel efficiencies can be adjusted within reasonable ranges, such that the agreement of the fit outcome is optimized. The signal composition of all included observables after this optimization is given in Tabs.~\ref{Tab:ATLAS_peakobs} and \ref{Tab:CMS_peakobs}. For the $H\to \gamma\gamma$ categories we use the published channel efficiencies.

Two-parameter fits were performed for each decay mode
to a signal strength modifier associated with the gluon fusion (ggf) and $t\bar{t}H$ production
mechanisms, $\mu_\mathrm{ggf+ttH}$, and a signal strength modifier for
the VBF and $VH$ production modes, $\mu_\mathrm{qqH+VH}$. 
The results of the same fits performed with \HS\ are shown in Fig.~\ref{Fig:CSscaling} in direct comparison with the results from ATLAS~\cite{ATLAS-CONF-2013-034,Aad:2013wqa} and CMS~\cite{CMS-PAS-HIG-13-005}, which are faintly overlaid in the figure.
  Using the ATLAS results, Fig.~\ref{Fig:CSscaling}(a), the derived
  $H\to WW$ ellipse is in perfect agreement with the official
  result. Also the $H\to \gamma\gamma$ and $H\to ZZ$ ellipses agree
  reasonably well.  The reproduced $H\to\gamma\gamma$ ellipse is
  slightly shifted towards larger values of $\mu_\mathrm{ggf+ttH}$.
  A potential source of this discrepancy may be the different mass positions at which the measurements are performed. Moreover, the inclusion of correlations
  among the experimental systematic uncertainties becomes more
  important, the more the measurements are divided into smaller
  subsets/categories. These correlations are not publicly known
    and hence not taken into account by \HS. In the $H\to ZZ$ result,
  a significant difference between the approximations in
    \HS\ and the full profile likelihood (PLL) treatment can be
  observed. The PLL has a longer tail at large signal strengths, thus
  leading to extended $68\%$ and $95\%$ C.L. regions at large values
  of $\mu_\mathrm{qqH+VH}$. This is partly due to the Gaussian approximation, which is more constraining
    at large values than a Poisson distribution with the same central value, as is used in the PLL. This is especially relevant for the very small event count for VBF $H\to ZZ$ candidates. In addition,
    missing information about correlations of experimental systematics
    might contribute to the observed difference at large
    $\mu_\mathrm{qqH+VH}$. Note also that one of the two $H\to ZZ$ category measurements that are publicly available~\cite{Aad:2013wqa}, cf.~Tab.~\ref{Tab:ATLAS_peakobs}, is a combination of the VBF and $VH$ production channels, whereas the ATLAS analysis internally treats these channels as separate categories. The requirement of a
  positive probability density function (pdf) leads to the edge at negative
  $\mu_\mathrm{qqH+VH}$ in the official ATLAS result. We checked
  that adding the requirement of a positive signal strength modifier
  in \HS this edge is reproduced quite well.

Using the CMS results, Fig.~\ref{Fig:CSscaling}(b), we find reasonably good agreement between \HS\ and the official results for $H\to WW,~bb,~\mbox{and}~\tau\tau$. The $H\to\gamma\gamma$ ellipses roughly agree in the $\mu_\mathrm{ggf+ttH}$ range as well as in the correlations of the fit parameters (seen in the tilt of the ellipses). However, our reproduced ellipse is shifted towards lower values of $\mu_\mathrm{qqH+VH}$. 
In order to investigate the influence of correlated experimental systematic uncertainties, we introduced a tunable degree of correlation among the VBF-tagged $H\to\gamma\gamma$ categories.
A much better agreement between \HS\ and the official result is obtained when around $30\%$ of the measured relative signal strength uncertainty of the VBF-tagged categories is treated as a fully correlated uncertainty. This indicates that including this type of (not public) information could potentially lead to an improvement of the \HS\ methodology in certain channels.
A similar effect from correlations of experimental systematics may lead to the differences observed in the $H\to\tau\tau$ ellipses. The $H\to ZZ$ ellipse can only be roughly reproduced using the publicly available data for the two $H\to ZZ$ observables. Even after adjusting their production mode efficiencies, cf. Tab.~\ref{Tab:CMS_peakobs}, differences remain due to the Gaussian approximation and possibly further (publicly unavailable) information on the VBF-likeness of the observed signal events~\cite{CMS-PAS-HIG-13-002}.

Using the results in Fig.~\ref{Fig:CSscaling}, we can estimate
  the typical differences between the official results from ATLAS and
  CMS and the \HS\ implementation. We classify the difference in two ways: first, the
  $\Delta\chi^2$ in our fit between the official best fit point from
  the collaboration and the best fit point from \HS, and second, the
  distance between the two best fit points in the parameter space
  relative to the $1\sigma$ uncertainty in the direction spanned by these two best-fit points. 

  For the comparison with the official ATLAS result, cf.~Fig.~\ref{Fig:CSscaling}(a), the $\Delta\chi^2$ is $0.158$,
  $3.5\times10^{-4}$ and $3.6\times10^{-3}$ for $H\to\gamma\gamma$,
  $H\to WW$ and $H\to ZZ$, respectively. 
  For $H\to\gamma\gamma$ the difference is small but non-negligible, as pointed out before.
  The latter two can be regarded as insignificant. The difference between the
  best fit points of ATLAS and \HS, relative to the corresponding $1\sigma$
  uncertainty is $24\%$, $6.6\%$ and $7.7\%$, respectively. Also
  here, a reasonable agreement well within $1\sigma$ is observed.

  For the comparison with the official CMS result, cf.~Fig.~\ref{Fig:CSscaling}(b), the differences in $\chi^2$ between the best fit points are
  $0.51$ for $H\to\gamma\gamma$, $0.34$ for $H\to ZZ$, and less than
  $0.05$ for the other channels. Plausible reasons for the differences in
  $H\to\gamma\gamma$ and $H\to ZZ$ are discussed above. For the
  remainder of channels there is very good agreement. The same picture
  arises for the relative distance of best fit points in parameter
  space with respect to the $1\sigma$ uncertainty measured in the same direction, where the largest deviation is observed for $H\to\gamma\gamma$ with $44\%$. Still, this
  is well within $1\sigma$ and should be sufficient for exploratory
  studies of new physics models. All other channels agree significantly better. 
  
\begin{figure}[t]
\centering
\subfigure[\HS\ result.]{\includegraphics[ width=8.0cm, height=6.2cm]{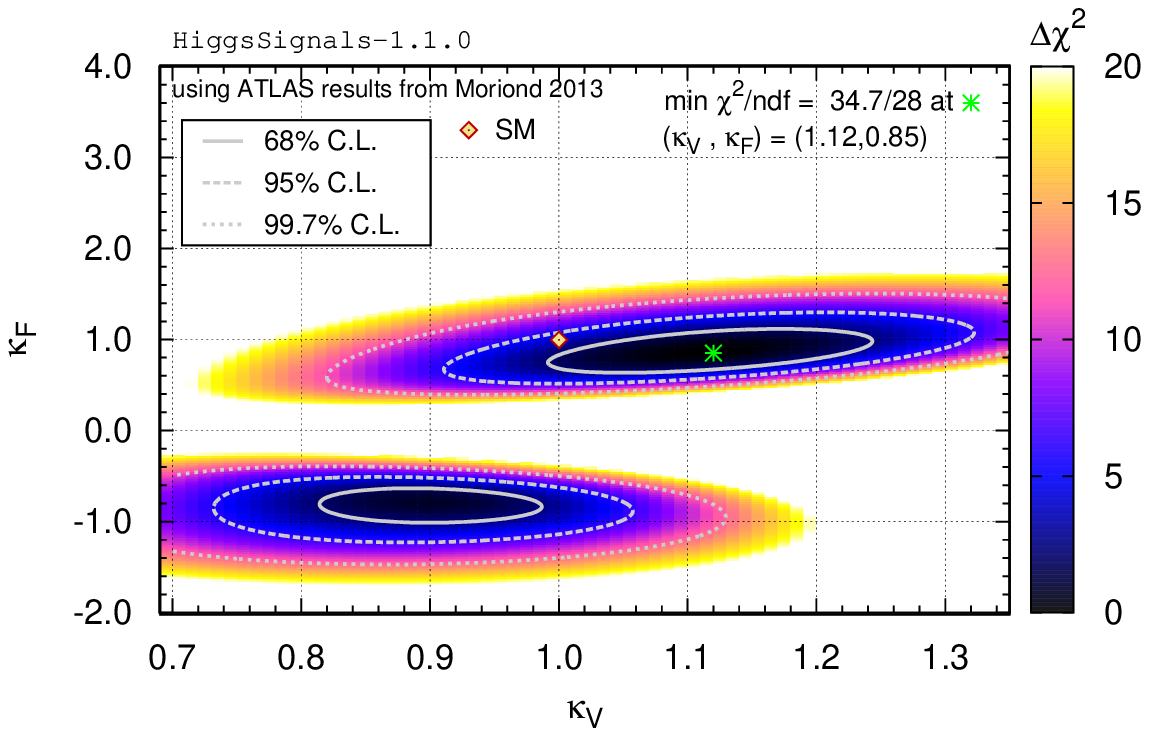}}\hfill
\subfigure[Official ATLAS result from Ref.~\cite{ATLAS-CONF-2013-034}.]{\includegraphics[trim = 0cm -1.4cm 0cm 0cm, clip,width=7.2cm]{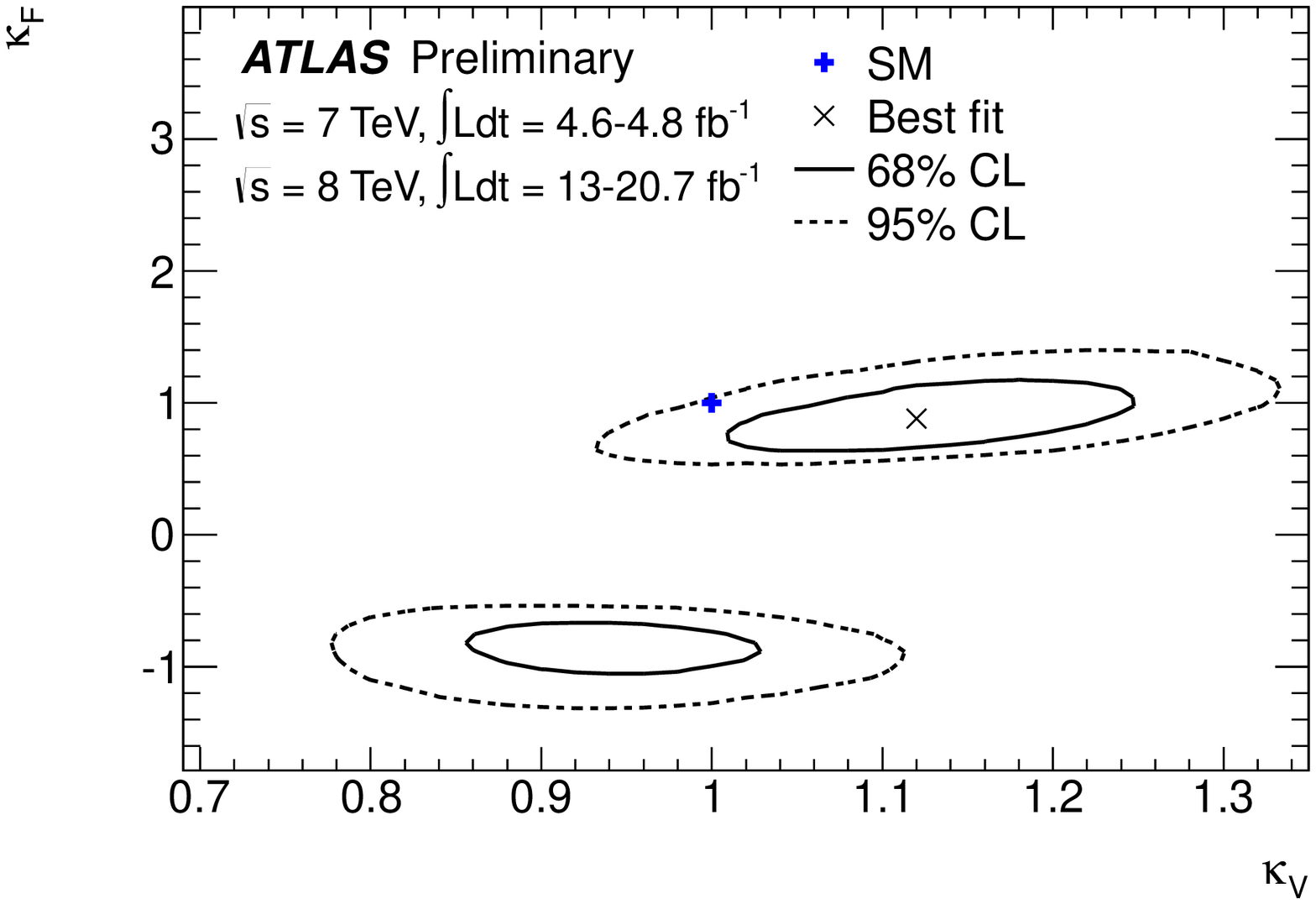}}
\caption{Comparison of the two-parameter fits probing different coupling strength scale factors for fermions, $\kappa_F$, and vector bosons, $\kappa_V$, derived by \HS\ (a) and ATLAS~\cite{ATLAS-CONF-2013-034}~(b). The signal strength measurements used for the \HS\ fit are listed in Tab.~\ref{Tab:ATLAS_peakobs}. The Higgs mass is chosen to be $m_H=125.5\gev$.
}
\label{Fig:ATLAS_FV-fit}
\end{figure}

We now turn to the discussion of global fits in the Higgs coupling scale factor benchmark scenarios.
Regarding the interpretation of the following benchmark fits, it should be kept in mind that only two parameters 
are allowed to deviate from their SM values, while all other Higgs couplings and partial decay widths have been fixed to their SM values. The way an
observed deviation from the SM manifests itself in the parameter space
of coupling strength modifiers $\kappa_i$ will sensitively depend on how 
general the basis of the $\kappa_i$ is that one has chosen. Furthermore 
the framework of the coupling strength modifiers $\kappa_i$ as defined
in Ref.~\cite{LHCHiggsCrossSectionWorkingGroup:2012nn} is designed for
the analysis of relatively small deviations from the SM. In case a firm
preference should be established in a parameter region that is very
different from the SM case (e.g. a different relative sign of Higgs
couplings), the framework of the coupling strength modifiers $\kappa_i$
would have to be replaced by a more general parametrization.

The first benchmark model we want to investigate is a two-dimensional fit to universal scale factors for the Higgs coupling to the massive SM vector bosons, $\kappa_V$, and to SM
fermions, $\kappa_F$. In this fit it is assumed that no other modifications to the total width than those induced by the 
coupling scale factors $\kappa_F$ and $\kappa_V$ are present, allowing for a fit to the coupling strength modifiers individually rather than to ratios of the scale factors~\cite{LHCHiggsCrossSectionWorkingGroup:2012nn}. Note that the loop-induced effective $H\gamma\gamma$ coupling is derived in this approximation from the (scaled) tree-level couplings $Ht\bar{t}$ and $HW^+W^-$ and thus exhibits a non-trivial scaling behavior. In particular the interference between the $t$ and $W$ boson loops introduces a dependence on the relative sign of the scale factors $\kappa_F$ and $\kappa_V$. In the case of a relative minus sign this interference term gives a positive contribution to the $H\gamma\gamma$ coupling.


\begin{figure}[t]
\centering
\subfigure[\HS~result.]{\includegraphics[trim=2.2cm 0.5cm 2.5cm 0.3cm, width=0.47\textwidth]{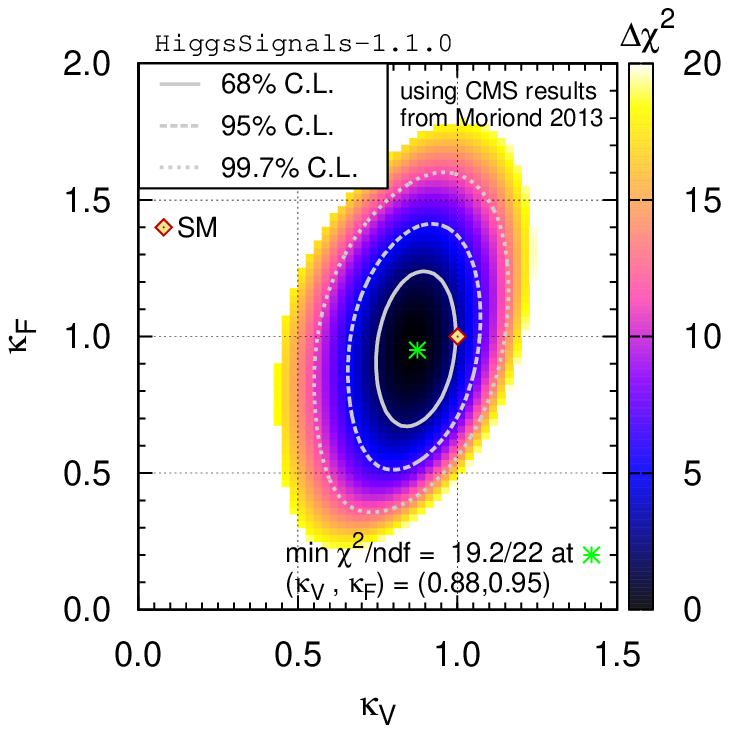}}\hfill
\subfigure[Official CMS result from Ref.~\cite{CMS-PAS-HIG-13-005}.]{\includegraphics[trim=0cm -1.3cm 0cm 0cm, height=7cm, width=0.47\textwidth]{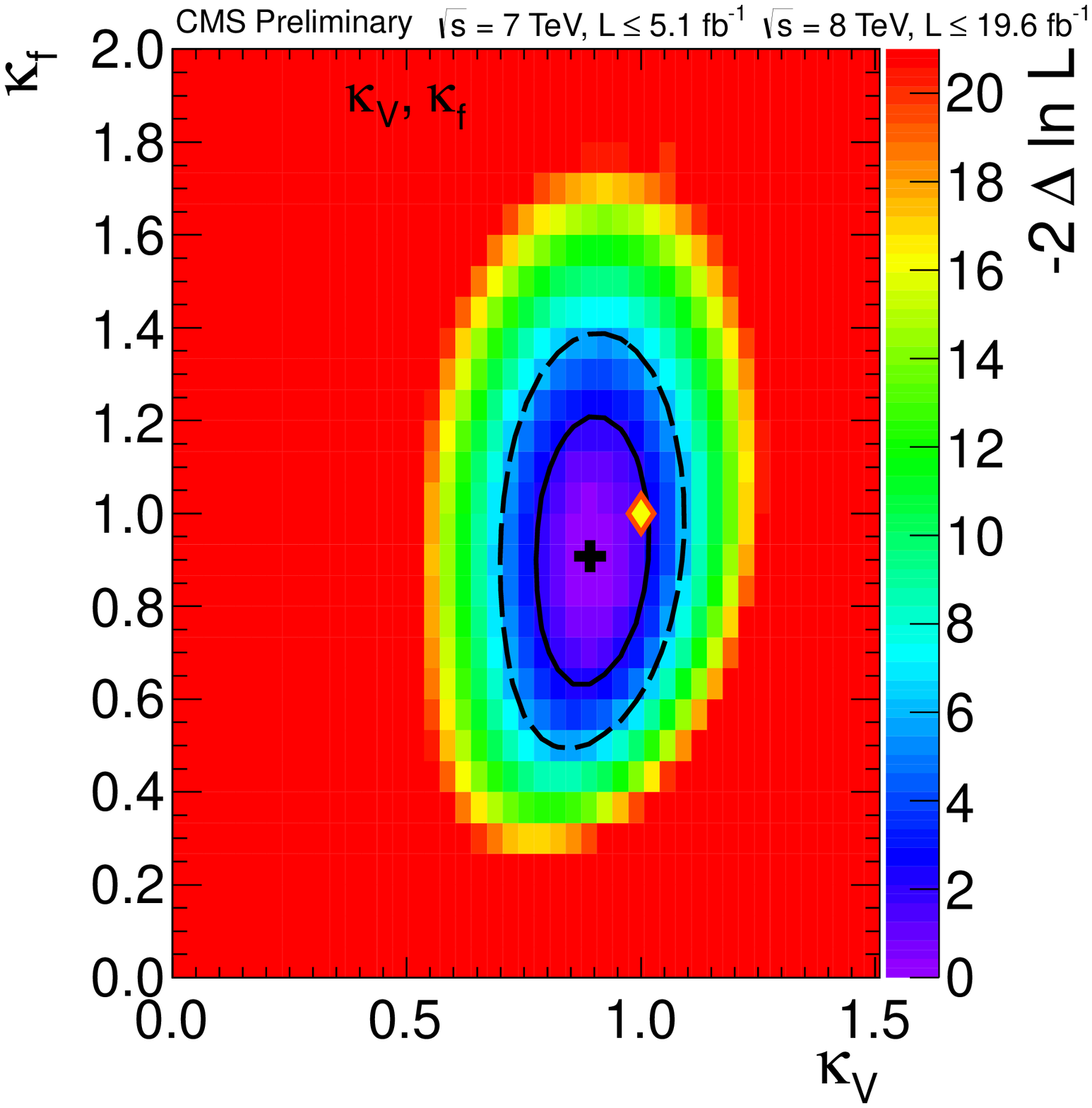}}
\caption{Comparison of the two-parameter fits probing different coupling strength scale factors for fermions, $\kappa_F$, and vector bosons, $\kappa_V$, obtained using \HS~(a), and by CMS~\cite{CMS-PAS-HIG-13-005} (b). The signal strength measurements used for the \HS\ fit are listed in Tab.~\ref{Tab:CMS_peakobs}. The Higgs mass is chosen to be $m_H=125.7\gev$.}
\label{Fig:CMS_FV-fit}
\end{figure}


The reconstructed ATLAS and CMS fits obtained with \HS\ are shown in Figs.~\ref{Fig:ATLAS_FV-fit}(a) and~\ref{Fig:CMS_FV-fit}(a), respectively. For comparison, we show the official fit results
from ATLAS~\cite{ATLAS-CONF-2013-034} and CMS~\cite{CMS-PAS-HIG-13-005} in
Figs.~\ref{Fig:ATLAS_FV-fit}(b) and~\ref{Fig:CMS_FV-fit}(b). 
We find overall very good agreement. The best points are located at
\begin{align}
(\kappa_V,\kappa_F) = \left\{ \begin{array}{c} (1.12, 0.85) \\ (0.88, 0.95) \end{array}\right.~\mbox{with}~\chi^2/\mathrm{ndf} = \left\{ \begin{array}{c}  34.7/28 \\  19.2/22 \end{array}\right.\quad \begin{array}{l} \mathrm{(ATLAS)} \\ \mathrm{(CMS)} \end{array}.
\end{align}
The (2D) compatibility with the SM hypothesis of these points is $11.1\%$ and $28.4\%$ for the reproduced ATLAS and CMS fit, respectively.

\begin{figure}[t]
\centering
\subfigure[\HS\ result.]{\includegraphics[ width=8.0cm, height=6.2cm]{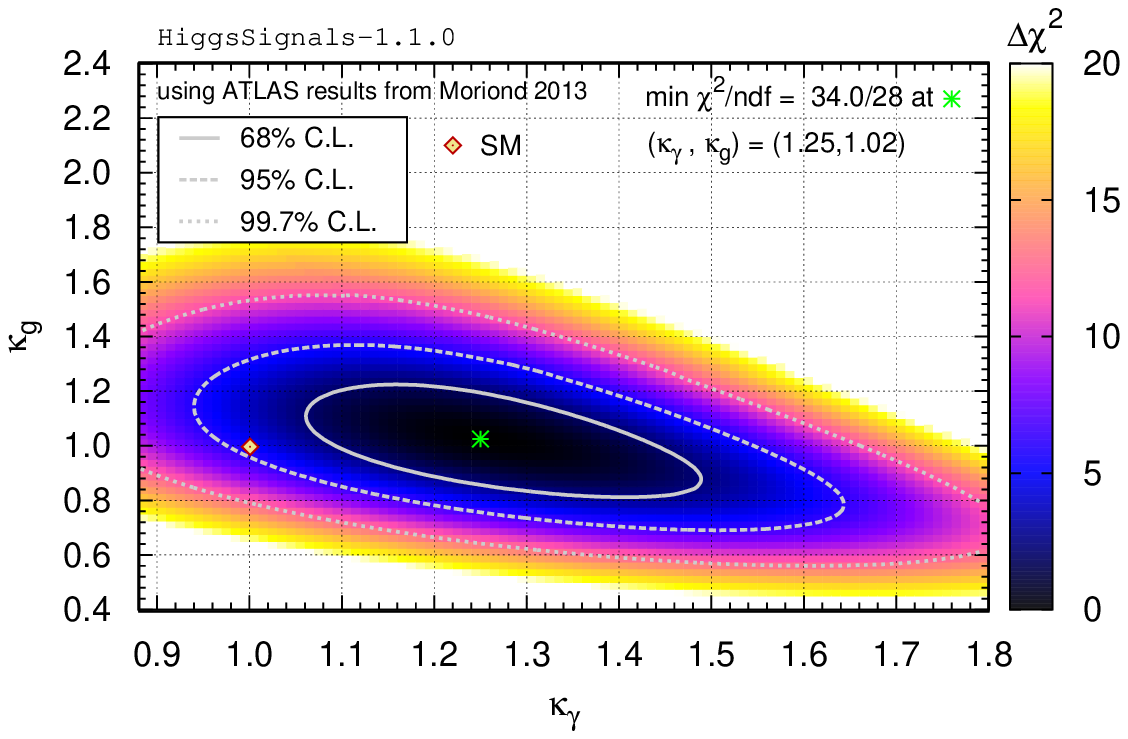}}\hfill
\subfigure[Official ATLAS result from Ref.~\cite{ATLAS-CONF-2013-034}.]{\includegraphics[trim = 0cm -1.4cm 0cm 0cm, clip,width=7.2cm]{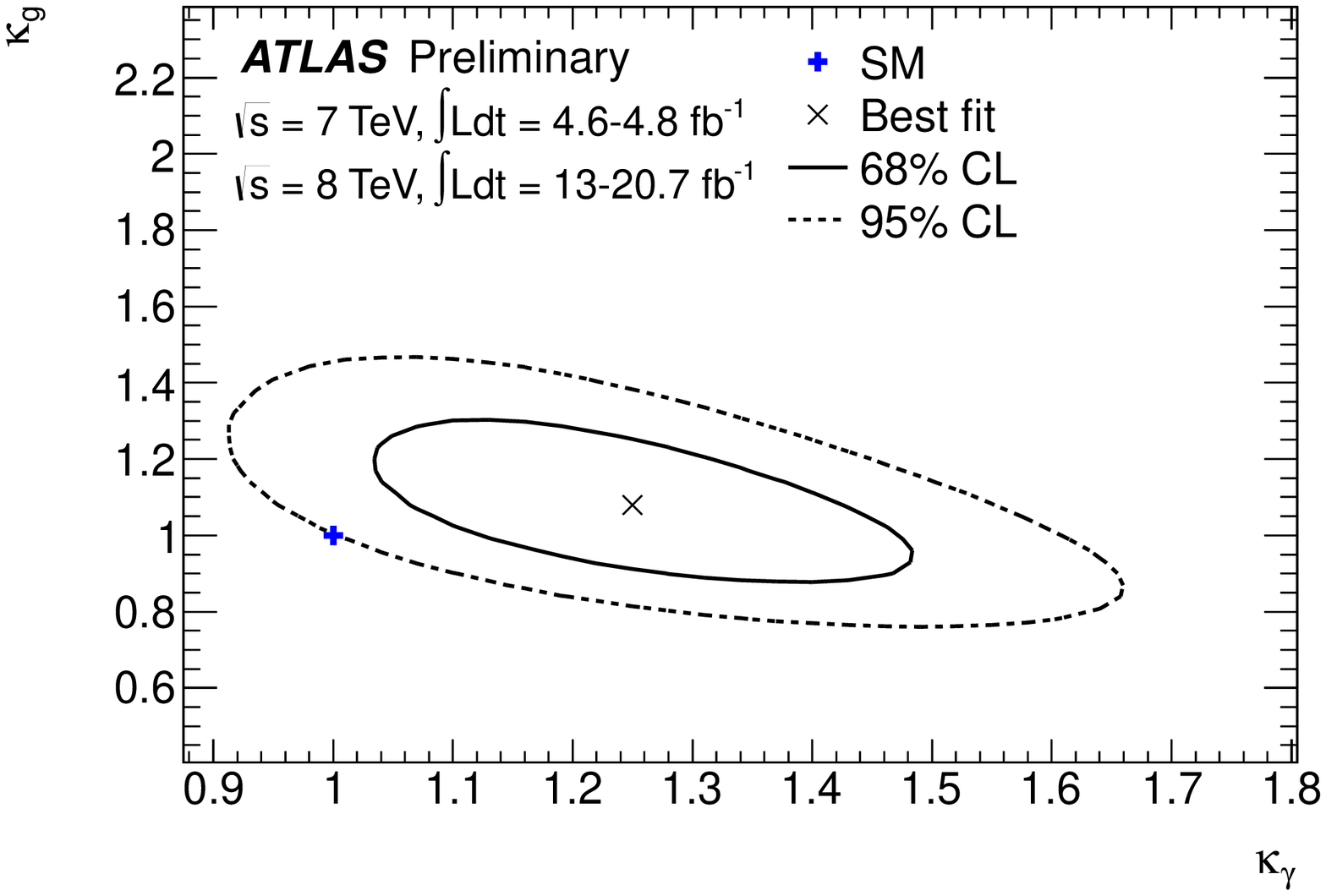}}
\caption{Comparison of the two-parameter fits probing different coupling
strength scale factors to gluons, $\kappa_g$, and photons,
$\kappa_\gamma$, obtained by \HS~(a), and
ATLAS~\cite{ATLAS-CONF-2013-034} (b). It is assumed that no new Higgs
boson decay modes are open, $\Gamma_\mathrm{BSM} = 0\gev$, and that
no other modifications of the couplings occur with respect to their SM values. The signal strength measurements used for the \HS\ fit are listed in Tab.~\ref{Tab:ATLAS_peakobs}. The Higgs mass is chosen to be $m_H=125.5\gev$.}
\label{Fig:ATLAS_kgkga-fit}
\end{figure}

\begin{figure}[t]
\centering
\subfigure[\HS~result.]{\includegraphics[trim=2.2cm 0.5cm 2.5cm 0.3cm, clip, width=0.47\textwidth]{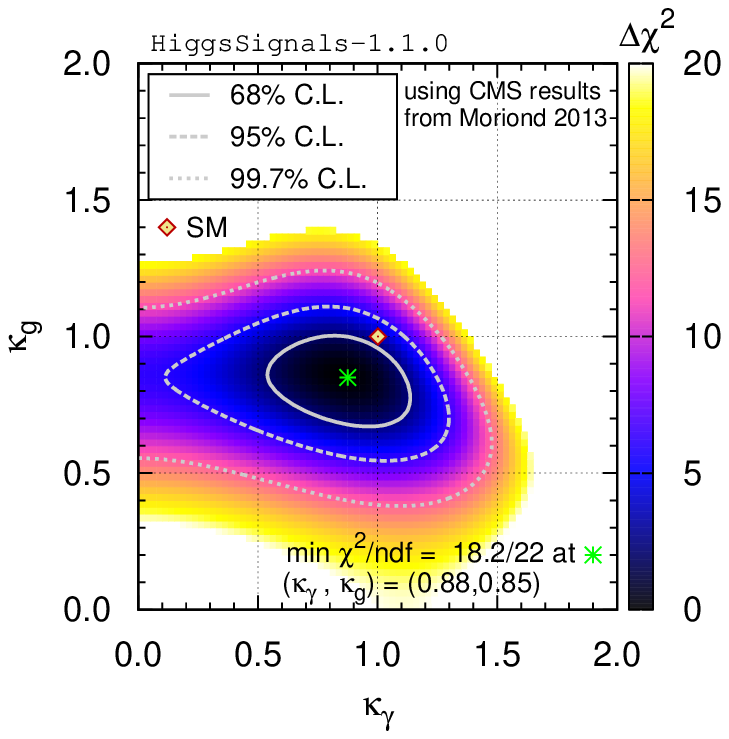}}\hfill
\subfigure[Official CMS result from Ref.~\cite{CMS-PAS-HIG-13-005}.]{\includegraphics[trim=0cm -1.3cm 0cm 0cm, clip, height=7cm, width=0.47\textwidth]{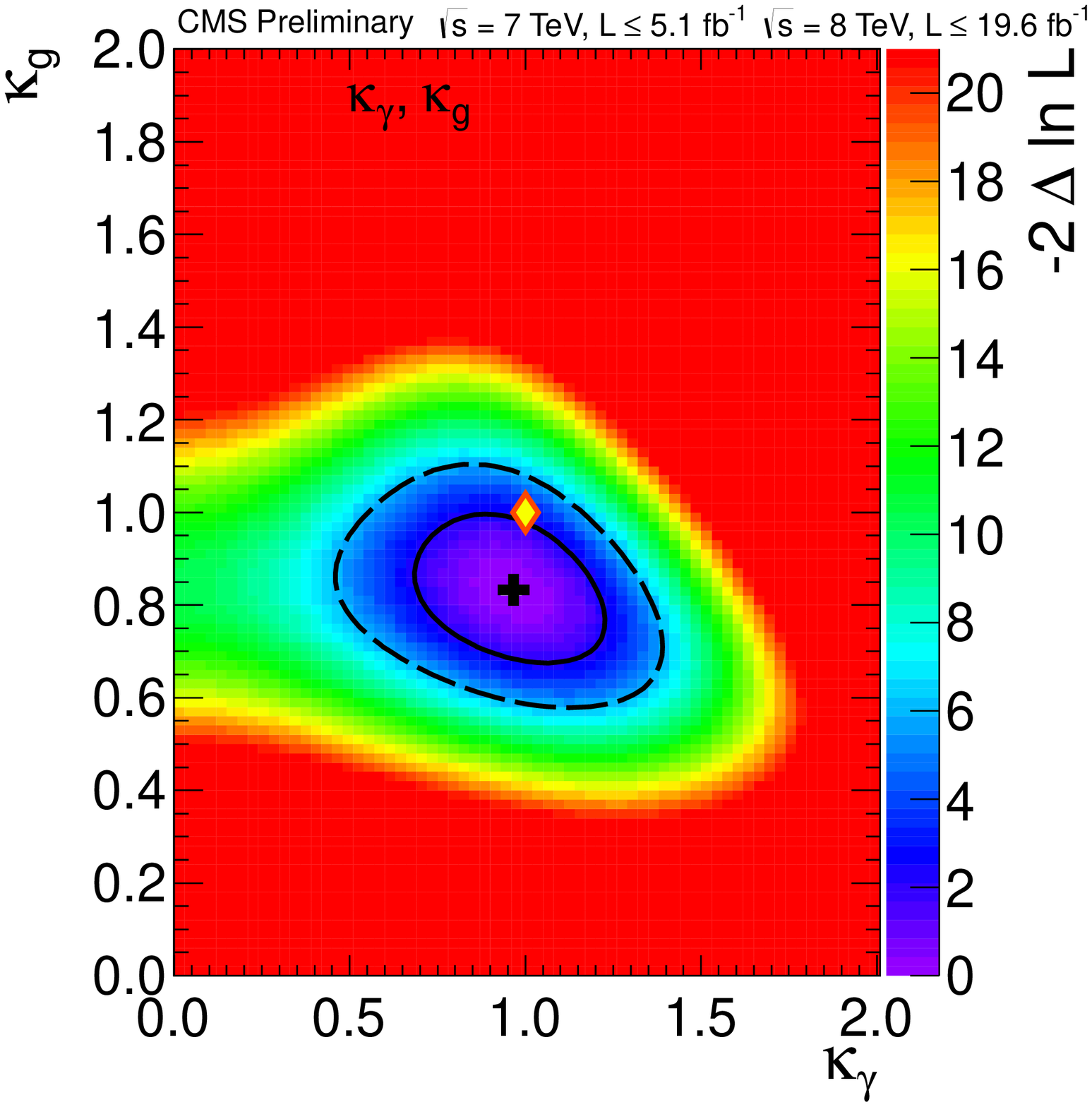}}
\caption{Comparison of the two-parameter fits probing different coupling
strength scale factors to gluons, $\kappa_g$, and photons,
$\kappa_\gamma$, obtained using \HS~(a), and by
CMS~\cite{CMS-PAS-HIG-13-005} (b). It is assumed that no new Higgs boson
decay modes are open, $\Gamma_\mathrm{BSM} = 0\gev$,
and that no other modifications of the couplings occur with respect to 
their SM values. The signal strength measurements used for the \HS\ fit are listed in Tab.~\ref{Tab:CMS_peakobs}.
The Higgs mass is chosen to be $m_H=125.7\gev$.}
\label{Fig:CMS_kgkga-fit}
\end{figure}

In order to probe the presence of BSM physics in the Higgs boson
phenomenology a fit to the loop-induced Higgs couplings to gluons,
$\kappa_g$, and photons, $\kappa_\gamma$, can be performed. In this fit
it is assumed that all other (tree-level) Higgs couplings are as in the SM and no new Higgs boson decay modes exist. 
Figs.~\ref{Fig:ATLAS_kgkga-fit}(a) and~\ref{Fig:CMS_kgkga-fit}(a) show the 2D likelihood maps in the $(\kappa_\gamma,~\kappa_g)$ parameter plane for the \HS~result using the ATLAS and CMS observables, respectively. The corresponding official ATLAS and CMS results are given in Figs.~\ref{Fig:ATLAS_kgkga-fit}(b) and~\ref{Fig:CMS_kgkga-fit}(b). Again, we observe reasonably good agreement with the official results. We find the best fit points at
\begin{align}
(\kappa_\gamma,\kappa_g) = \left\{ \begin{array}{c} (1.25, 1.02) \\ (0.88, 0.85) \end{array}\right.~\mbox{with}~\chi^2/\mathrm{ndf} = \left\{ \begin{array}{c}  34.0/28 \\  18.2/22 \end{array}\right.\quad \begin{array}{l} \mathrm{(ATLAS)} \\ \mathrm{(CMS)} \end{array}.
\end{align}
These are (2D) compatible with the SM at the level of $7.6\%$ and $17.1\%$, respectively.
In the ATLAS fit, the best-fit region obtained by \HS\ is slightly shifted with respect to the official result towards lower values of $\kappa_g$ by roughly $\Delta\kappa_g \sim 0.05 - 0.10$, whereas the agreement in $\kappa_\gamma$ direction is very good. In the CMS fit, the agreement is better. Here, the \HS\ $\Delta\chi^2$ distribution is slightly shallower than the official CMS likelihood at low values of $\kappa_\gamma$, leading to slightly larger C.L. contours.

We conclude this section by pointing out that, despite some discrepancies that are observed in fits to single decay modes using subsets of the available measurements, Fig.~\ref{Fig:CSscaling}, the combination of all available channels from each experiment reproduces the official results quite well. We are therefore confident that the accuracy of the \HS\ method is sufficient for surveys of new physics parameter spaces compatible with the Higgs measurements, and for simple coupling scale factor fits.
For a more precise determination of the Higgs boson coupling structure with \HS, however, it would be desirable if the experimental collaborations made information on efficiencies, correlated experimental uncertainties and all category measurements publicly available in a more complete way. We would expect a significant reduction of the observed remaining discrepancies if this information was included in \HS.

\subsection{Example applications of~\HS}
\label{sect:Combinedfits}

\begin{figure}[t]
\centering
\subfigure[($\kappa_V, \kappa_F$) fit.]{\includegraphics[width=0.48\textwidth]{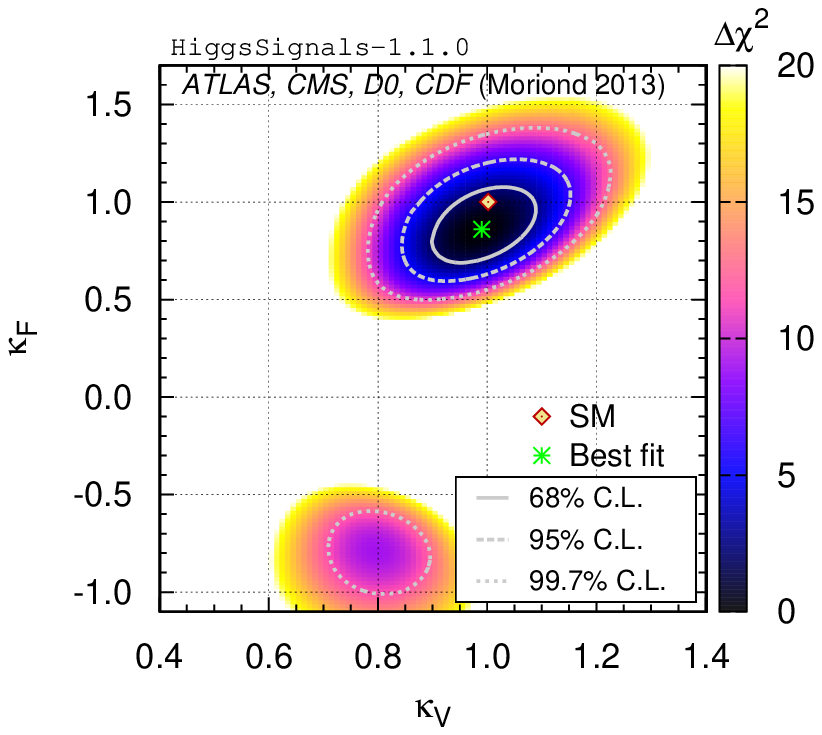}}
\hfill
\subfigure[($\kappa_\gamma, \kappa_g$) fit.]{\includegraphics[width=0.48\textwidth]{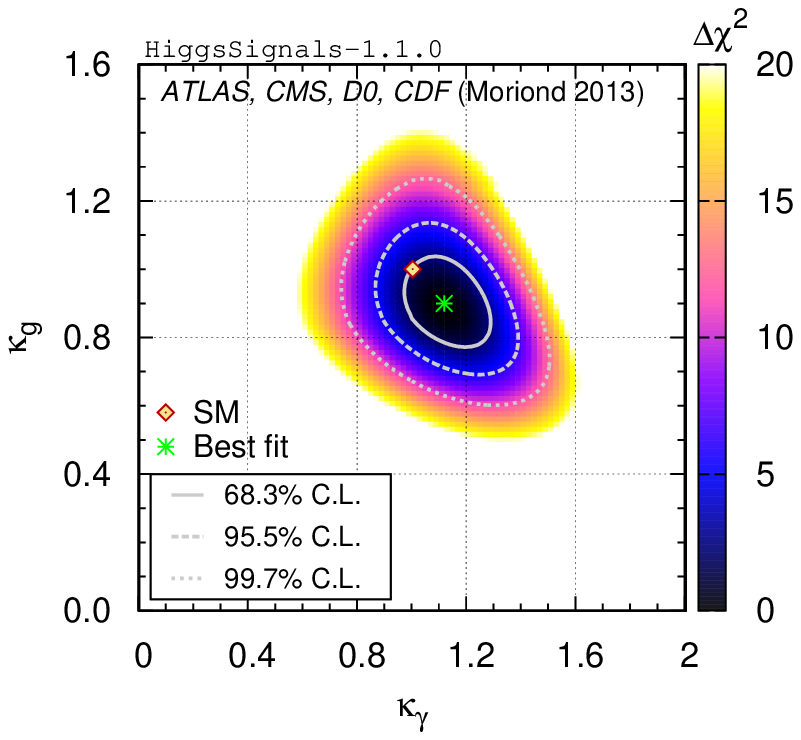}}
\caption{Two-dimensional fit results for the two different benchmark scenarios of Higgs coupling scaling factors discussed
above. (a) Common
    scale factors for the vector boson and fermion couplings, $\kappa_V$
    and $\kappa_F$, respectively; (b) Scale factors for the
    loop-induced Higgs couplings to photons, $\kappa_\gamma$, and
    gluons, $\kappa_g$. In these fits, the Higgs boson mass is assumed
    to be $126\gev$. The full available data from the Tevatron and LHC
    experiments as presented at the Moriond 2013 conference (and shortly after) is used. This data is 
    summarized in Fig.~\ref{Fig:peakobservables}.} 
\label{Fig:Moriond2013_couplingfit}
\end{figure}

We now go beyond validation and repeat the two discussed Higgs coupling scaling factor fits including the full presently available data from the LHC and Tevatron experiments, as listed in Fig.~\ref{Fig:peakobservables}. This includes data presented up until shortly after the Moriond 2013 conference. We assume a Higgs boson mass of $126\gev$. The fit results for the Higgs coupling scale factors  ($\kappa_V,~\kappa_F$), defined in Sect.~\ref{Sec:ValidationOffEff} and~\cite{LHCHiggsCrossSectionWorkingGroup:2012nn}, are shown in Fig.~\ref{Fig:Moriond2013_couplingfit}(a). The best-fit point is found at
  \begin{equation}
\kappa_V = 0.99\substack{+0.06\\-0.06},\quad \kappa_F =
0.86\substack{+0.14\\-0.10}, \quad \mbox{with}\quad \chi^2/\mathrm{ndf} = 68.7/61,
  \end{equation}
  where the profiled one-dimensional $68\%$ C.L.\ uncertainties are given. 
For this fit the SM point is found to be
located well within the 68\% C.L.\ contour, with a (2D) $\chi^2$ compatibility with the best fit point of $59.5\%$. Compared to the individual results from ATLAS~\cite{ATLAS-CONF-2013-034} and CMS~\cite{CMS-PAS-HIG-13-005} presented in Fig.~\ref{Fig:ATLAS_FV-fit} and \ref{Fig:CMS_FV-fit}, a significant degradation of
  the fit quality of the non-SM minimum (i.e.~for negative $\kappa_F$) is observed, which highlights
  the power of such simultaneous global analyses.


  A similar improvement is seen for the ($\kappa_\gamma, \kappa_g$)
  fit, shown in Fig.~\ref{Fig:Moriond2013_couplingfit}(b), where the best fit point is found at
  \begin{equation}
  \kappa_\gamma = 1.12\substack{+0.10\\-0.08},\quad \kappa_g = 0.90\substack{+0.09\\-0.08}, \quad\mbox{with}\quad \chi^2/\mathrm{ndf} = 67.6/61,
  \end{equation}
  which can be compared with Fig.~\ref{Fig:ATLAS_kgkga-fit} and
  \ref{Fig:CMS_kgkga-fit}. Here, the SM is compatible with the
fit result at the level of $31.8\%$. The fit shows a weak tendency towards slightly reduced $\kappa_g$ and slightly enhanced $\kappa_\gamma$. The discrimination power on $\kappa_g$ will increase only slowly with more data, since the large uncertainty of the rate prediction for single Higgs production is already the dominant limitation of the precision of the combined fit~\cite{Djouadi:2012rh}.

As a further example application we performed fits
in three of the MSSM benchmark
  scenarios recently proposed for the interpretation of the SUSY Higgs
  search results at the LHC~\cite{Carena:2013qia}. These scenarios
  are defined in terms of two free parameters, $\tb=v_2/v_1$ (the ratio of the vacuum expectation values of the two Higgs doublets), and either $\MA$ (the $\cp$-odd Higgs boson mass) or $\mu$ (the Higgsino mass parameter). The other parameters are fixed to their default values as specified in~\cite{Carena:2013qia} to exhibit certain features of the MSSM Higgs phenomenology. For each parameter point in these two-dimensional planes we calculated the model predictions with \texttt{FeynHiggs-2.9.4} and evaluated the
  total $\chi^2$, comprised of the LEP Higgs exclusion $\chi^2$ value \cite{Barate:2003sz,Schael:2006cr} obtained from \texttt{HiggsBounds-4}~\cite{Bechtle:2013wla,Bechtle:2013gu}, as well as the total $\chi^2$ from \texttt{HiggsSignals} using the \emph{peak-centered} $\chi^2$ method. The theoretical mass uncertainty of the lightest Higgs boson is set to $2\gev$ when treated as a Gaussian uncertainty (i.e. in the LEP exclusion $\chi^2$ from \HB\ and in \HS), and to $3\gev$ in the evaluation of $95\%~\mathrm{C.L.}$ LHC exclusions with \HB.

\begin{figure}[t]
\centering
\includegraphics[width=0.8\textwidth]{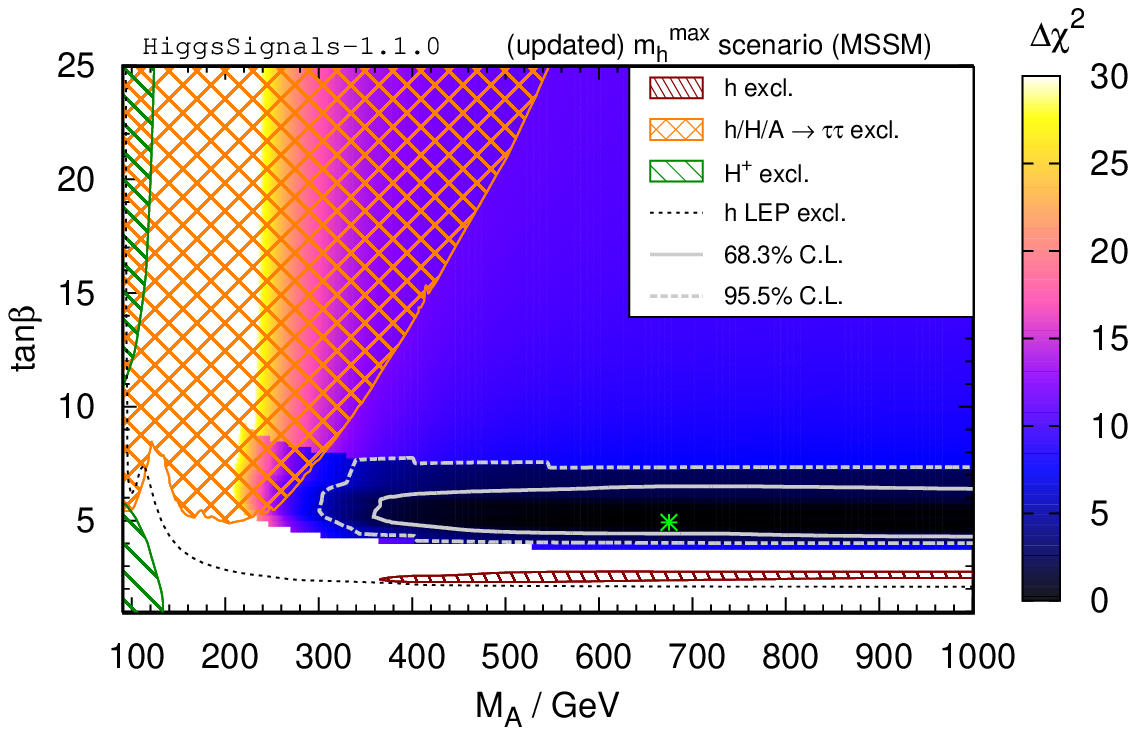} 
\caption{Distribution of $\Delta\chi^2$ in the (updated) \mhmax\ benchmark scenario of the MSSM~\cite{Carena:2013qia}. The result from \HS\ and the LEP exclusion $\chi^2$ of \HB\ are added.
The patterned areas indicate parameter regions excluded at $95\%~\mathrm{C.L.}$ from the following LHC Higgs searches: CMS $h/H/A \to \tau\tau$~\cite{Chatrchyan:2012vp} (orange, checkered), ATLAS $t\to H^+ b \to \tau^+ \nu_\tau b$~\cite{Aad:2012tj} (green, coarsely striped), CMS SM Higgs combination~\cite{CMS-PAS-HIG-12-045} (red, striped). The $95\%~\mathrm{C.L.}$ LEP excluded region~\cite{Barate:2003sz,Schael:2006cr}, corresponding to $\chi^2_\mathrm{LEP,HB} = 4.0$, is below the black dashed line. The best-fit point, $(\MA, \tan\beta) = (674\gev, 5.0)$ with $\chi^2/\mathrm{ndf} = 70.2/66$, is indicated by a green star. The $68\%$ and $95\%$ C.L. preferred regions (based on the 2D $\Delta\chi^2$ probability w.r.t.~the best fit point) are shown as solid and dashed gray lines, respectively.}
\label{Fig:mhmax}
\end{figure}

The first scenario is an updated version of the well-known 
\mhmax\ benchmark scenario~\cite{Carena:2013qia,Carena:1999xa}, where
the masses of the gluino and the squarks of the first and second
generation were set to higher values in view of the latest bounds from
SUSY searches at the LHC, see~\cite{Carena:2013qia} for details. The results are shown in
Fig.~\ref{Fig:mhmax} in the ($\MA$, $\tb$) plane. Besides the colors
indicating the $\De\chi^2 = \chi^2 - \chi^2_\mathrm{best-fit}$ distribution relative to the best-fit
point (shown as a green star) we also show the parameter regions
that are excluded at $95\%~\mathrm{C.L.}$ by LHC searches for a light charged Higgs boson (dark-green, coarsely striped)~\cite{Aad:2012tj}, neutral Higgs boson(s) in the $\tau\tau$ final state (orange, checkered)~\cite{Chatrchyan:2012vp} and the combination of SM search channels (red, striped)~\cite{CMS-PAS-HIG-12-045},
as obtained using \HB. As an indication for the parameter regions that are $95\%~\mathrm{C.L.}$ excluded by neutral Higgs searches at LEP~\cite{Barate:2003sz,Schael:2006cr} we include a corresponding contour (black, dashed) for the value $\chi^2_\mathrm{LEP,HB} = 4.0$. Conversely, the parameter regions favored by the fit are shown as $68\%$ and $95\%$ C.L. regions (based on the 2D $\Delta\chi^2$ probability w.r.t. the best fit point) by the solid and dashed gray lines, respectively.

As can be seen in the figure, the best fit regions are obtained in a
strip at relatively small values of $\tb \approx 4.5 - 7$, where in this scenario 
$\Mh \sim 125.5 \gev$ is found. At larger $\tb$ values the light Higgs
mass in this benchmark scenario (which was designed to 
maximise $\Mh$ for a given $\tb$ in the region of large $\MA$) 
turns out to be {\em higher\/} than the measured mass of the observed
signal, resulting in a corresponding $\chi^2$ penalty. At
very low $\tb$ values the light Higgs mass is found to be below the
preferred mass region, again resulting in a $\chi^2$ penalty. Here, the $\chi^2$ steeply rises (for $\Mh \lesssim 122\gev$), because the mass-sensitive observables ($H\to \gamma\gamma, ZZ^{(*)}$) cannot be explained by the light Higgs boson anymore, \cf Sect.~\ref{Sect:pc_performance}. Values of $\MA>300\gev$ are preferred in
this scenario, and thus the light Higgs has mainly SM-like
couplings. Consequently, the $\chi^2$ contribution from the rate
measurements is similar to the one for a SM Higgs boson. In this regime, the Higgs mass dependence of the total $\chi^2$ (from \HS) is comparable to the results shown in Fig.~\ref{Fig:SM}(d).
We find the best fit point at $(M_A, \tanb) = (674\gev, 5.0)$ with $\chi^2/\mathrm{ndf} = 70.2/66$.
The number of degrees of freedom (ndf) comprises $63$ signal strengths and $4$ mass measurements presented in Fig.~\ref{Fig:peakobservables}, as well as one LEP exclusion observable from \HB.

\begin{figure}[t]
\centering
\includegraphics[width=0.8\textwidth]{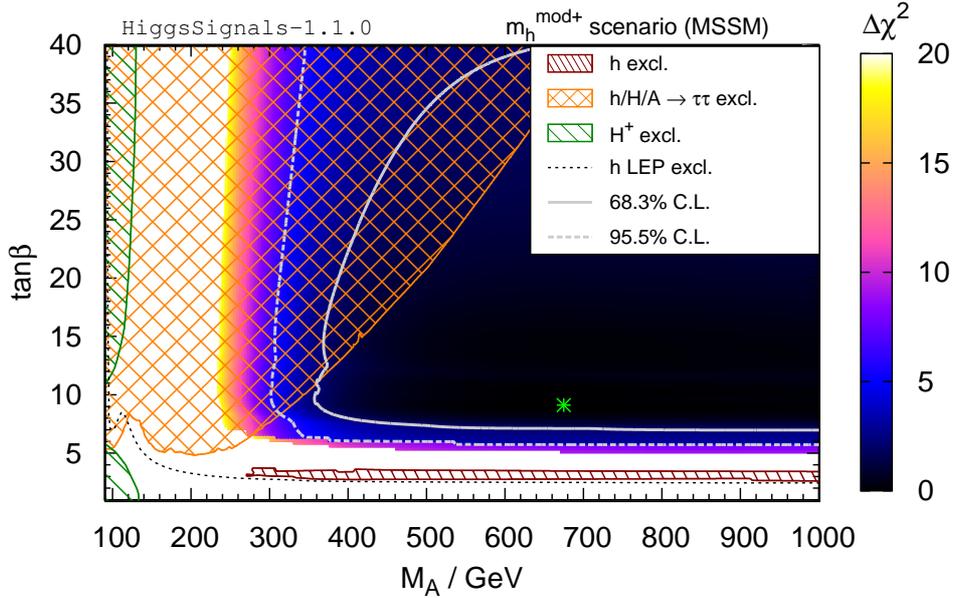}
\caption{$\Delta\chi^2$ distribution (\HS\ and \HB\ LEP exclusion $\chi^2$ added) in the $\mhmodp$ benchmark scenario of the MSSM~\cite{Carena:2013qia}. The excluded regions and contour lines have the same meaning as in Fig.~\ref{Fig:mhmax}. The best-fit point (indicated by a green star) is found at $(\MA, \tan\beta) = (674\gev, 9.3)$ with $\chi^2/\mathrm{ndf} = 70.7/66$}. 
\label{Fig:mhmod}
\end{figure}

The second scenario that we discuss here is a modification of the \mhmax\ scenario with a
lower value of $\Xt$, leading to $\Mh \sim 125.5 \gev$ over nearly the
whole ($\MA$, $\tb$) plane~\cite{Carena:2013qia}. This so-called $\mhmodp$ scenario is shown in
Fig.~\ref{Fig:mhmod} (with the same colors and meaning of theÊcontours as for the \mhmax\ scenario, Fig.~\ref{Fig:mhmax}). 
The best fit point is
found at $(\MA,\tb) = (674\gev, 9.3)$ with $\chi^2 = 70.7/66$.
 Only slightly larger $\chi^2$ values are found over the rest
of the plane, except for the lowest $\MA$ and $\tb$ values, where $\Mh$ 
is found to be below the preferred mass region.
As in the preferred region for the \mhmax\ scenario the lightest Higgs boson is mostly
SM-like here, and the $\chi^2$ from the rates is close to the one found in the
\mhmax\ scenario. 

\begin{figure}[t]
\centering
\includegraphics[width=0.9\textwidth]{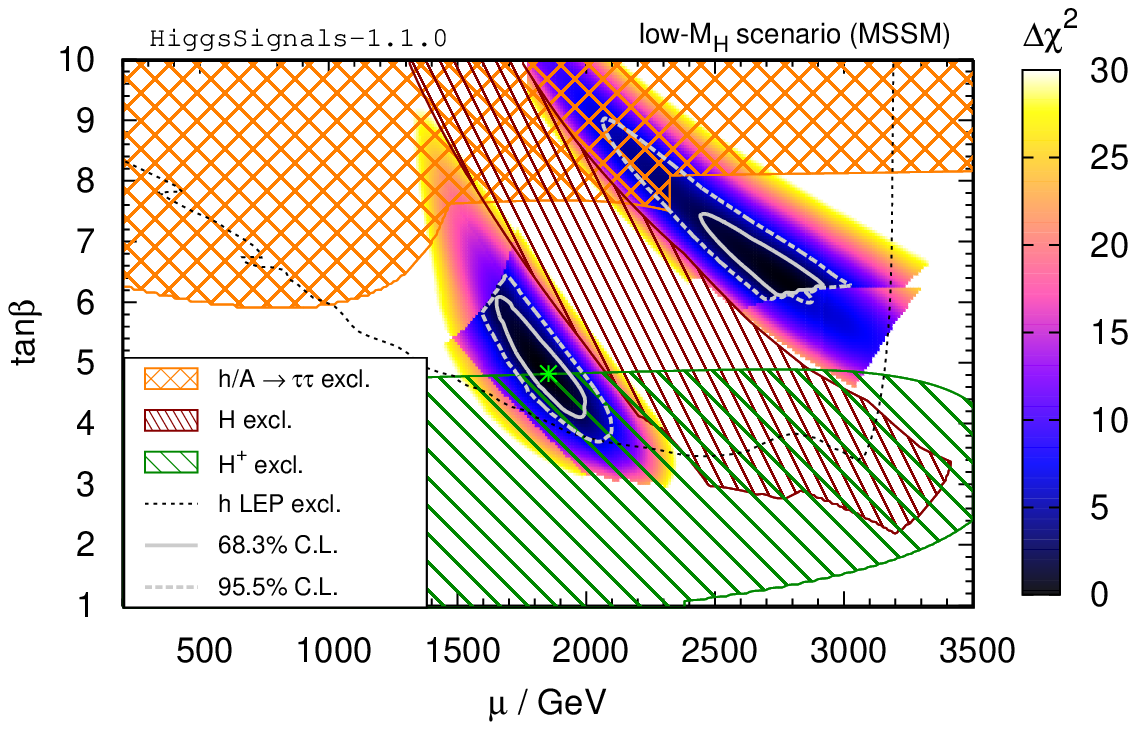}
\caption{$\Delta\chi^2$ distribution (\HS\ and \HB\ LEP exclusion $\chi^2$ added) in the $\mHlow$ benchmark scenario of the MSSM~\cite{Carena:2013qia}. The excluded regions and contour lines have the same meaning as in Fig.~\ref{Fig:mhmax}, except the red, finely striped region, which gives the $95\%~\mathrm{C.L.}$ exclusion from the CMS Higgs search $H\to ZZ^{(*)}\to 4\ell$~\cite{CMS-PAS-HIG-13-002}, applied to the SM-like heavy $\cp$ even Higgs boson. The best-fit point (indicated by a green star) is found at $(\mu, \tan\beta) = (1850\gev, 4.9)$ with $\chi^2/\mathrm{ndf} = 80.3/66$.} 
\label{Fig:lowmH}
\end{figure}

As a final example, we performed a fit in the $\mHlow$
benchmark scenario of the MSSM~\cite{Carena:2013qia}.
This scenario is based on the assumption that the Higgs observed at
$\sim 125.5 \gev$ is the heavy $\cp$-even Higgs boson of the MSSM. 
In this case the light $\cp$-even Higgs has a mass below the LEP limit
for a SM Higgs boson of $114.4 \gev$~\cite{Barate:2003sz}, but is effectively decoupled from
the SM gauge bosons. The other states of the Higgs spectrum are
also rather light, with masses around $\sim 130 \gev$, so that this
scenario offers good prospects for the searches for additional Higgs
bosons~\citeheavyH.
Since $\MA$ must be relatively small in this case the ($\mu$, $\tb$)
plane is scanned~\cite{Carena:2013qia}, where only $\tb \lesssim 10$ is
considered. The $\cp$-odd Higgs boson mass is fixed to $\MA = 110 \gev$.
Our results are shown in Fig.~\ref{Fig:lowmH}. The $95\%~\mathrm{C.L.}$ excluded regions are obtained from the same Higgs searches as in Fig.~\ref{Fig:mhmax}, except for the red patterned region, which results from applying the limit from the CMS SM Higgs search $H\to ZZ^{(*)}\to 4\ell$~\cite{CMS-PAS-HIG-13-002} to the SM-like, heavy $\cp$-even Higgs boson (see below).

Two distinct best-fit regions are found~\cite{Carena:2013qia}: The parameter space with $\mu\sim (1.6-2.0)\tev$ and $\tb \sim 4-6$ predicts a heavy $\cp$-even Higgs boson with a well compatible mass value $\MH \approx 126\gev$ and SM-like couplings. However, large parts (at low $\tanb\lesssim 4.9$) of this region favored by the rate and mass measurements are severely constrained by charged Higgs searches~\cite{Aad:2012tj}. The best-fit point is found at the edge of the excluded region at $(\mu,~\tb)=(1850\gev,~4.9)$. The second region favored by the fit is located at large values of $\mu \sim (2.4 - 2.9)\tev$ and $\tb \sim 6-7$. Here, the masses of the $\cp$-even Higgs bosons are generally lower. For instance, at $(\mu,~\tb)\sim (3070 \gev, 6.0)$, we have $\Mh\approx 76.1\gev$ and $\MH\approx 122.8\gev$. For slightly larger (lower) values of $\mu$ ($\tb$) we find a steep edge in the \HS\ $\chi^2$ distribution, because $\MH$ becomes too low to allow for an assignment of the heavy $\cp$-even Higgs boson to all mass-sensitive peak observables, cf.~the results shown in Fig.~\ref{Fig:SM}(d),  Sect.~\ref{Sect:pc_performance}. Due to the low mass of the light $\cp$-even Higgs boson in this region, the LEP channel $e^+e^- \to h A$~\cite{Schael:2006cr} is kinematically accessible and contributes a non-negligible $\chi^2$ which increases with $\mu$.
The parameter space between the two preferred regions suffers a rather large $\chi^2$ penalty, since in particular the predicted rates for the $H \to ZZ^{(*)}, WW^*$ channels are above the rates measured at the LHC, as can also be seen from the $95\%$ C.L.~exclusion by \HB\ in this region.

At the best-fit point we find a $\chi^2/\mathrm{ndf}=80.3/66$. Compared with the light $\cp$-even Higgs interpretation of the observed signal, as discussed in the \mhmax\ and $\mhmodp$ scenarios, the fit quality is only slightly worse.


\section{Conclusions} 
\label{sect:Conclusions}

We have presented \HS, a public {\tt Fortran} code to test the
predictions of models with arbitrary Higgs sectors against measurements
obtained from Higgs searches at the LHC, the Tevatron, and any potential
future experiment. The code is publicly available at\\[.3em]
\centerline{\url{http://higgsbounds.hepforge.org}}\\[.3em]
The code features two statistical
tests, one which determines the compatibility of the model with
experimentally observed Higgs signals, and a second which tests for
general compatibility with the observed Higgs data at the predicted
mass(es) of the Higgs boson(s) in the theory. Since the two tests are
complementary, we also provide a method to perform both simultaneously
and use the combined results for models with multiple Higgs bosons. 

The main experimental results used by \HS\ are the signal strength 
modifiers, $\muobs$, as a
function of the Higgs mass in the various search channels. 
These results have to be supplemented by their respective experimental
uncertainties, $\dmuobs$, and (preferably, if this information is
available) with the experimental efficiencies and
correlations. The information on $\muobs$ and $\dmuobs$ channel by
channel constitutes the most general and robust experimental input 
for testing the theoretical predictions of different models, and we strongly
encourage the experimental collaborations to continue to make them
public with as much details provided as possible.

The default implementation of \HS\ uses the $\muobs$ results available
from the LHC and the Tevatron, and it is planned to continuously
update these results in forthcoming versions of \HS. However, it is
easily possible for the user to include additional experimental
data. For assessing possible future projections it is also possible to
implement hypothetical future experimental results.

The input that has to be provided by the user (and which is similar to the \HB\ input) consists of the Higgs boson masses, preferably the corresponding theory uncertainties, the Higgs production cross sections and decay
branching ratios, where several levels of approximation are possible. In
case of the MSSM also the SLHA~\cite{Skands:2003cj,SLHA2}
can be used as input/output format. 

We presented in detail the two statistical methods provided by \HS: the 
\textit{peak-centered $\chi^2$ method}, in which each observable is
defined by a Higgs signal rate measured at a specific hypothetical
Higgs mass, corresponding to a tentative Higgs signal. In the second, 
the \textit{mass-centered $\chi^2$ method}, the $\chi^2$
is evaluated by comparing the signal rate measurement to the theory
prediction at the Higgs mass predicted by the model. It was described
how these two methods can be combined, as it is an option of \HS, to
yield the most reliable consistency test. In this combination, the
mass-centered $\chi^2$ method is applied only to those Higgs bosons which have not yet been tested with the peak-centered $\chi^2$ method against the same data. Similarly, in order to include a more complete set of constraints on the Higgs sector, it is recommended
to use \HS\ together with \HB\ to test the model under consideration
also against the existing Higgs exclusion bounds.

The installation, usage and subroutines of \HS\ were explained in
detail, together with the various input and output formats. It was
explained how the user can add new (hypothetical) experimental
data. Several pre-defined example codes were presented that permit the
user to get familiar with \HS\ and, by modifying the example
codes, analyze own models of interest. As an example, by linking
\HS\ to {\tt FeynHiggs}, the consistency of any MSSM parameter point with the
observed LHC signal can be analyzed in a simple way. Furthermore, some example codes demonstrate how to use \HB\ and \HS\ simultaneously in an efficient way.

We have presented several examples of the use of \HS. As a first example the combined best-fit signal strength has been determined. For the peak-centered $\chi^2$ method the mass dependence, the impact of correlations
between the systematic uncertainties and the treatment of theoretical
uncertainties has been discussed in detail. For the case of a SM-like Higgs boson, we demonstrated how the mass can be determined from a fit to the signal rate measurements as a function of the mass using the mass-centered $\chi^2$ method. Moreover, we employed this method for a combination of different search channels over the full investigated mass range. Various fits for coupling strength
modifiers have been carried out using the peak-centered $\chi^2$ method. Their results have been compared for validation purposes with official results from the ATLAS and CMS
collaborations, and very good agreement has been found.

It is expected that the agreement with the official results published by
ATLAS and CMS could be improved even further if relative signal
efficiencies of different production modes in all search channels
would be publicly provided by the experimental collaborations. The same
applies to a more complete description of the impact of individual
experimental systematic uncertainties and their correlations amongst
search channels. In particular, it would be useful if
systematic uncertainties were
given as a relative error on the quoted signal strength. 
We would furthermore welcome the publication of the full $\muobs$ plot for every analysis to allow a $\chi^2$ test at various Higgs masses.

Going beyond just a validation of \HS\ results, we have also given
a few examples of \HS\ applications. In particular, we have performed
fits of Higgs coupling scaling factors including the full presently
available data from both the LHC and the Tevatron.  
Furthermore we have investigated benchmark scenarios recently proposed for the SUSY Higgs search at the LHC, where we have
taken into account both the limits obtained from the searches at LEP,
the Tevatron and the LHC, as well as the information about the observed
signal at about $126\gev$. The provided examples give only a first
glimpse of the capabilities of \HS. The applicability of \HS\ goes far
beyond those examples, and in particular it should be a useful tool for
taking into account Higgs sector information in global fits.


\ack{
We thank Oliver Brein and Karina Williams for their great contributions to the \HB\ project, which was the basis for the development of \HS. We thank the Fittino collaboration, in particular Sebastian Heer, Xavier Prudent, Bj\"orn Sarrazin and Mathias Uhlenbrock, for comments and suggestions on the code development. We are grateful for helpful discussions with Andr\'e David, Michael D\"uhrssen, Michael Kr\"amer, Stefan Liebler, Alex Read, Jana Schaarschmidt, Florian Staub and Lisa Zeune. T.S. would like to thank the Bonn-Cologne-Graduate-School for financial support and is grateful for the hospitality of the Oskar Klein Centre at Stockholm University, where part of the concepts of \HS\ were developed. This work is supported by the Helmholtz Alliance ``Physics at the Terascale'' and the Collaborative Research Center SFB676 of the DFG, ``Particles, Strings, and the Early Universe''. The work of S.H.~was supported in part by CICYT (grant FPA 2010--22163-C02-01) and by the Spanish MICINN's Consolider-Ingenio 2010 Program under grant MultiDark CSD2009-00064. The work of O.S.~is supported by the Swedish Research Council (VR) through the OKC.
}


\section*{Appendix}
\appendix
\renewcommand{\sectionmark}[1]{\markright{Appendix \thesection}{} }

\section{Theory mass uncertainties in the mass-centered $\chi^2$ method}
\label{App:MCuncertainties}

In order to illustrate the two possible treatments of theoretical mass uncertainties in the mass-centered $\chi^2$ method we first discuss a constructed toy example (Example 1). Then we show how a typical $\muobs$ plot changes if it is convolved with a Higgs mass pdf, which parametrizes the theoretical mass uncertainty (Example 2).

\subsection*{Example 1: Variation of the predicted Higgs mass}

We look at a simple toy model with three neutral Higgs bosons $h_i$ ($i=1,2,3$) with masses $m_1 = 125\gev$, $m_2 = 135\gev$, $m_3 = 140\gev$. For every Higgs boson the theoretical mass uncertainty is set to $2\gev$. We test this model using the experimental data from the four $\muobs$ plots of the ATLAS searches for $H\to \gamma \gamma$~\cite{ATLAS-CONF-2012-091} ($7$ and $8\tev$ separately), $H\to ZZ^{(*)}\to 4 \ell$~\cite{ATLAS-CONF-2012-092} and $H\to WW^{(*)}\to \ell\nu\ell\nu$~\cite{ATLAS-CONF-2012-098} (both $7+8\tev$ combination). The predicted signal strength modifiers are set for every analysis to $\mu_1 = 1.0$, $\mu_2 = 0.5$ and $\mu_3 = 0.2$ for the three neutral Higgs bosons, respectively. Note that the experimental mass resolution of the $H\to WW$ search is estimated to $8\gev$, while the $H\to ZZ$ and $H\to \gamma \gamma$ searches have a lower experimental mass uncertainty of $\lesssim 2\gev$. All $\muobs$ plots include the mass region between $120\gev$ and $150\gev$, thus all three Higgs bosons can be tested with all four analyses.

In the first step of the mass-centered $\chi^2$ method, \HS~constructs possible Higgs boson combinations following the \textit{Stockholm clustering scheme}. In our example, $h_2$ and $h_3$ are combined in a Higgs cluster, denoted by $h_{23}$, for the $H\to WW$ analysis since their mass difference is lower than the experimental mass resolution. In all other cases, the Higgs bosons are tested singly, thus we have in total 11 observables. The mass and its uncertainty associated with the Higgs cluster $h_{23}$ are derived from Eq.~\eqref{Eq:mc_cluster_m} and~\eqref{Eq:mc_cluster_dm} to $m_{23} = 137.5\gev$ and $\Delta m_{23} = 1.4\gev$. Its predicted signal strength is $\mu_{23} = 0.7$.

\begin{figure}
\subfigure[Box-shaped parametrization of the theory mass uncertainties. The light gray striped regions show the scanned mass regions $M_i$ of the three Higgs bosons, whereas the darker gray striped region corresponds to $M_k$ of the Higgs boson cluster $k$.\label{Fig:mc_box}]{\includegraphics[width=0.48\textwidth]{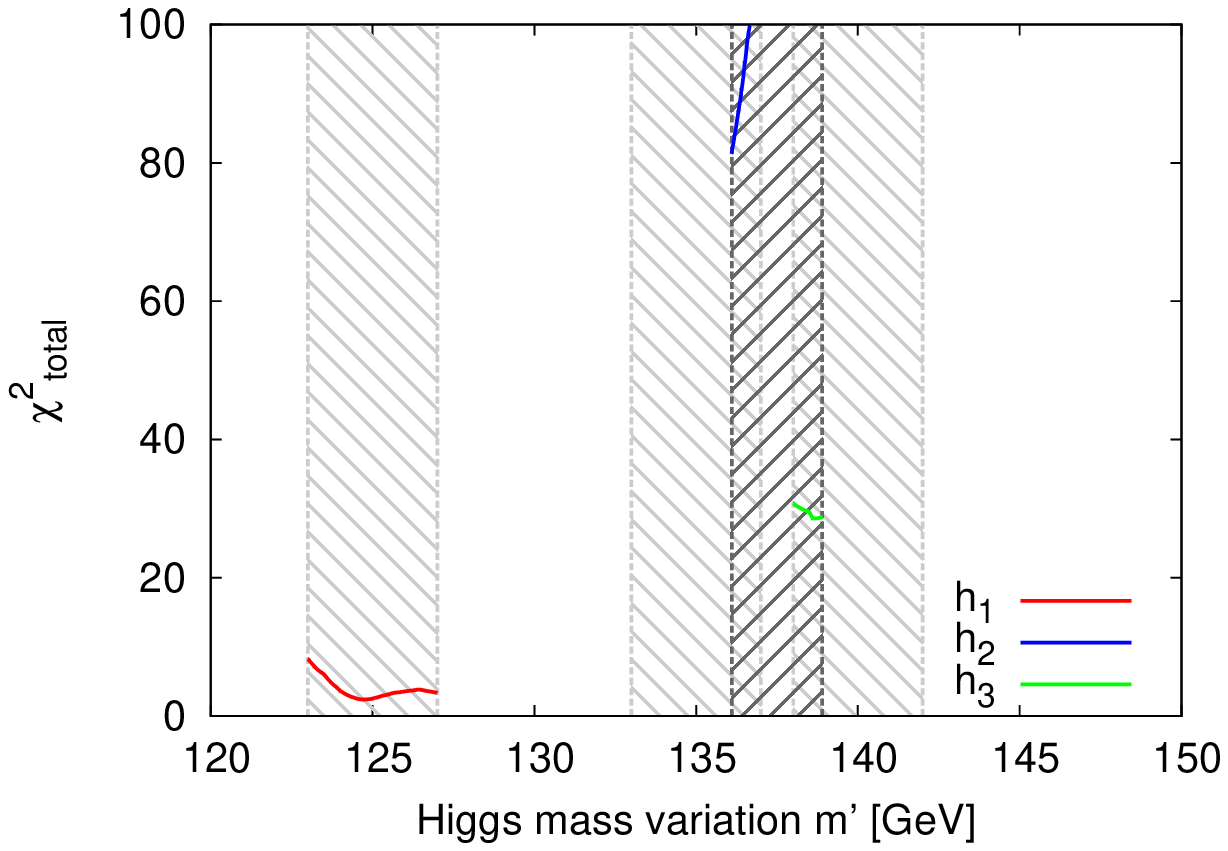}}\hfill
\subfigure[Gaussian parametrization of the theory mass uncertainties. The light gray striped regions now indicate the $\chi^2$ contribution to the tentative total $\chi_i^2$ from the Higgs mass, \cf \refeq{Eq:chisq_dmth_variation_gaussian}.\label{Fig:mc_gauss}]{\includegraphics[width=0.48\textwidth]{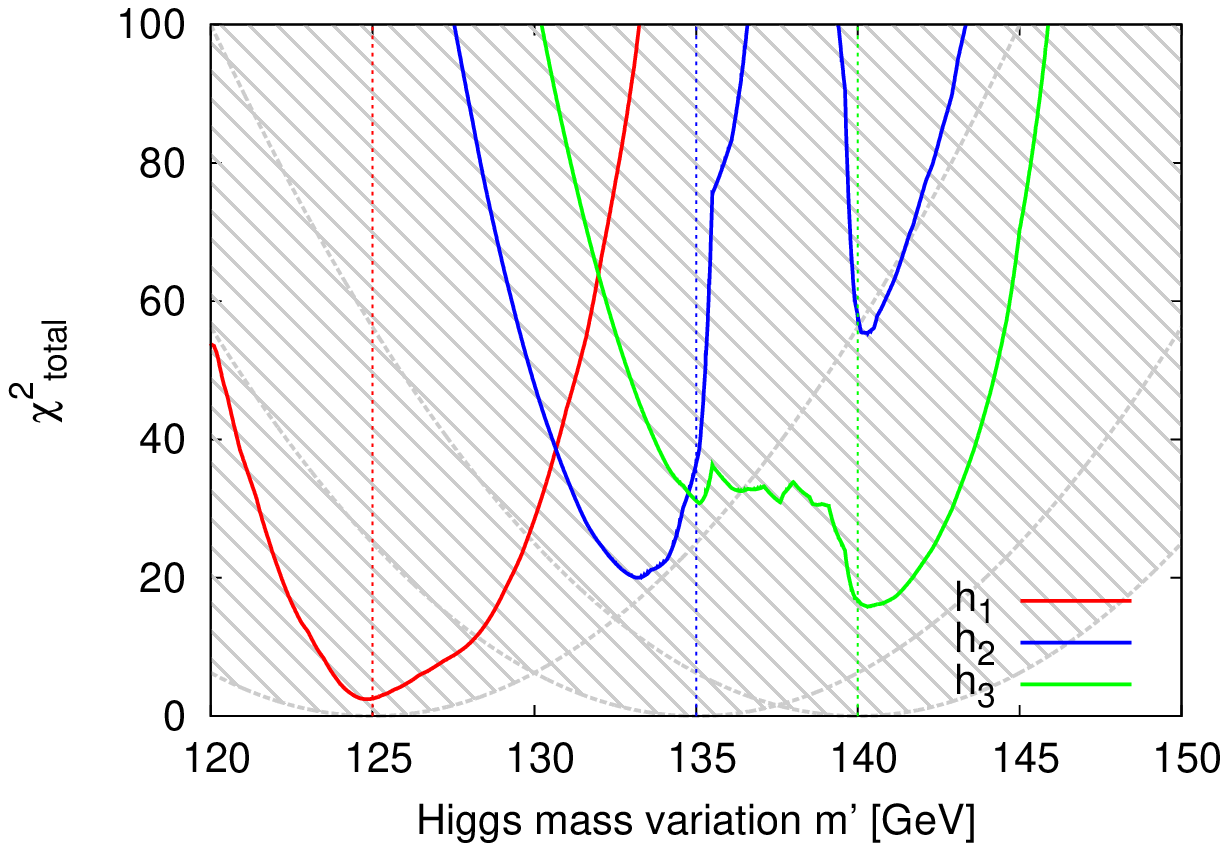}}
\caption{Illustration of the treatment of the theoretical mass uncertainties by variation of the predicted Higgs boson masses (\textit{first option}) for the toy model and observables discussed (see text).
For the $H\to WW$ analysis, $h_2$ and $h_3$ are combined in a Higgs cluster $h_{23}$ with $m_{23}=137.5\gev$ and $\dmth_{23} = 1.4\gev$. We show the tentative total $\chi_i^2(m')$ distributions for each Higgs boson $h_i$ for (a) the box-shaped and (b) the Gaussian parametrization.}
\label{Fig:mcmethod_dmth_variation}
\end{figure}

In the second step, the observed quantities $\muobs_\alpha$ and $\dmuobs_\alpha$ have to be determined from the $\muobs$ plots for each observable $\alpha$. In order to take into account the theoretical mass uncertainties, the relevant mass region is scanned to construct the tentative total $\chi_i^2(m')$ distribution for each Higgs boson $h_i$, as described in Sect.~\ref{Sect:mc_chisq}. For this example, the $\chi_i^2(m')$ distributions for the box-shaped and Gaussian parametrization of the theoretical mass uncertainty are shown in Fig.~A\ref{Fig:mc_box} and Fig.~A\ref{Fig:mc_gauss}, respectively. At the mass position $\mobs_i$, where $\chi^2_i (m')$ is minimal, the observed quantities $\muobs_\alpha$ and $\dmuobs_\alpha$ are extracted from the $\muobs$ plots for those observables $\alpha$, which test the Higgs boson $i$. 

In the box-shaped parametrization, the measured signal strengths of all \textit{mass-centered} observables which test $h_1$ are defined at $\mobs_1 = 124.7\gev$, where $\chi^2_1$ is minimal. In contrast, the Higgs bosons $h_2$ and $h_3$ form the Higgs cluster $h_{23}$ in the $H\to WW$ analysis, therefore their allowed mass variations are restricted to the overlap regions $M_2 \cap M_{23}$ and $M_3 \cap M_{23}$, \cf Eq.~\eqref{Eq:massrange}, respectively. In those observables, where $h_2$ ($h_3$) is tested singly, the measured quantities are defined at $\mobs = 136.1\gev~(138.9\gev)$. For the observable testing the Higgs cluster $h_{23}$ the observable is defined by the minimum of the joint $\chi^2$ distribution, which is located at $\mobs = 138.9\gev$.

In the Gaussian parametrization the mass variation is less restricted. In contrast to the box-shaped parametrization, each mass variation is allowed over the full available mass range of the analyses, however, the additional contribution of the Higgs mass to the tentative $\chi^2$, \cf~Eq.~\eqref{Eq:chisq_dmth_variation_gaussian}, tries to keep the varied mass close the its original predicted value. From the minimum of each tentative $\chi^2$ distribution, the observed quantities of analyses, which test either $h_1$, $h_2$ or $h_3$ singly, are defined at $\mobs = 124.8\gev,~133.2\gev$ and $140.3\gev$, respectively. For the Higgs cluster $h_{23}$ the position $\mobs = 140.3\gev$ is chosen.

\begin{figure}[p]
\centering
\subfigure[Original $\mu$-plot (from \cite{ATLAS-CONF-2012-092}) after the convolution with zero mass theory uncertainty.]{\includegraphics[width=5.5cm, angle=270]{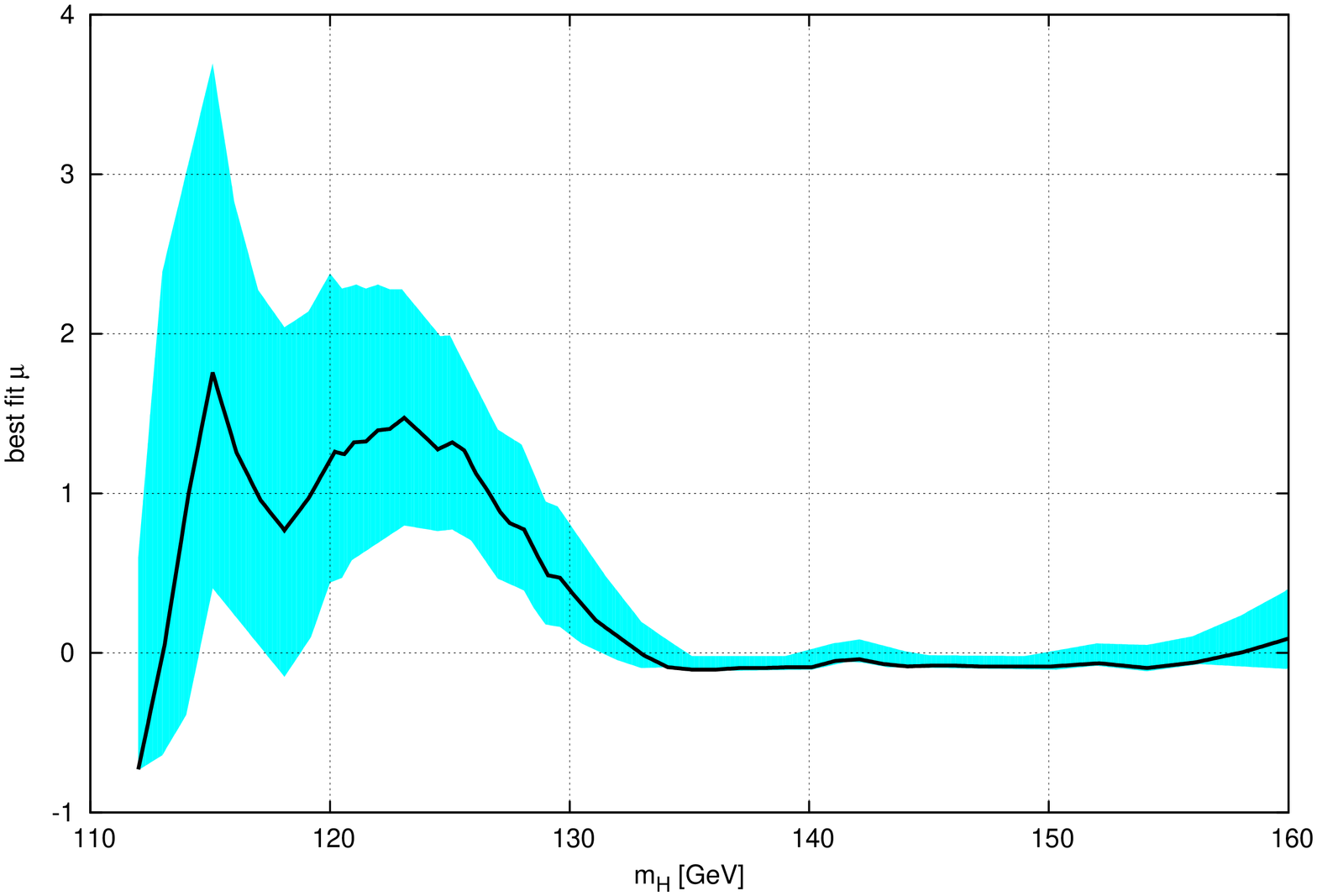}}\hfill
\subfigure[$\mu$-plot after the convolution with a box-shaped mass pdf with $\dmth= 2\gev$.]{\includegraphics[width=5.5cm, angle=270]{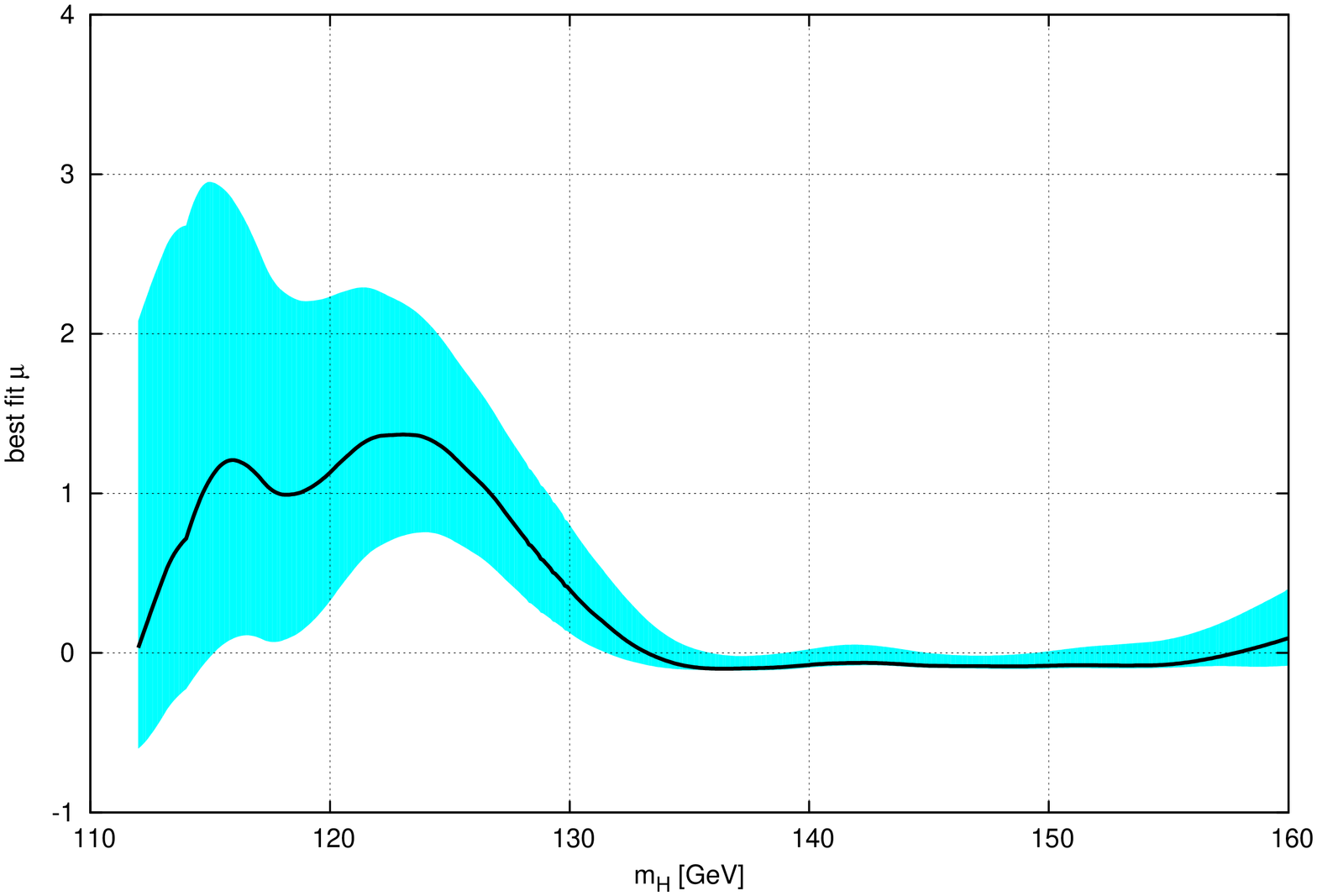}}\hfill
\subfigure[$\mu$-plot after the convolution with a Gaussian mass pdf with $\dmth = 2\gev$.]{\includegraphics[width=5.5cm, angle=270]{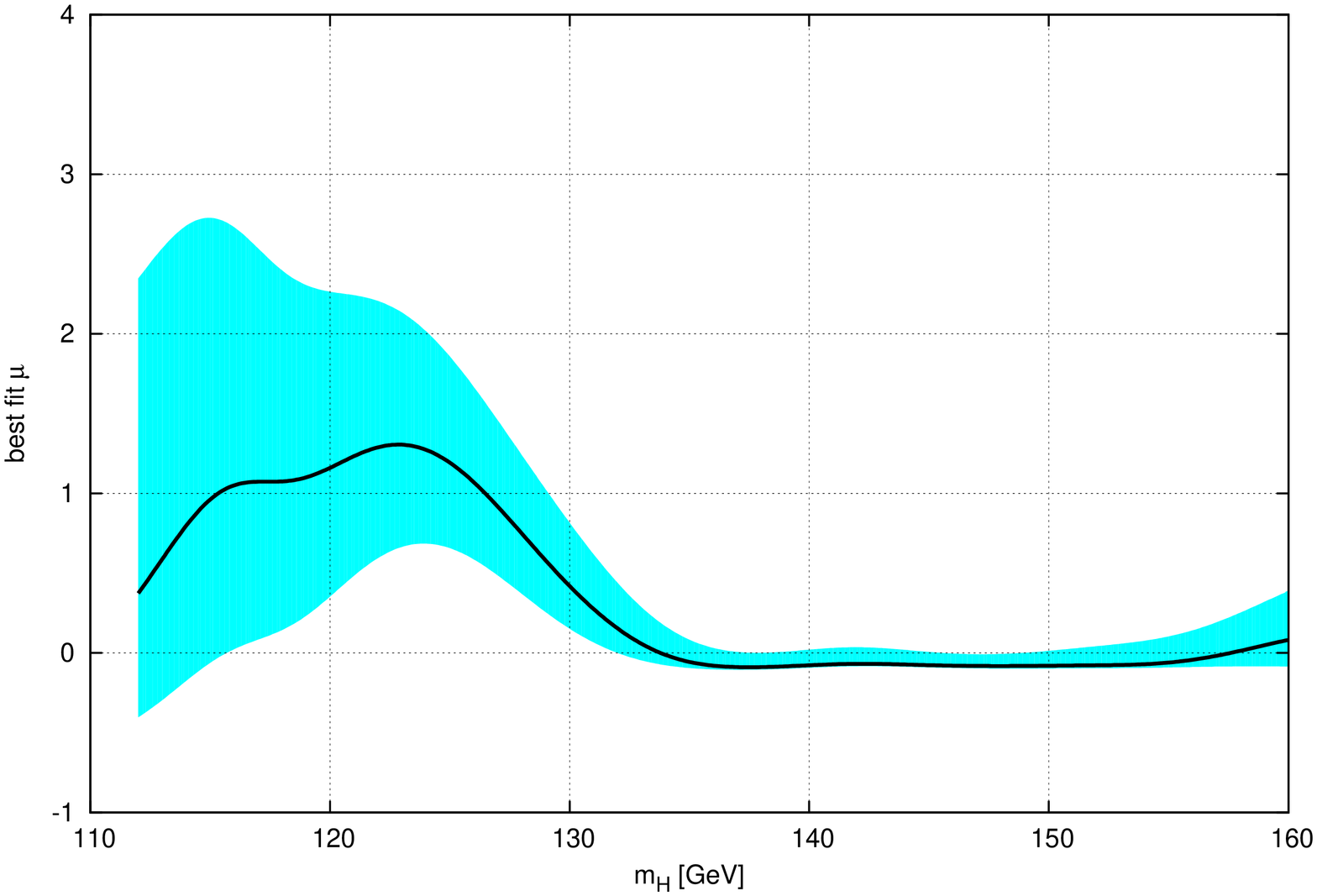}}\hfill
\subfigure[$\mu$-plot after the convolution with a box-shaped mass pdf with $\dmth= 5\gev$.]{\includegraphics[width=5.5cm, angle=270]{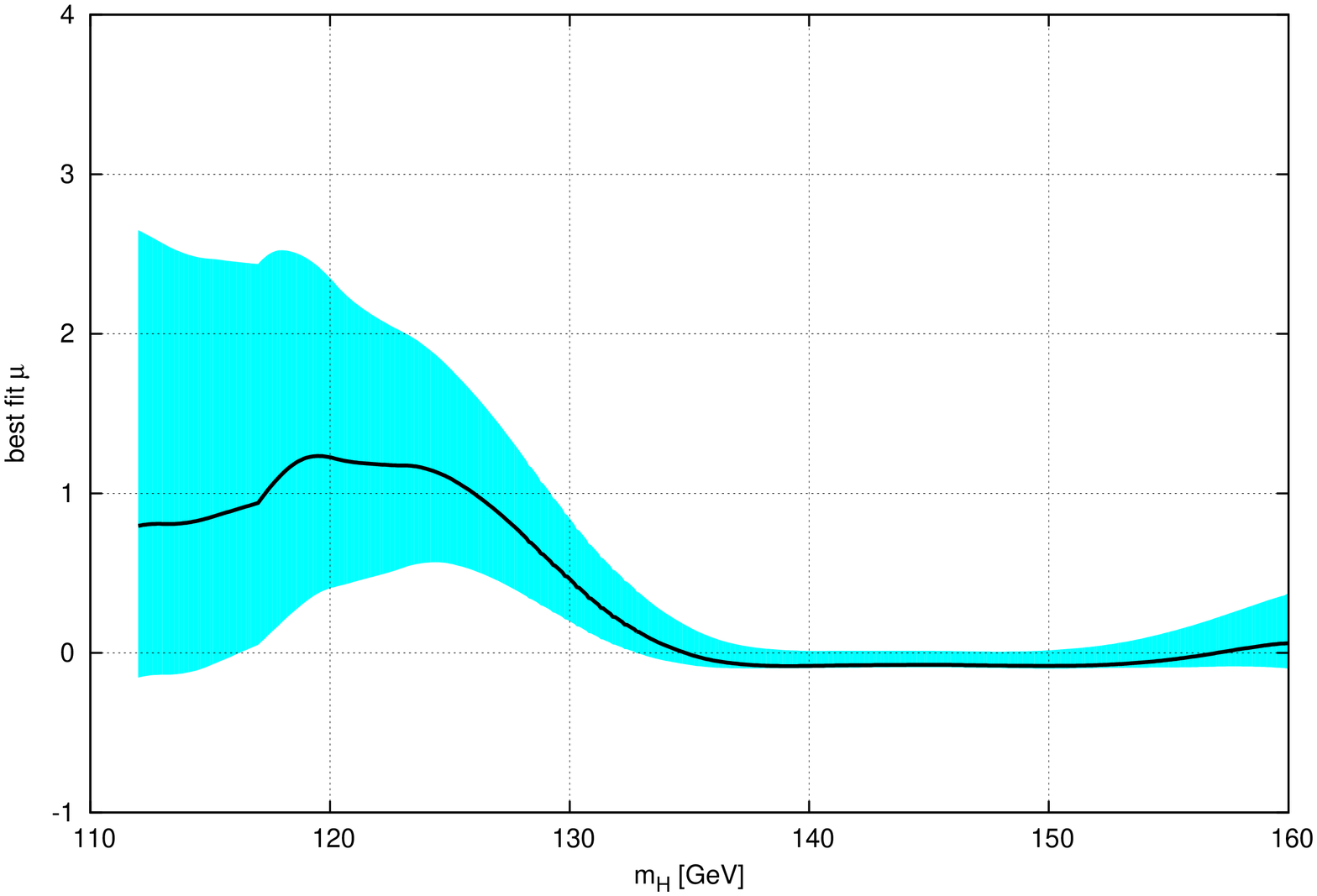}}\hfill
\subfigure[$\mu$-plot after the convolution with a Gaussian mass pdf with $\dmth = 5\gev$.]{\includegraphics[width=5.5cm, angle=270]{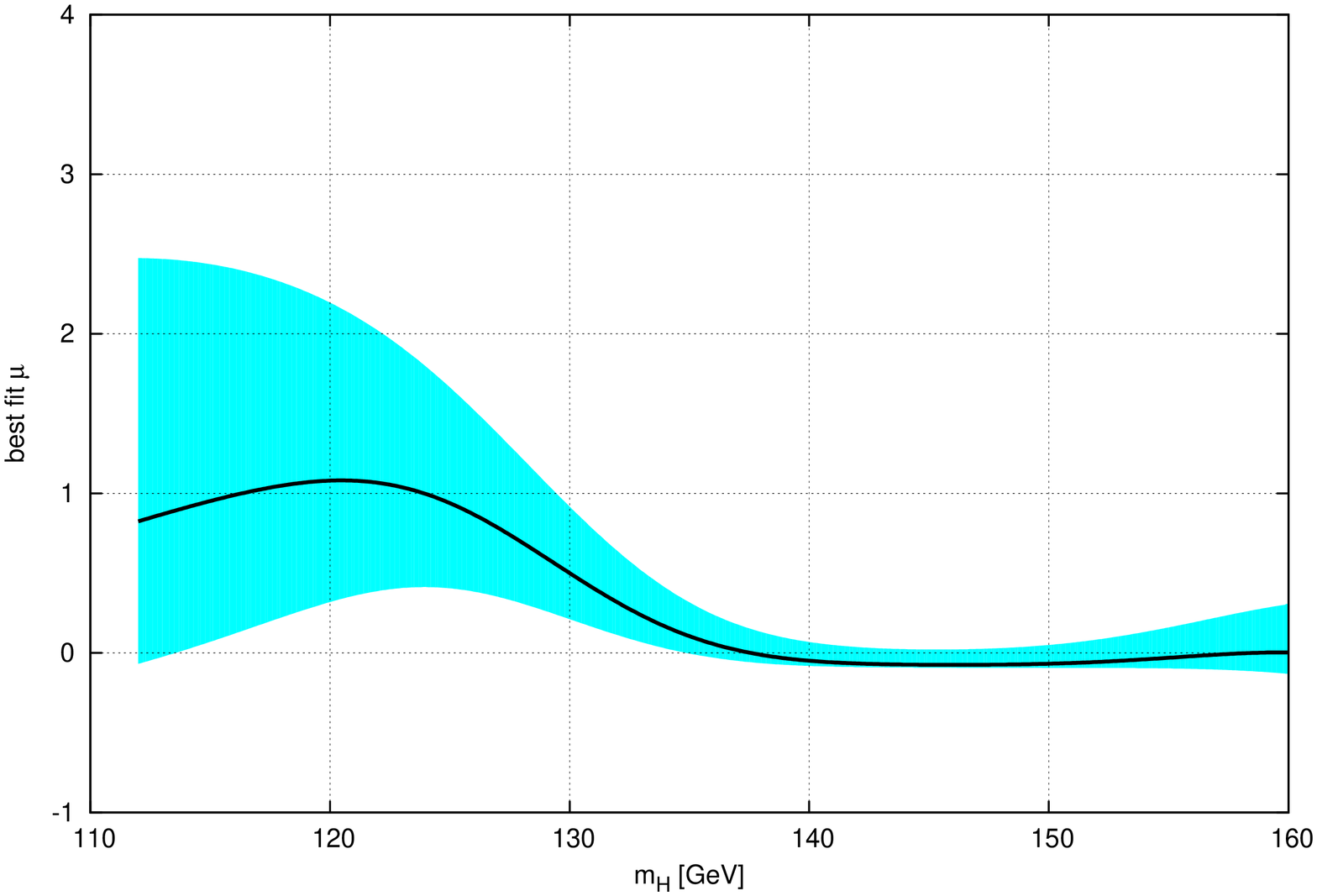}}\hfill
\caption{Plots for the ATLAS $H\to ZZ$ analysis~\cite{ATLAS-CONF-2012-092} after convolution with the Higgs mass pdf for $\dmth=0\gev$ (a), $\dmth=2\gev$ (b),(c), and $\dmth =5\gev$ (d),(e), respectively. In (b) and (d) a uniform (box) pdf is used for the theoretical Higgs mass uncertainty, whereas a Gaussian parametrization was used in (c) and (e).}
\label{Fig:mcmethod_smearing}
\end{figure}

\clearpage
\subsection*{Example 2: Smearing of the $\muobs$-plot with $\dmth$}  
  
We want to illustrate how the experimental data changes, if we choose to fold the theoretical Higgs mass uncertainty, $\dmth$, into the original $\muobs$ plot, as discussed in Sect.~\ref{Sect:mc_chisq}. For this, we look at the $\muobs$ plot published by ATLAS for the $H\to ZZ^{(*)}\to 4\ell$ search~\cite{ATLAS-CONF-2012-092} and convolve it with a uniform (box) or Gaussian Higgs mass pdf, centered at $m_H$, for various theoretical mass uncertainties $\dmth=(0,~2,~5)\gev$, following Eq.~\eqref{Eq:muobssm} and \eqref{Eq:dmuobssm}. This is done over the full mass range, $m_H\in[112,~160]\gev$, to obtain the results shown in Fig.~\ref{Fig:mcmethod_smearing}. For $\dmth=0\gev$, the $\muobs$ plot is unchanged, whereas for increasing $\dmth$ it becomes smoother and fluctuations tend to vanish. This happens faster for the Gaussian pdf.


\section*{References}

\bibliographystyle{JHEP}

\bibliography{HiggsSignals}

\end{document}